\documentclass[aps,pre,twocolumn,superscriptaddress]{revtex4-2}
\usepackage{amsfonts}
\usepackage{graphicx}
\usepackage{epstopdf}
\usepackage{epsfig}
\usepackage{amsmath}
\usepackage[normalem]{ulem}
\usepackage{multirow}
\usepackage{color}
\usepackage{makecell}
\usepackage{array}
\usepackage{booktabs}
\usepackage{mathtools}
\usepackage{bm}
\usepackage{tabularx}

\definecolor{blue}{rgb}{0,0,1}

\definecolor{red}{rgb}{1,0,0}

\newcolumntype{Y}{>{\centering\arraybackslash}X}

\newcommand{\Rmnum}[1]{\expandafter\@slowromancap\romannumeral #1@}
\makeatother

\begin{document}

\title{Digital twins of nonlinear dynamical systems}

\author{Ling-Wei Kong}
\affiliation{School of Electrical, Computer and Energy Engineering, Arizona State University, Tempe, Arizona 85287, USA}

\author{Yang Weng}
\affiliation{School of Electrical, Computer and Energy Engineering, Arizona State University, Tempe, Arizona 85287, USA}

\author{Bryan Glaz}
\affiliation{Vehicle Technology Directorate, CCDC Army Research Laboratory, 2800 Powder Mill Road, Adelphi, MD 20783-1138, USA}

\author{Mulugeta Haile}
\affiliation{Vehicle Technology Directorate, CCDC Army Research Laboratory, 2800 Powder Mill Road, Adelphi, MD 20783-1138, USA}

\author{Ying-Cheng Lai} \email{Ying-Cheng.Lai@asu.edu}
\affiliation{School of Electrical, Computer and Energy Engineering, Arizona State University, Tempe, Arizona 85287, USA}
\affiliation{Department of Physics, Arizona State University, Tempe, Arizona 85287, USA}

\date{\today}

\begin{abstract}

% 150 words
We articulate the design imperatives for machine-learning based digital twins for nonlinear dynamical systems subject to external driving, which can be used to monitor the ``health'' of the target system and anticipate its future collapse. We demonstrate that, with single or parallel reservoir computing configurations, the digital twins are capable of challenging forecasting and monitoring tasks. Employing prototypical systems from climate, optics and ecology, we show that the digital twins can extrapolate the dynamics of the target system to certain parameter regimes never experienced before, make continual forecasting/monitoring with sparse real-time updates under non-stationary external driving, infer hidden variables and accurately predict their dynamical evolution, adapt to different forms of external driving, and extrapolate the global bifurcation behaviors to systems of some different sizes. These features make our digital twins appealing in significant applications such as monitoring the health of critical systems and forecasting their potential collapse induced by environmental changes.

\end{abstract}

\date{\today}

\maketitle

\section{Introduction} \label{sec:intro}

The concept of digital twins originated from aerospace engineering for aircraft
structural life prediction~\cite{TIES:2011}. In general, a digital twin can be 
used for predicting dynamical systems and generating solutions of emergent 
behaviors that can potentially be catastrophic~\cite{TQ:2019}. Digital twins 
have attracted a great deal of attention from a wide range of 
fields~\cite{RSK:2020} including medicine and health 
care~\cite{BSvdH:2018,SWBB:2020}. For example, the idea of developing medical 
digital twins in viral infection through a combination of mechanistic 
knowledge, observational data, medical histories, and artificial intelligence 
has been proposed recently~\cite{LSG:2021}, which can potentially lead to 
a powerful addition to the existing tools to combat future pandemics.
In a more dramatic development, the European Union plans to fund the 
development of digital twins of Earth for its green 
transition~\cite{Voosen:2020,BSH:2021}.

The physical world is nonlinear. Many engineering systems, such as complex
infrastructural systems, are governed by nonlinear dynamical rules, too. In 
nonlinear dynamics, various bifurcations leading to chaos and system
collapse can take place~\cite{LT:book}. For example, in ecology, environmental
deterioration caused by global warming can lead to slow parameter drift towards
chaos and species extinction~\cite{MY:1994,HACFGLMPSZ:2018}. In an electrical 
power system, voltage collapse can occur after a parameter shift that lands
the system in transient chaos~\cite{DL:1999}. The various climate systems in 
different geographic regions of the world are also nonlinear and the emergent
catastrophic behaviors as the result of increasing human activities are of 
grave concern. In all these cases, it is of interest to develop a digital twin 
of the system of interest to monitor its ``health'' in real time as well as for 
predictive problem solving in the sense that, if the digital twin indicates
a possible system collapse in the future, proper control strategies should and
can be devised and executed in time to prevent the collapse. 

What does it take to create a digital twin for a nonlinear dynamical system? 
For natural and engineering systems, there are two general approaches: one 
is based on mechanistic knowledge and another is based on observational data. 
In principle, if the detailed physics of the system is well understood, it 
should be possible to construct a digital twin through mathematical modeling. 
However, there are two difficulties associated with this modeling approach. 
First, a real-world system can be high-dimensional and complex, preventing the 
rules governing its dynamical evolution from being known at a sufficiently 
detailed level. Second, the hallmark of chaos is sensitive dependence on 
initial conditions. Because no mathematical model of the underlying physical 
system can be perfect, the small deviations and high dimensionality of the 
system coupled with environmental disturbances can cause the model predictions 
of the future state of the system to be inaccurate and completely 
irrelevant~\cite{LGK:1999,LG:1999}. These difficulties motivate the 
proposition that data-based approach can have advantages in many realistic
scenarios and a viable method to develop a digital twin is
through data.
While in certain cases, approximate system equations can be found
from data through sparse optimization~\cite{WYLKG:2011,WLG:2016,Lai:2021}, the 
same difficulties with the modeling approach arise. These considerations have 
led us to exploit machine learning to create digital twins for nonlinear 
dynamical systems.

Given a nonlinear dynamical system, its digital twin is also a dynamical system,
rendering appropriate exploitation of recurrent neural networks that can be 
designed to generate self-dynamical evolution with memory. In this regard,
reservoir computers (RC)~\cite{Jaeger:2001,MNM:2002,JH:2004} that have been extensively studied in recent years~\cite{HSRFG:2015,LBMUCJ:2017,PLHGO:2017,LPHGBO:2017,PHGLO:2018,Carroll:2018,NS:2018,ZP:2018,GPG:2019,JL:2019,TYHNKTNNH:2019,FJZWL:2020,ZJQL:2020,KKGGM:2020,KFGL:2021a,PCGPO:2021,KLNPB:2021,FKLW:2021,KFGL:2021b,Bollt:2021,GBGB:2021,HR:2021,Carroll:2022optimizing}
provide a starting point, which can be trained from observational data to 
generate closed-loop dynamical evolution that follows the evolution of the 
target system for a finite amount of time. Another advantage of RC is that
no back-propagation is needed for optimizing the parameters - only a linear
regression is required in the training so it is computationally efficient. A
common situation is that the target system is subject to external driving,
such as a driven laser, a regional climate system, or an ecosystem under
external environmental disturbances. Accordingly, the digital twin must
accommodate a mechanism to control or steer the dynamics of the RC
neural network to account for the external driving. Introducing a control
mechanism into the RC structure with an exogenous control signal
acting directly onto the RC network distinguishes our work from existing
ones in the literature of RC as applied to nonlinear dynamical systems. 
Of particular interest is whether the collapse of the target chaotic system
can be anticipated from the digital twin. The purpose of this paper is to
demonstrate that the digital twin so created can accurately produce the
bifurcation diagram of the target system and faithfully mimic its dynamical 
evolution from a statistical point of view. The digital twin can then be used 
to monitor the present and future ``health'' of the system. More importantly, 
with proper training from observational data the twin can reliably anticipate 
system collapses, providing early warnings of potentially catastrophic failures
of the system.

More specifically, using three prototypical systems from optics, ecology, and 
climate, respectively, we demonstrate that the RC based digital twins
developed in this paper solve the following challenging problems: 
(1) extrapolation of the dynamical evolution of the target system into 
certain ``uncharted territories'' in the parameter space, (2) long-term 
continual forecasting of nonlinear dynamical systems subject to non-stationary 
external driving with sparse state updates, (3) inference of hidden variables 
in the system and accurate prediction of their dynamical evolution into the 
future, (4) adaptation to external driving of different waveform, and 
(5) extrapolation of the global bifurcation behaviors of network systems to 
some different sizes. These features make our digital twins appealing in 
applications.

\section{Methods} \label{sec:methods}

\begin{figure*}[htbp]
\centering
\includegraphics[width=\linewidth]{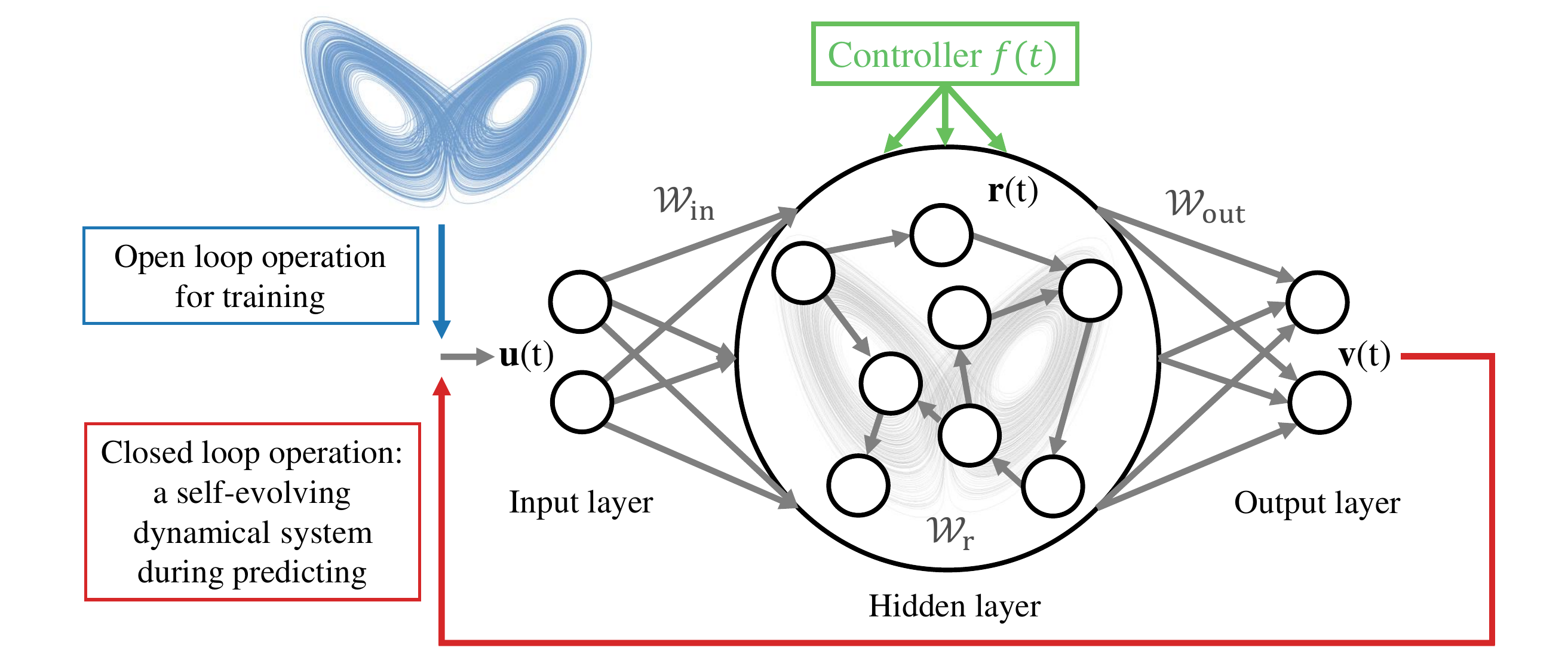}
\caption{Basic structure of the digital twin of a chaotic system. It consists 
of three layers: the input layer, the hidden recurrent layer, an output layer, 
as well as a controller component. The input matrix $\mathcal{W}_\text{in}$
maps the $D_\text{in}$-dimensional input chaotic data to a vector of much higher
dimension $D_r$, where $D_r \gg D_\text{in}$. The recurrent hidden layer is
characterized by the $D_r\times D_r$ weighted matrix $\mathcal{W}_r$. The
dynamical state of the $i^{th}$ neuron in the reservoir is $r_i$, for 
$i=1,\ldots,D_r$. The hidden-layer state vector is $\mathbf{r}(t)$, which is 
an embedding of the input~\cite{HHD:2020}. The output matrix 
$\mathcal{W}_{\rm out}$ readout the hidden state into the 
$D_\text{out}$-dimensional output vector. The controller provides an external 
driving signal $f(t)$ to the neural network. During training, the vector 
$\mathbf{u}(t)$ is the input data, and the blue arrow exists during the 
training phase only. In the predicting phase, the output vector $\mathbf{v}(t)$
is directly fed back to the input layer, generating a closed-loop, 
self-evolving dynamical system, as indicated by the red arrow connecting 
$\mathbf{v}(t)$ to $\mathbf{u}(t)$. The controller remains on in both the 
training and predicting phases.}
\label{fig:DT}
\end{figure*}

The basic construction of the digital twin of a nonlinear dynamical 
system~\cite{codes} is illustrated in Fig.~\ref{fig:DT}. It is essentially a 
recurrent RC neural network with a control mechanism, which requires two
types of input signals: the observational time series for training and the
control signal $f(t)$ that remains in both the training and self-evolving phase.
The hidden layer hosts a random or complex network of artificial neurons.
During the training, the hidden recurrent layer is driven by both the input
signal $\mathbf{u}(t)$ and the control signal $f(t)$. The neurons in the
hidden layer generate a high-dimensional nonlinear response signal. Linearly
combining all the responses of these hidden neurons with a set of trainable
and optimizable parameters yields the output signal. 
Specifically, the digital twin consists of four components: (i) an input 
subsystem that maps the low-dimensional ($D_\text{in}$) input signal into a 
(high) $D_r$-dimensional signal through the weighted $D_r\times D_\text{in}$ 
matrix $\mathcal{W}_\text{in}$, (ii) a reservoir network of $N$ neurons 
characterized by $\mathcal{W}_r$, a weighted network matrix of dimension 
$D_r\times D_r$, where $D_r \gg D_\text{in}$, (iii) an readout subsystem that 
converts the $D_r$-dimensional signal from the reservoir network into an 
$D_\text{out}$-dimensional signal through the output weighted matrix 
$\mathcal{W}_{out}$, and (iv) a controller with the matrix $\mathcal{W}_c$. 
The matrix $\mathcal{W}_r$ defines the structure of the reservoir neural 
network in the hidden layer, where the dynamics of each node are described 
by an internal state and a nonlinear hyperbolic tangent activation function.

The matrices $\mathcal{W}_\text{in}$, $\mathcal{W}_c$, and $\mathcal{W}_r$
are generated randomly prior to training, whereas all elements of 
$\mathcal{W}_\text{out}$ are to be determined through training. Specifically,
the state updating equations for the training and self-evolving phases are,
respectively,
\begin{align}
\mathbf{r}(t+&\Delta t) = (1-\alpha)\mathbf{r}(t) \nonumber \\ 
&+ \alpha \tanh{[\mathcal{W}_r\mathbf{r}(t)+\mathcal{W}_\textrm{in}\mathbf{u}(t) +\mathcal{W}_c f(t)]}, \label{eq1} \\ 
\mathbf{r}(t+&\Delta t) = (1-\alpha)\mathbf{r}(t) \nonumber \\
&+ \alpha \tanh{[\mathcal{W}_r\mathbf{r}(t)+\mathcal{W}_\textrm{in}\mathcal{W}_\textrm{out}\mathbf{r'}(t)+\mathcal{W}_c f(t)]}, \label{eq2}
\end{align}
where $\mathbf{r}(t)$ is the hidden state, $\mathbf{u}(t)$ is the vector of 
input training data, $\Delta t$ is the time step, the vector 
$\tanh{(\mathbf{p})}$ is defined to be $[\tanh{(p_1)},\tanh{(p_2)},\ldots]^T$ 
for a vector $\mathbf{p} = [p_1,p_2,...]^T$, and $\alpha$ is the leakage factor.
During the training, several trials of data are typically used under different 
driving signals so that the digital twin can ``sense, learn, and mingle'' the 
responses of the target system to gain the ability to extrapolate a response
to a new driving signal that has never been encountered before. We input these
trials of training data, i.e., a few pairs of $\mathbf{u}(t)$ and the associated $f(t)$, 
through the matrices $\mathcal{W}_\text{in}$ and $\mathcal{W}_c$ sequentially. 
Then we record the state vector $\mathbf{r}(t)$ of the neural network during 
the entire training phase as a matrix $\mathcal{R}$. We also record all the 
desired output, which is the one-step prediction result 
$\mathbf{v}(t)=\mathbf{u}(t+\Delta t)$, as the matrix $\mathcal{V}$. To make 
the readout nonlinear and to avoid unnecessary symmetries in the 
system~\cite{LPHGBO:2017,HR:2020}, we change the matrix $\mathcal{R}$ into 
$\mathcal{R}'$ by squaring the entries of even dimensions in the states of 
the hidden layer. [The vector ($\mathbf{r'}(t)$ in Eq.~\eqref{eq2} is defined 
in a similar way.] We carry out a linear regression between $\mathcal{V}$ and 
$\mathcal{R}'$, with a $\ell$-2 regularization coefficient $\beta$, to 
determine the readout matrix: 
\begin{align}
\mathcal{W}_\textrm{out} = \mathcal{V} \cdot \mathcal{R}'^T(\mathcal{R}'\cdot \mathcal{R}'^T+\beta \mathcal{I})^{-1}.
\end{align}
To achieve acceptable learning performance, optimization of hyperparameters
is necessary. The four widely used global optimization methods are genetic 
algorithm~\cite{Goldberg:2006,CGT:1991,CGT:1997}, particle swarm 
optimization~\cite{KE:1995,MC:2011}, Bayesian 
optimization~\cite{GSA:2014,SLA:2012}, and surrogate 
optimization~\cite{Gutmann:2001,RS:2007,WS:2014}.
We use the surrogate optimization (the algorithm \textit{surrogateopt} in 
Matlab). The hyperparameters that are optimized include $d$ - the average 
degree of the recurrent network in the hidden layer, $\lambda$ - the spectral 
radius of the recurrent network, $k_{in}$ - the scaling factor of 
$\mathcal{W}_\textrm{in}$, $k_c$ - the scaling of $\mathcal{W}_\textrm{c}$,
$c_0$ - the bias in Eq.~\eqref{eq1} and \eqref{eq2}, $\alpha$ - the leakage 
factor, and $\beta$ - the $\ell$-2 regularization coefficient.
In this paper, the validation of the RC networks are done with the same driving
signals $f(t)$ as in the training data. We test driving signals $f(t)$ that are
different from those generating the training data (e.g., with different 
amplitude, frequency, or waveform). To generate the predicted bifurcation
diagrams, we let the RC networks make predictions for long enough periods
to approach the asymptotic behavior. During the warming-up process to 
initialize the RC networks prior to making the predictions, we feed randomly
chosen short segments of the training time series to feed into the RC network. 
That is, no data from the target system under the testing driving signals 
$f(t)$ are required for making the predictions.

\section{Results} \label{sec:results_1}

For clarity, we present results on the digital twin for a prototypical 
nonlinear dynamical systems with adjustable phase-space dimension: the 
Lorenz-96 climate network model~\cite{Lorenz:1996}. In the appendix, 
we present two additional examples: a chaotic laser 
(Appendix A) and a driven ecological system (Appendix B), 
together with a number of pertinent issues.

\begin{figure*}[ht!bp]
\centering
\includegraphics[width=0.95\linewidth]{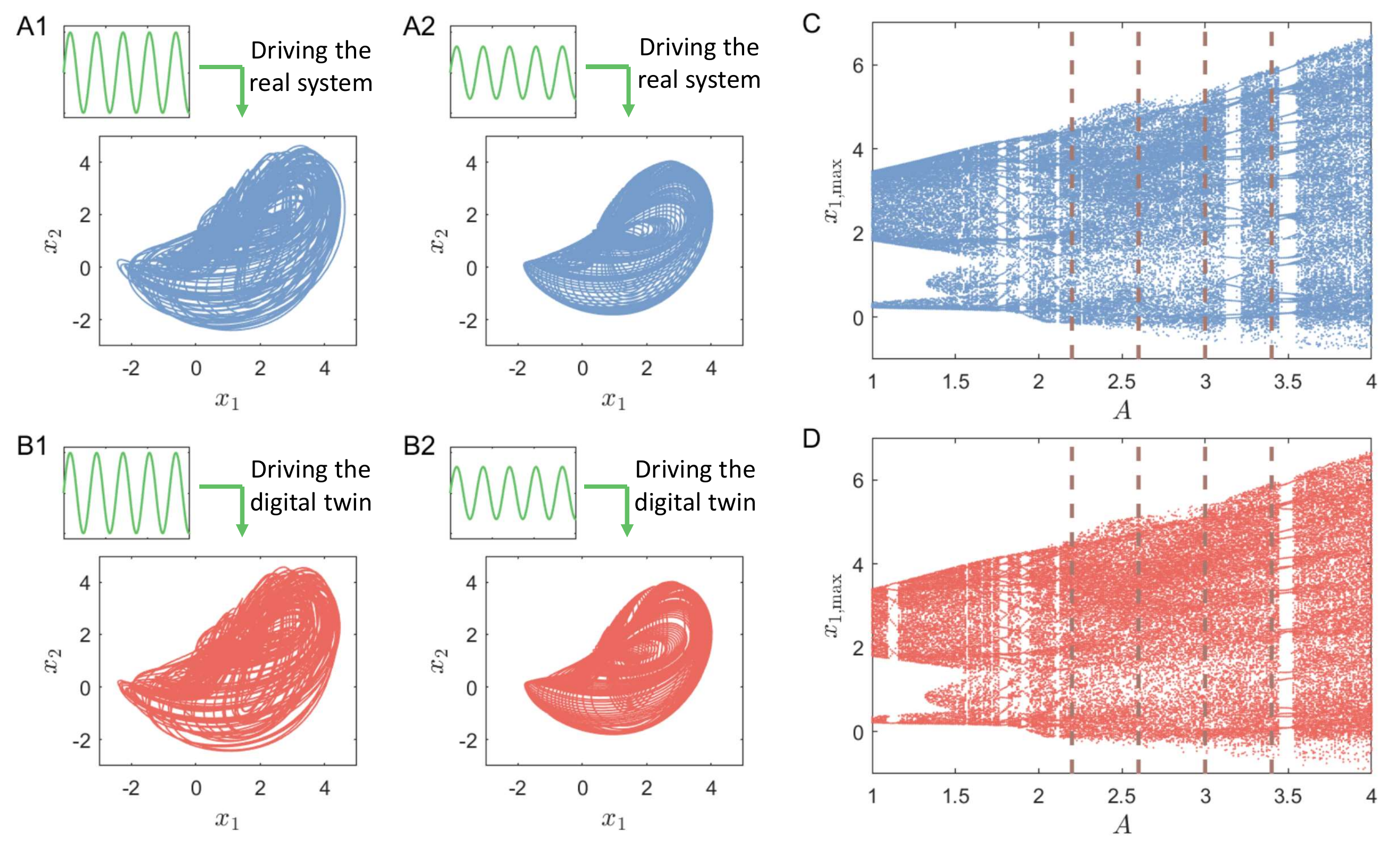}
\caption{Digital twin of the Lorenz-96 climate system. The toy climate system 
is described by six coupled first-order nonlinear differential equations 
(phase-space dimension $m=6$), which is driven by a sinusoidal signal 
$f(t)= A\sin(\omega t)+F$. (A1,A2) Ground truth: chaotic and quasi-periodic 
dynamics in the system for $A=2.2$ and $A=1.6$, respectively, for $\omega = 2$ 
and $F = 2$. The sinusoidal driving signals $f(t)$ are schematically 
illustrated. (B1, B2) The corresponding dynamics of the digital twin under the 
same driving signal $f(t)$. Training of the digital twin is conducted using 
time series from the chaotic regime. The result in (B2) indicates that the 
digital twin is able to extrapolate outside the chaotic regime to generate 
the unseen quasi-periodic behavior. (C, D) True and digital-twin generated 
bifurcation diagrams of the toy climate system, where the four vertical red 
dashed lines indicate the values of driving amplitudes $A$, from which the 
training time series data are obtained. The remarkable agreement between the 
two bifurcation diagrams attests to the strong ability of the digital twin to 
reproduce the distinct dynamical behaviors of the target climate system in 
different parameter regimes, even with training data only in the chaotic 
regime. Note that there are mismatches in the details such as the positions 
of some periodic windows.}
\label{fig:m6}
\end{figure*}

\subsection{A low-dimensional Lorenz-96 climate network and its digital twin} \label{subsec:Lorenz96_LD}

The Lorenz-96 system~\cite{Lorenz:1996} is an idealized atmospheric climate 
model. Mathematically, the toy climate system is described by $m$ coupled 
first-order nonlinear differential equations subject to external periodic 
driving $f(t)$: 
\begin{align} \label{eq:Lorenz_96}
\frac{dx_i}{dt} = x_{i-1}(x_{i+1}-x_{i-2}) - x_i + f(t),
\end{align}
where $i=1,\ldots,m,$ is the spatial index. Under the periodic boundary 
condition, the $m$ nodes constitute a ring network, where each node is coupled
to three neighboring nodes. To be concrete, we set $m = 6$ (more complex 
high-dimensional cases are treated below).
%in Secs.~\ref{subsec:Lorenz96_HD} and \ref{subsec:Lorenz96_DD}). 
The driving force is sinusoidal with a bias $F$:
$f(t)=A\sin(\omega t)+F$. We fix $\omega = 2$ and $F=2$, and use the forcing
amplitude $A$ as the bifurcation parameter. For relatively large values of $A$,
the system exhibits chaotic behaviors, as exemplified in Fig.~\ref{fig:m6}(A1)
for $A = 2.2$. Quasi-periodic dynamics arise for smaller values of $A$, as
exemplified in Fig.~\ref{fig:m6}(A2). As $A$ decreases from a large value, a 
critical transition from chaos to quasi-periodicity occurs at $A_c \approx 1.9$.
We train the digital twin with time series from four values of $A$, all in the 
chaotic regime: $A=2.2, 2.6, 3.0,$ and $3.4$. The size of the random reservoir 
network is $D_r=1,200$. For each value of $A$ in the training set, the training
and validation lengths are $t=2,500$ and $t=12$, respectively, where the latter
corresponds to approximately five Lyapunov times. The warming-up length is 
$t=20$ and the time step of the reservoir dynamical evolution is 
$\Delta t=0.025$. The hyperparameter values (See Sec.~\ref{sec:methods}
for their meanings) are optimized to be $d=843$, $\lambda=0.48$, 
$k_\textrm{in}=0.29$, $k_c=0.113$, $\alpha=0.41$, and $\beta=1\times10^{-10}$. 
Our computations reveal that, for the deterministic version of the Lorenz-96 
model, it is difficult to reduce the validation error below a small threshold. 
However, adding an appropriate amount of noise into the training time 
series~\cite{Jaeger:2001} can lead to smaller validation errors. We add
an additive Gaussian noise with standard deviation $\sigma_{\rm noise}$ to
each input data channel to the reservoir network [including the driving channel
$f(t)$]. The noise amplitude $\sigma_{\rm noise}$ is treated as an additional
hyperparameter to be optimized. For the toy climate system, we test several
noise levels and find the optimal noise level giving the best validating
performance: $\sigma_{\rm noise} \approx 10^{-3}$.

Figures~\ref{fig:m6}(B1) and \ref{fig:m6}(B2) show the dynamical behaviors 
generated by the digital twin for the same values of $A$ as in 
Figs.~\ref{fig:m6}(A1) and \ref{fig:m6}(A2), respectively. It can be seen that 
not only does the digital twin produce the correct dynamical behavior in the 
same chaotic regime where the training is carried out, it can also 
extrapolate beyond the training parameter regime to correctly predict the unseen
system dynamics there (quasiperiodicity in this case). To provide support in 
a broader parameter range, we calculate true bifurcation diagram, as shown in 
Fig.~\ref{fig:m6}(C), where the four vertical dashed lines indicate the four 
values of the training parameter. The bifurcation generated by the digital twin
is shown in Fig.~\ref{fig:m6}(D), which agrees remarkably well with the true 
diagram even at a detailed level. Note that there are mismatches in the 
details such as the positions of some periodic windows in Figs.~\ref{fig:m6}(C)
and \ref{fig:m6}(D). To predict all the features in a bifurcation diagram 
requires extensive interpolation and extrapolation of the system dynamics in 
the phase space. 

% Extrapolation can be a dangerous topic. It is considered to be extremely
% difficult for many researchers. We need more supporting if we want to
% declare out model has certain extrapolation capability, and avoid possible
% troubles with the referees.
Previously, it was suggested that RC can have a certain degree of 
extrapolability~\cite{KKGGM:2020,KFGL:2021a,PCGPO:2021,KLNPB:2021,FKLW:2021,KFGL:2021b}.
Figure~\ref{fig:m6} represents an example where the target system's response
is extrapolated to external sinusoidal driving with unseen amplitudes. In 
general, extrapolation is a difficult problem. Some limitations of the 
extrapolability with respect to the external driving signal is discussed in 
Appendix A, where the digital twin can predict the crisis point but cannot 
extrapolate the asymptotic behavior after the crisis.

In the following, we systematically study the applicability of the digital 
twin in solving forecasting problems in more complicated situations than the 
basic settings demonstrated in Fig.~\ref{fig:m6}. The issues to be addressed 
are high dimensionality, the effect of the waveform of the driving on 
forecasting, and the generalizability across Lorenz-96 networks of different
sizes. Results of continual forecasting and inferring hidden dynamical 
variables using only rare updates of the observable are presented in Appendices
C and D, respectively.

\begin{figure*}[ht!bp]
\centering
\includegraphics[width=0.9\linewidth]{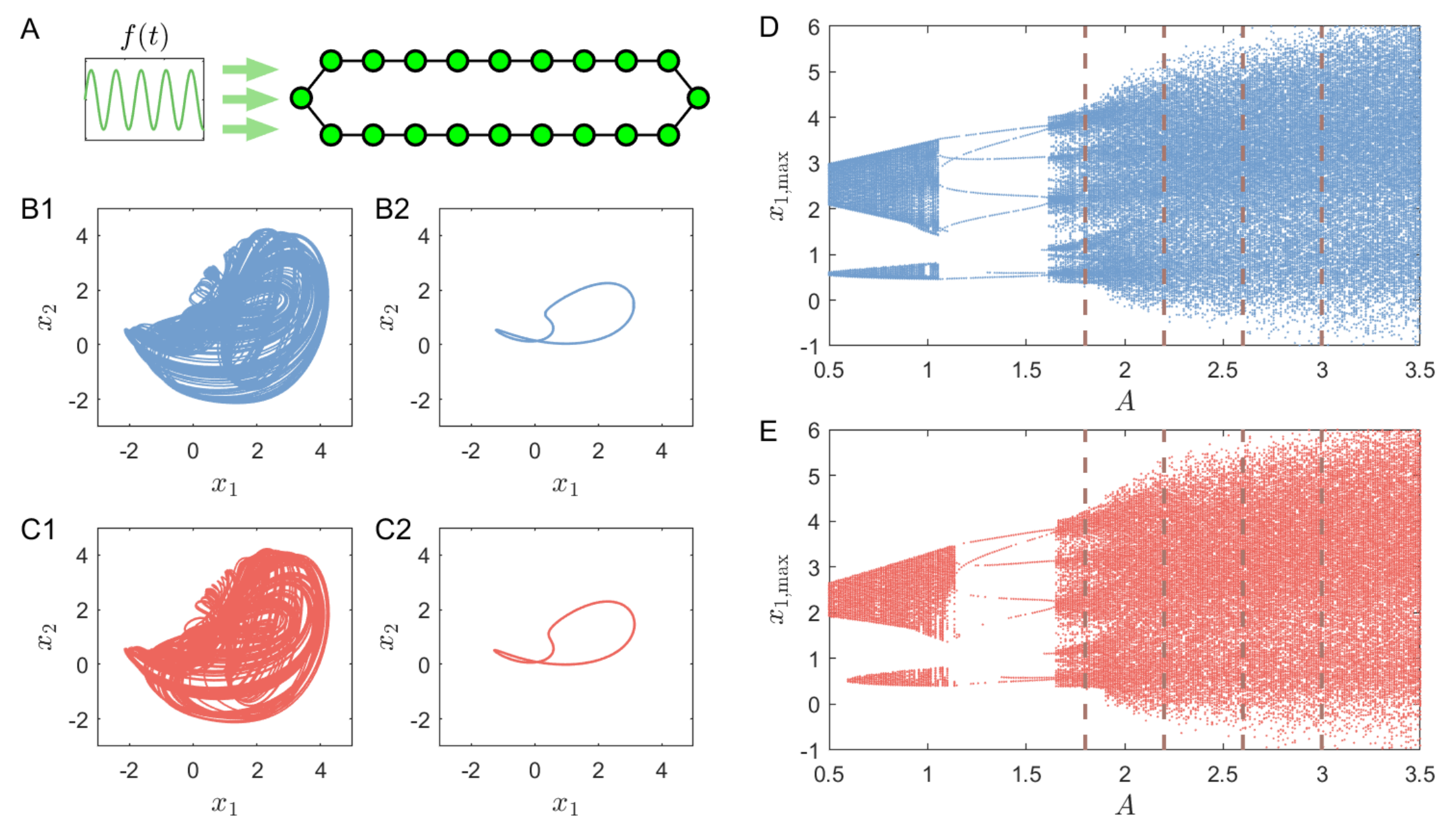}
\caption{Digital twin consisting of a number of parallel RC neural networks
for high-dimensional chaotic systems. The target system is the Lorenz-96
climate network of $m=20$ nodes, subject to a global periodic driving 
$f(t)=A\sin(\omega t) + F$. (A) The structure of the digital twin, where each
filled green circle represents a small RC network with the input 
dimension $D_{in} = 5$ and output dimension $D_\textrm{out}=2$. (B1, B2) 
A chaotic and periodic attractor in a two-dimensional subspace of the target
system for $A = 1.8$ and $A=1.6$, respectively, for $\omega = 2$ and 
$F = 2$. (C1, C2) The attractors generated by the digital twin corresponding
to those in (B1, B2), respectively, where the training is done using four time 
series from four different values of forcing amplitude $A$, all in the chaotic
regime. The digital twin with a parallel structure is able to successfully 
extrapolate the unseen periodic behavior with completely chaotic training data.
(D, E) The true and digital-twin generated bifurcation diagrams, respectively,
where the four vertical dashed lines in (c) specify the four values of $A$ 
from which the training time series are obtained. The remarkable agreement 
between the two bifurcation diagrams indicates that the digital twin so trained
can faithfully generate the dynamical behaviors of the high-dimensional target 
system.}
\label{fig:m20}
\end{figure*}

\subsection{Digital twins of parallel RC neural networks for high-dimensional Lorenz-96 climate networks} \label{subsec:Lorenz96_HD}

We extend the methodology of digital twin to high-dimensional Lorenz-96 climate
networks, e.g., $m=20$. To deal with such a high-dimensional target system, if 
a single reservoir system is used, the required size of the neural network in 
the hidden layer will be too large to be computationally efficient. We thus 
turn to the parallel configuration~\cite{PHGLO:2018} that consists of many
small-size RC networks, each ``responsible'' for a small part of the target
system. For the Lorenz-96 network with $m=20$ coupled nodes, our digital
twin consists of ten parallel RC networks, each monitoring and forecasting
the dynamical evolution of two nodes ($D_\textrm{out}=2$). Because each node 
in the Lorenz-96 network is coupled to three nearby nodes, we set 
$D_\textrm{in} = D_\textrm{out} + D_\textrm{couple} = 2+3= 5$ to ensure that 
sufficient information is supplied to each RC network. 

The specific parameters of the digital twin are as follows. The size of the 
recurrent layer is $D_r=1,200$. For each training value of the forcing 
amplitude $A$, the training and validation lengths are $t=3,500$ and $t=100$, 
respectively. The ``warming up'' length is $t=20$ and the time step of the 
dynamical evolution of the digital twin is $\Delta t=0.025$. The optimized 
hyperparameter values are $d=31$, $\lambda=0.75$, $k_\textrm{in}=0.16$, 
$k_c=0.16$, $\alpha=0.33$, $\beta=1\times10^{-12}$, and 
$\sigma_{\textrm{noise}}=10^{-2}$.

The periodic signal used to drive the Lorenz-96 climate network of 20 nodes
is $f(t)=A\sin(\omega t) + F$ with $\omega = 2$, and $F = 2$. The structure
of the digital twin consists of 20 small RC networks as illustrated in
Fig.~\ref{fig:m20}(A). Figures~\ref{fig:m20}(B1) and \ref{fig:m20}(B2) 
show a chaotic and a periodic attractor for $A = 1.8$ and $A = 1.6$, 
respectively, in the $(x_1,x_2)$ plane. Training of the digital twin is 
conducted by using four time series from four different values of $A$, all in 
the chaotic regime. The attractors generated by the digital twin for $A = 1.8$
and $A = 1.6$ are shown in Figs.~\ref{fig:m20}(C1) and \ref{fig:m20}(C2),
respectively, which agree well with the ground truth. Figure~\ref{fig:m20}(D)
shows the bifurcation diagram of the target system (the ground truth),
where the four values of $A$: $A=1.8$, 2.2, 2.6, and 3.0, from which the 
training chaotic time series are obtained, are indicated by the four respective
vertical dashed lines. The bifurcation diagram generated by the digital 
twin is shown in Fig.~\ref{fig:m20}(E), which agrees well with the ground 
truth in Fig.~\ref{fig:m20}(D).

\begin{figure}[ht!bp]
\centering
\includegraphics[width=0.9\linewidth]{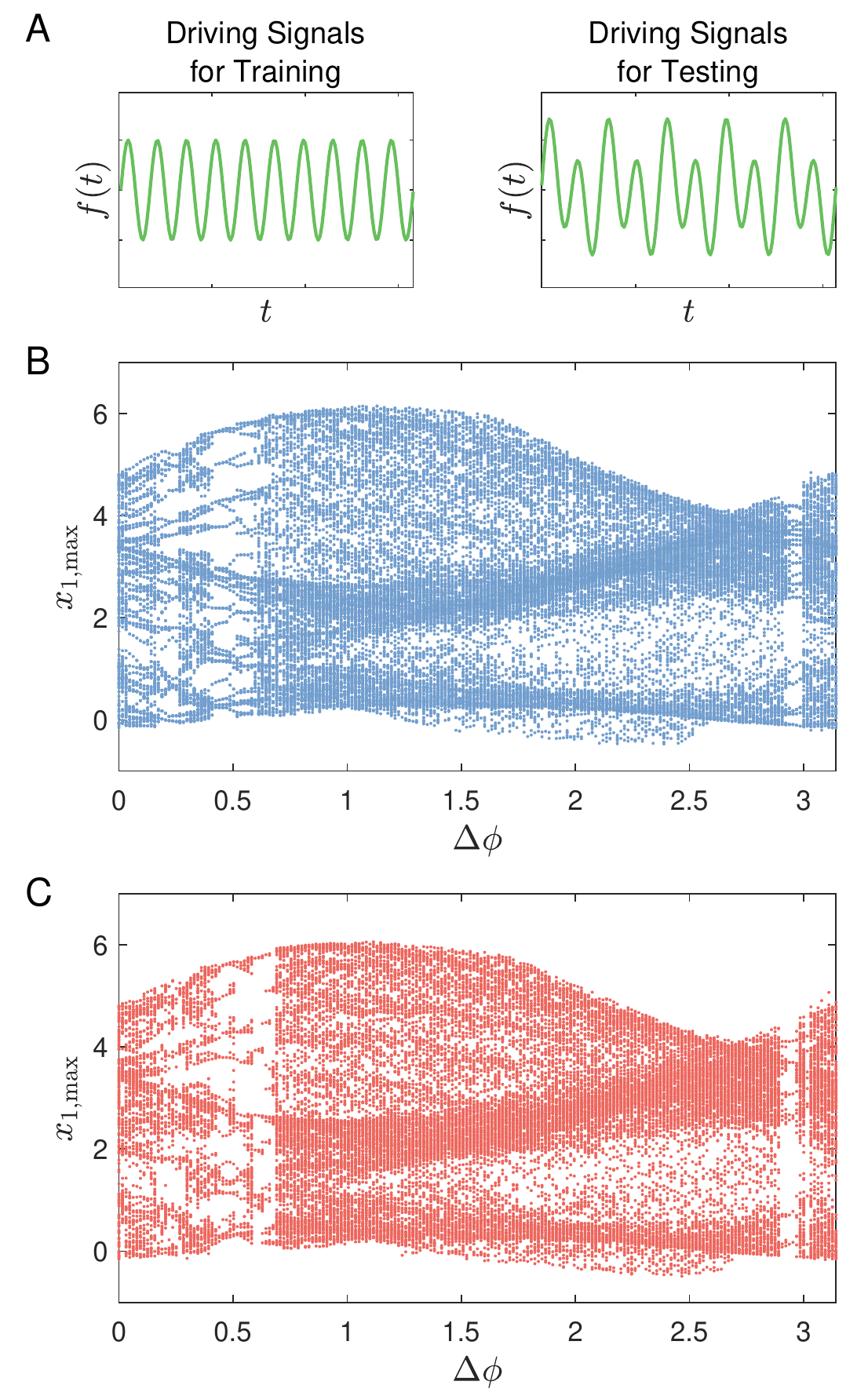}
\caption{Effects of waveform change in the external driving on the performance 
of the digital twin. The time series used to train the digital twin are from
the target system subject to external driving of a particular waveform. A 
change in the waveform occurs subsequently, leading to a different driving 
signal during the testing phase. (A) During the training phase, the driving 
signal is of the form $f(t) = A\sin(\omega t) + F$ and time series from four 
different values of $A$ are used for training the digital twin. The right panel 
illustrates an example of the changed driving signal during the testing phase. 
(B) The true bifurcation diagram of the target system under a testing driving 
signal. (C) The bifurcation diagram generated by the digital twin, facilitated
by an optimal level of training noise determined through hyperparameter 
optimization.}
\label{fig:phi}
\end{figure}

\subsection{Digital twins under external driving with varied waveform} \label{subsec:Lorenz96_waveform}

The external driving signal is an essential ingredient in our articulation of
the digital twin, which is particularly relevant to critical systems of 
interest such as the climate systems. In applications, the mathematical
form of the driving signal may change with time. Can a digital twin produce
the correct system behavior under a driving signal that is different than the 
one it has ``seen'' during the training phase? Note that, in the examples 
treated so far, it has been demonstrated that our digital twin can extrapolate
the dynamical behavior of a target system under a driving signal of the same
mathematical form but with a different amplitude. Here, the task is more 
challenging as the form of the driving signal has changed. 

As a concrete example, we consider the Lorenz-96 climate network of $m=6$ 
nodes, where a digital twin is trained with a purely sinusoidal signal
$f(t)=A\sin(\omega t) + F$, as illustrated in the left column of 
Fig.~\ref{fig:phi}(A). During the testing phase, the driving signal has the 
form of the sum of two sinusoidal signals with different frequencies:
$f(t)=A_1\sin(\omega_1 t)+A_2\sin(\omega_2 t+\Delta \phi)+F$, as illustrated
in the right panel of Fig.~\ref{fig:phi}(A). We set $A_1=2$, $A_2=1$, 
$\omega_1=2$, $\omega_2=1$, $F=2$, and use $\Delta \phi$ as the bifurcation
parameter. The RC parameter setting is the same as that in Fig.~\ref{fig:m6}.
%Sec.~\ref{subsec:Lorenz96_LD}. 
The training and validating lengths for each 
driving amplitude $A$ value are $t=3,000$ and $t=12$, respectively. We fine 
that this setting prevents the digital twin from generating an accurate 
bifurcation diagram, but a small amount of dynamical noise to the target 
system can improve the performance of the digital twin. To demonstrate this, 
we apply an additive noise term to the driving signal $f(t)$ in the training 
phase: $df(t)/dt = \omega A \cos(\omega t) +\delta_{\textrm{DN}}\xi(t)$, 
where $\xi(t)$ is a Gaussian white noise of zero mean and unit variance, and 
$\delta_{\textrm{DN}}$ is the noise amplitude (e.g., 
$\delta_{\textrm{DN}}=3\times 10^{-3}$). We use the 2nd-order Heun 
method~\cite{BPTK:1997} to solve the stochastic differential equations 
describing the target Lorenz-96 system. Intuitively, the noise serves to 
excite different modes of the target system to instill richer information into 
the training time series, making the process of learning the target dynamics 
more effective. Figures~\ref{fig:phi}(B) and \ref{fig:phi}(C) show the actual 
and digital-twin generated bifurcation diagrams. Although the digital twin
encountered driving signals in a completely ``uncharted territory,'' it is 
still able to generate the bifurcation diagram with a reasonable accuracy.
The added dynamical noise is creating small fluctuations in the driving signal
$f(t)$. This may yield richer excited dynamical features of the target system
in the training data set, which can be learned by the RC network. This should
be beneficial for the RC network to adapt to different waveform in the testing.
Additional results with varying testing waves $f(t)$ are presented in
Appendix E.

\begin{figure}[ht!]
\centering
\includegraphics[width=\linewidth]{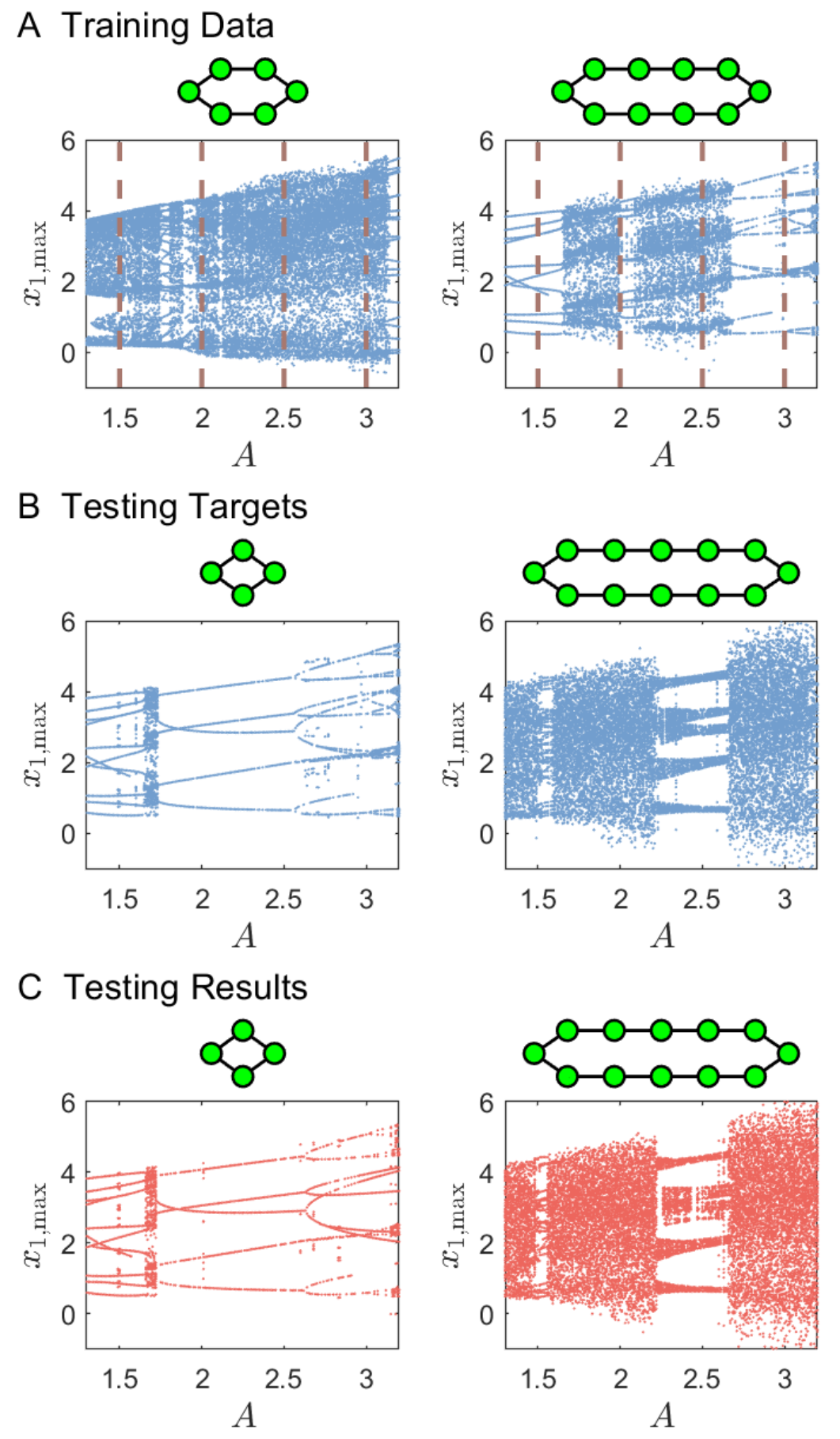}
\caption{Demonstration of extrapolability of digital twin in system size. 
(A) The digital twin is trained using time series from the Lorenz-96 climate
networks of size $m=6$ and $m=10$. The target climate system is subject to a 
sinusoidal driving $f(t)=A\sin(\omega t)+F$, and the training time series data 
are from the $A$ values marked by the eight vertical orange dashed lines. 
(B) The true bifurcation diagrams of the target climate network of size $m=4$ 
and $m=12$. (C) The corresponding digital-twin generated bifurcation diagrams,
where the twin consists of $m/2$ parallel RC networks, each taking input from
two nodes in the target system and from the nodes in the network that are
coupled to the two nodes.} 
\label{fig:m}
\end{figure}

\subsection{Extrapolability of digital twin with respect to system size} \label{subsec:Lorenz96_DD}

In the examples studied so far, it has been demonstrated that our RC based
digital twin has a strong extrapolability in certain dimensions of the 
parameter space. Specifically, the digital twin trained with time series data 
from one parameter region can follow the dynamical evolution of the target 
system in a different parameter regime. One question is whether the digital 
twin possesses certain extrapolability in the system size. For example, 
consider the Lorenz-96 climate network of size $m$. In Fig.~\ref{fig:m20}, we 
use an array of parallel RC networks to construct a digital twin for the 
climate network of a fixed size $m$, where the number of parallel RCs is $m/2$
(assuming that $m$ is even), and training and testing/monitoring are 
carried out for the same system size. We ask, if a digital twin is trained for 
climate networks of certain sizes, will it have the ability to generate the 
correct dynamical behaviors for climate networks of different sizes? If yes, we
say that the digital twin has the extrapolability with respect to system size.

As an example, we create a digital twin with a parallel structure based on time 
series data from the Lorenz-96 climate networks of sizes $m=6$ and $m=10$, 
i.e., with $m/2=3$ and $m/2=5$ numbers of identical RC networks coupled in a 
parallel fashion. Testing is done with the same individual RC networks that
are coupled together to simulate the target system of different system sizes. 
We also test if the digital twins can make predictions of the system dynamics 
under driving signals with unseen amplitudes.
The training data with $m=6$ and $m=10$ is shown in Fig.~\ref{fig:m}(A). 
For each system size in the training set, four values of the forcing
amplitude $A$ are used to generate the training time series: $A=$1.5, 2.0,
2.5, and 3.0, as marked by the vertical orange dashed lines in
Figs.~\ref{fig:m}(A) and \ref{fig:m}(B). 
As in Fig.~\ref{fig:m20}, 
%Sec.~\ref{subsec:Lorenz96_HD}, 
the digital twin consists of $m/2$ parallel
RC networks, each of size $D_r=1,500$. The optimized hyperparameter 
values are determined to be $d=927$, $\lambda=0.71$, $k_\textrm{in}=0.076$, 
$k_c=0.078$, $\alpha=0.27$, $\beta=1\times10^{-11}$, and 
$\sigma_{\rm noise} = 3\times10^{-3}$. Then we consider climate networks of 
two different sizes: $m=4$ and $m=12$, and test if the trained digital twin  
can be adapted to the new systems. For the network of size $m = 4$, we keep
only two parallel RC networks for the digital twin. For $m=12$, we add one 
additional RC network to the trained digital twin for $m=10$, so the new twin
consists of six parallel RC networks of the same hyperparameter values. The
true bifurcation diagrams for the climate system of sizes $m=4$ and $m=12$ are 
shown in Fig.~\ref{fig:m}(B) (the left and right panels, respectively). The 
corresponding bifurcation diagrams generated by the adapted digital twins are 
shown in Fig.~\ref{fig:m}(C), which agree with the ground truth reasonably well,
demonstrating that our RC based digital twin possesses certain extrapolability 
in system size.

\section{Discussion} \label{sec:discussion}

We have articulated the principle of creating digital twins for nonlinear
dynamical systems based on RCs that are recurrent neural networks.
In general, RC is a powerful neural network framework that does not require
backpropagation during training but only a linear regression is needed.
This feature makes the development of digital twins based on RC computationally
efficient. We have demonstrated that a well-trained RC network is able to serve 
as a digital twin for systems subject to external, time-varying driving. The 
twin can be used to anticipate possible critical transitions or regime shifts 
in the target system as the driving force changes, thereby providing early 
warnings for potential catastrophic collapse of the system. We have used a 
variety of examples from different fields to demonstrate the workings and the 
anticipating power of the digital twin, which include the Lorenz-96 climate
network of different sizes (in the main text), a driven chaotic ${\rm CO}_2$ 
laser system (Appendix A), and an ecological system (Appendix B). For low-dimensional 
nonlinear dynamical systems, a single RC network is sufficient for the digital 
twin. For high-dimensional systems such as the climate network of a relatively 
large size, parallel RC networks can be integrated to construct the 
digital twin. At the level of the detailed state evolution, our recurrent 
neural network based digital twin is essentially a dynamical twin system that 
evolves in parallel to the real system, and the evolution of the digital twin 
can be corrected from time to time using sparse feedback of data from the 
target system (Appendix C). In cases where direct measurements of the target system
are not feasible or are too costly, the digital twin provides a way to assess 
the dynamical evolution of the target system. At the qualitative level, the 
digital twin can faithfully reproduce the attractors of the target system, 
e.g., chaotic, periodic, or quasiperiodic, without the need of state updating. 
In addition, we show that the digital twin is able to accurately predict 
a critical bifurcation point and the average lifetime of transient chaos that 
occurs after the bifurcation, even under a driving signal that is different 
from that during the training (Appendix F). The issue of robustness against 
dynamical and observational noises in the training data has also been
treated (Appendix G).

To summarize, our RC based digital twins are capable of performing the
following tasks: (1) extrapolating certain dynamical evolution of the target
system beyond the training parameter regime, (2) making long-term 
continual forecasting of nonlinear dynamical systems under nonstationary 
external driving with sparse state updates, (3) inferring the existence of 
hidden variables in the system and reproducing/predicting their dynamical 
evolution, (4) adapting to external driving of different waveform, and 
(5) extrapolating the global bifurcation behaviors to systems of different 
sizes. 

Our design of the digital twins for nonlinear dynamical systems can be extended
in a number of ways. 

\paragraph*{1. Online learning.} 
Online or continual learning is a recent trend in machine-learning
research. Unlike the approach of batch learning, where one gathers all the 
training data in one place and does the training on the entire data set (the
way by which training is conducted for our work), in an online learning 
environment, one evolves the machine learning model incrementally with the 
flow of data. For each training step, only the newest inputted training data 
is used to update the machine learning model. When a new data set is available,
it is not necessary to train the model over again on the entire data set 
accumulated so far, but only on the new set. This can result in a significant
reduction in the computational complexity. Previously, an online learning 
approach to RC known as the FORCE learning was developed~\cite{SA:2009}.
An attempt to deal with the key problem of online learning termed 
``catastrophic forgetting'' was made in the context of RC~\cite{KS:2019}. 
Further investigation is required to see if these methods can be exploited for 
creating digital twins through online learning.

\paragraph*{2. Beyond reservoir computing.}
Second, the potential power of recurrent neural network based digital twin
may be further enhanced by using more sophisticated recurrent neural network
models depending on the target problem. We use the RC networks because they 
are relatively simple yet powerful enough for both low- and high-dimensional
dynamical systems. Schemes such as knowledge-based hybrid 
RC~\cite{PWFCHGO:2018} or ODE-nets~\cite{CRBD:2018} are worth investigating.

\paragraph*{3. Reinforcement learning.}
Is it possible to use digital twins to make reinforcement learning 
feasible in situations where the target system cannot be ``disturbed''? 
Particularly, reinforcement learning requires constant interaction with the 
target system during training so that the machine can learn from its mistakes 
and successes. However, for a real-world system, these interactions may be 
harmful, uncontrollable, and irreversible. As a result, reinforcement learning 
algorithms are rarely applied to safety-critical systems~\cite{BTSK:2017}.
In this case, digital twins can be beneficial. By building a digital twin, the 
reinforcement learning model does not need to interact with the real system, 
but with its simulated replica for efficient training. This area of research
is called model-based reinforcement learning~\cite{MBJ:2020}. 

\paragraph*{4. Potential benefits of noise.}
A phenomenon uncovered in our study is the beneficial role of dynamical 
noise in the target system. As briefly discussed in Fig.~\ref{fig:phi}, 
%Sec.~\ref{subsec:Lorenz96_waveform}, 
adding dynamic noise in the training 
dataset enhances the digital twin's ability to extrapolate the dynamics of the 
target system with different waveform of driving. Intuitively, noise can 
facilitate the exploration of the phase space of the target nonlinear system. 
A systematic study of the interplay between dynamical noise and the performance
of the digital twin is worthy.

\paragraph*{5. Extrapolability.}
The demonstrated extrapolability of our digital twin, albeit limited,
may open the door to forecasting the behavior of large systems using twins 
trained on small systems. Much research is needed to address this issue. 

\paragraph*{6. Spatiotemporal dynamical systems with multistability.}
We have considered digital twins for a class of coupled dynamical systems: 
the Lorenz-96 climate model. When developing digital twins for spatiotemporal
dynamical systems, two issues can arise. One is the computational complexity 
associated with such high-dimensional systems. We have demonstrated that 
parallel reservoir computing provides a viable solution. Another issue is 
multistability. Spatiotemporal dynamical systems in general exhibit extremely 
rich dynamical behaviors such as chimera states~\cite{KB:2002,AS:2004,OMHS:2011,TNS:2012,HMRHOS:2012,OOHS:2013,OZHSS:2015,OOZS:2018,KL:2020}. 
To develop digital twins of spatiotemporal dynamical systems with multiple
coexisting states requires that the underlying recurrent neural networks 
possess certain memory capabilities. To develop methods to incorporate 
memories into digital twins is a problem of current interest. 

\section*{Data Availability}

All relevant data are available from the authors upon request.

\section*{Code Availability}

All relevant computer codes are available from the authors upon request.

\section*{Acknowledgment}

We thank Z.-M. Zhai for discussions. This work was supported by the Army
Research Office through Grant No.~W911NF-21-2-0055 and by the U.S.-Israel 
Energy Center managed by the Israel-U.S. Binational Industrial Research and 
Development (BIRD) Foundation. 

\section*{Author Contributions}

All authors designed the research project, the models, and methods. L.-W.K. 
performed the computations. All analyzed the data. L.-W.K. and Y.-C.L wrote 
the paper.

\section*{Competing Interests}

The authors declare no competing interests.

\section*{Correspondence}

To whom correspondence should be addressed. E-mail: Ying-Cheng.Lai@asu.edu.

\appendix

\section{A driven chaotic laser system} \label{sec:laser}

\begin{figure*}[ht!]
\centering
\includegraphics[width=\linewidth]{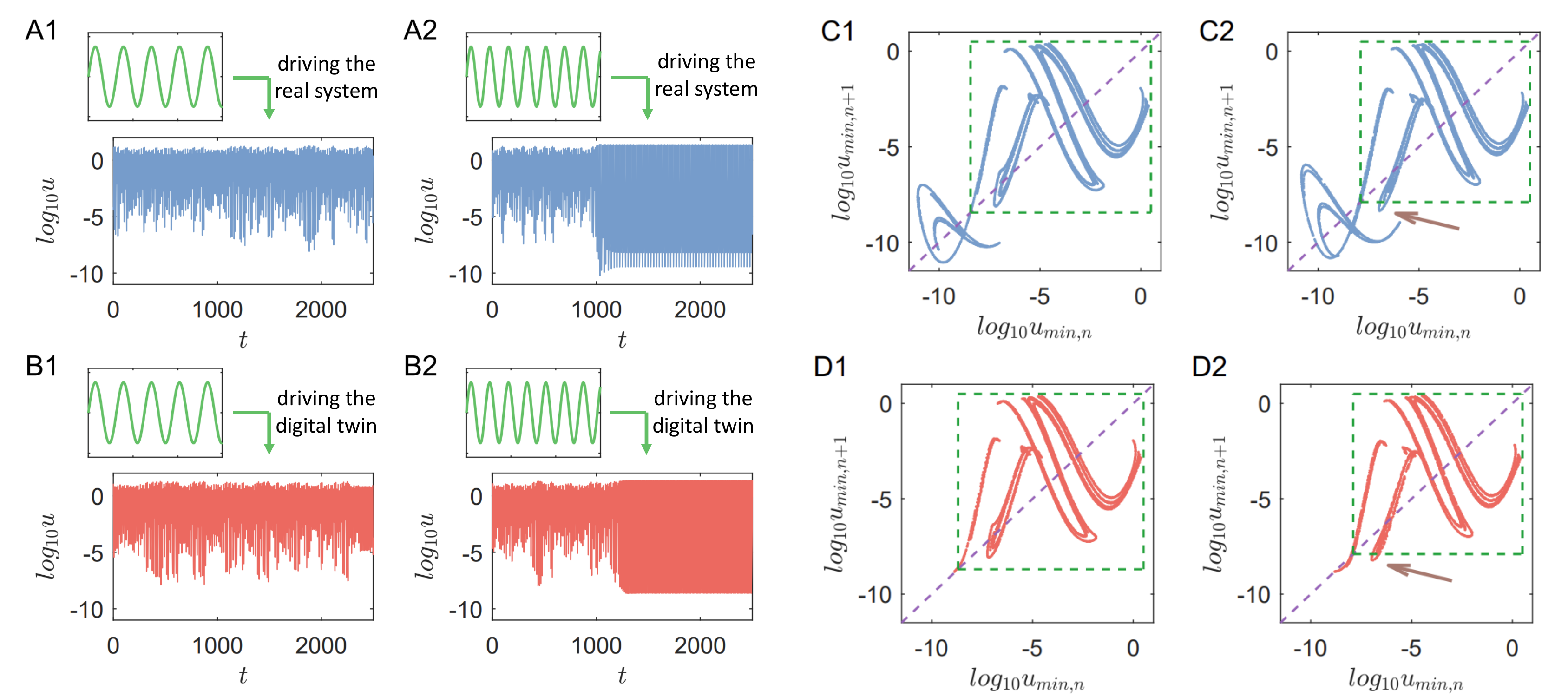}
\caption{\footnotesize
Performance of the digital twin of the driven ${\rm CO_2}$ laser system to 
extrapolate system dynamics under different driving frequencies. (A1, A2) True 
sustained and transient chaotic time series of $\log_{10}{u(t)}$ of the target 
system, for driving frequencies $\Omega=0.905<\Omega_c$ and 
$\Omega=0.925>\Omega_c$, respectively, where the sinusoidal driving signal
$f(t)$ is schematically illustrated. In (A1), the system exhibits sustained 
chaos. In (A2), the system settles into a periodic state after transient chaos. 
(B1, B2) The corresponding time series generated by the digital twin. In both
cases, the dynamical behaviors generated by the digital twin agree with the 
ground truth in (A1, A2): sustained chaos in (B1) and transient chaos to a 
periodic attractor in (B2). (C1, C2) The return maps constructed from the local 
minima of $u(t)$ from the true dynamics, where the green dashed square defines 
an interval that contains the chaotic attractor (C1) or a nonattracting chaotic
set due to the escaping region (marked by the red arrow) leading to transient 
chaos (C2). (D1, D2) The return maps generated by the digital twin for the same 
values of $\Omega$ as in (C1, C2), respectively, which agree with the ground 
truth.}
\label{fig:LaserMain}
\end{figure*}

We consider the single-mode, class B, driven chaotic ${\rm CO_2}$ laser 
system~\cite{DGH:1986,DGH:1987,SEGT:1987,Schwartz:1988} described by
\begin{align}
\frac{du}{dt} &= -u[f(t)-z], \\
\frac{dz}{dt} &= \epsilon_1z-u-\epsilon_2zu+1,
\end{align}
where the dynamical variables $u$ and $z$ are proportional to the normalized
intensity and the population inversion, $f(t)=A\cos(\Omega t+\phi)$ is the 
external sinusoidal driving signal of amplitude $A$ and frequency $\Omega$, 
$\epsilon_1$ and $\epsilon_2$ are two parameters. Chaos is common in this laser 
system~\cite{DGH:1986,DGH:1987,Schwartz:1988}. For example, for 
$\epsilon_1=0.09$, $\epsilon_2=0.003$, and $A=1.8$, there is a chaotic 
attractor for $\Omega < \Omega_c \approx 0.912$, as shown by a sustained 
chaotic time series in Fig.~\ref{fig:LaserMain}(a1). The chaotic attractor is 
destroyed by a boundary crisis~\cite{GOY:1983} at $\Omega_c$. For 
$\Omega > \Omega_c$, there is transient chaos, after which the system settles 
into periodic oscillations, as exemplified in Fig.~\ref{fig:LaserMain}(a2). 
Suppose chaotic motion is desired. The crisis bifurcation at $\Omega_c$ can 
then be regarded as a kind of system collapse. 

To build a digital twin for the chaotic laser system, we use the external
driving signal as the natural control signal for the RC 
network. Different from the examples in the main text, here the driving
frequency $\Omega$, instead of the driving amplitude $A$, serves as the
bifurcation parameter. 
Assuming observational data in the form of time series are available
for several values of $\Omega$ in the regime of a chaotic attractor, we train 
the RC network using chaotic time series collected from four values 
of $\Omega < \Omega_c$: $\Omega=0.81$, 0.84, 0.87, and 0.90. The training 
parameter setting is as follows. For each $\Omega$ value in the training set, 
the training and validation lengths are $t=2,000$ and $t=83$, respectively, 
where the latter corresponds to approximately five Lyapunov times. The 
``warming up'' length is $t=0.5$. The time step of the reservoir system is
$\Delta t=0.05$. The size of the random RC network is $D_r=800$. The 
optimal hyperparameter values are determined to be $d=151$, $\lambda=0.0276$, 
$k_\textrm{in}=1.18$, $k_c=0.113$, $\alpha=0.33$, and $\beta=2\times10^{-4}$.

Figures~\ref{fig:LaserMain}(A1) and \ref{fig:LaserMain}(A2) show two 
representative time series from the laser model (the ground truth)
for $\Omega=0.905<\Omega_c$ and $\Omega=0.925>\Omega_c$, respectively.
The one in panel (A1) is associated with sustained chaos (pre-critical)
and the other in panel (A2) is characteristic of transient chaos with a
final periodic attractor (post-critical). The 
corresponding time series generated by the digital twin are shown in 
Figs.~\ref{fig:LaserMain}(B1) and \ref{fig:LaserMain}(B2), respectively. It 
can be seen that the training aided by the control signal enables the digital 
twin to correctly capture the dynamical climate of the target system, e.g., 
sustained or transient chaos. The true return maps in the pre-critical and 
post-critical regimes are shown in Figs.~\ref{fig:LaserMain}(C1) and 
\ref{fig:LaserMain}(C2), respectively, and the corresponding maps generated by 
the digital twin are shown in Figs.~\ref{fig:LaserMain}(D1) and 
\ref{fig:LaserMain}(D2). In the pre-critical regime, an invariant region (the 
green dashed square) exists on the return map in which the trajectories are 
confined, leading to sustained chaotic motion, as shown in 
Figs.~\ref{fig:LaserMain}(C1) and \ref{fig:LaserMain}(D1). Within the 
invariant region in which the chaotic attractor lives, the digital twin 
captures the essential dynamical features of the attractor. Because the 
training data are from the chaotic attractor of the target system, the digital 
twin fails to generate the portion of the real return map that lies outside 
the invariant region, which is expected because the digital twin has never 
been exposed to the dynamical behaviors that are not on the chaotic attractor. 
In the post-critical regime, a ``leaky'' region emerges, as indicated by the 
red arrows in Figs.~\ref{fig:LaserMain}(C2) and \ref{fig:LaserMain}(D2), which 
destroys the invariant region and leads to transient chaos. The remarkable 
feature is that the digital twin correctly assesses the existence of the leaky 
region, even when no such information is fed into the twin during training. 
From the point of view of predicting system collapse, the digital twin is able
to anticipate the occurrence of the crisis and transient chaos. A quantitative
result of these predictions are demonstrated in \ref{sec:Q}.

As indicated by the predicted return maps in Figs.~\ref{fig:LaserMain}(D1)
and \ref{fig:LaserMain}(D2), the digital twin is unable to give the final 
state after the transient, because such state must necessarily lie outside 
the invariant region from which the training data are originated. In 
particular, the digital twin is trained with time series data from the chaotic
attractors prior to the crisis. With respect to Figs.~\ref{fig:LaserMain}(D1)
and \ref{fig:LaserMain}(D2), the digital twin can learn the dynamics within 
the dash green box in the plotted return maps, but is unable to predict the 
dynamics outside the box, as it has never been exposed to these dynamics.  

\begin{figure*}[ht!]
\centering
\includegraphics[width=0.5\linewidth]{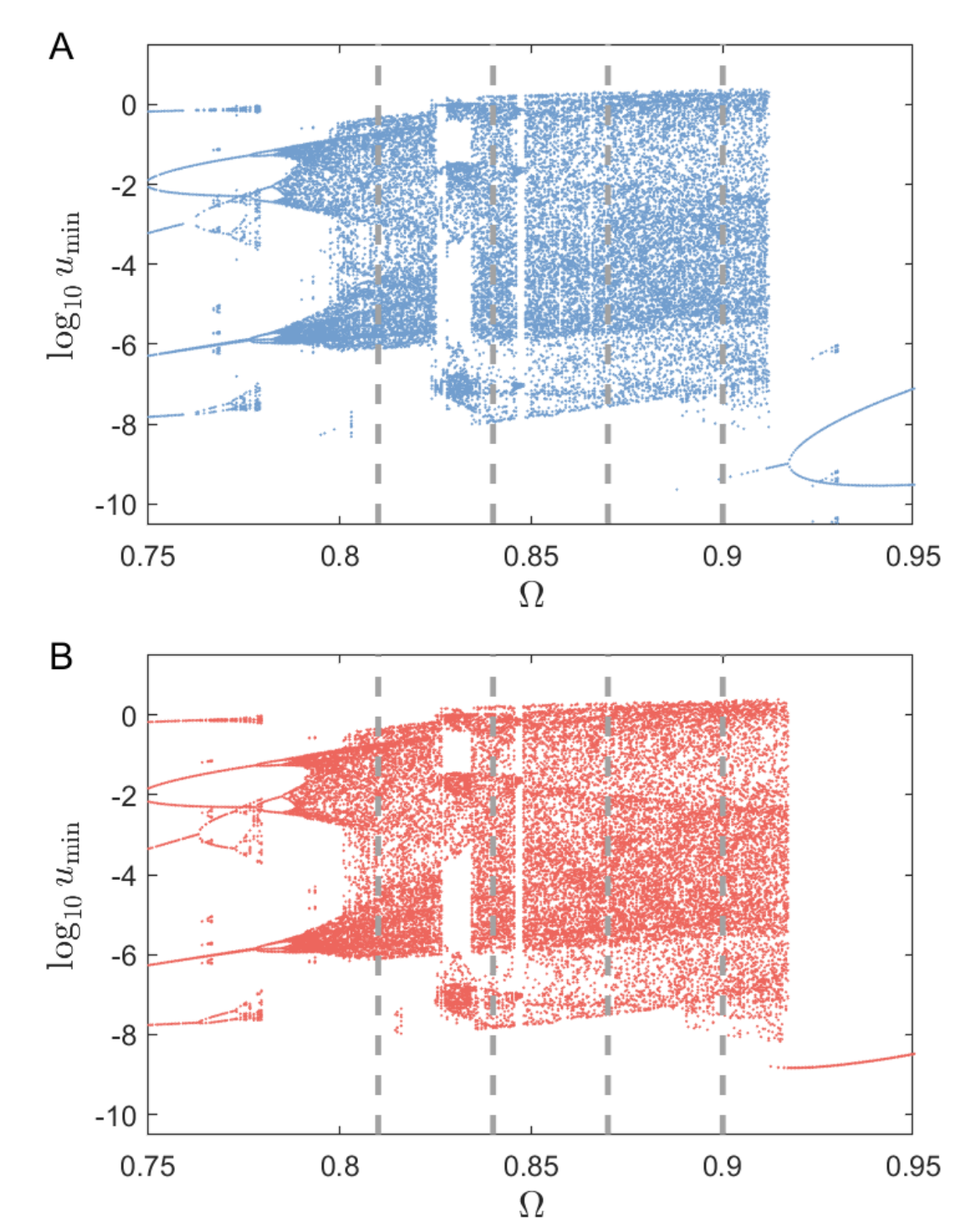}
\caption{\footnotesize Comparison of the real (A) and predicted (B) bifurcation
diagrams of the driven laser system with varying driving frequencies.
The four vertical grey dashed lines indicate the values of driving frequencies
$\Omega$ used for training the RC neural network. The strong resemblance
between the two bifurcation diagrams indicates the power of the digital twin in
extrapolating the correct global behavior of the target system, and demonstrates
that not only can this approach extrapolate system dynamics to various driving
amplitudes $A$, but also to varying driving frequency $\Omega$. 
}
\label{fig:LaserBifur}
\end{figure*}

A comparison of the real and predicted bifurcation diagram is demonstrated in 
Fig.~\ref{fig:LaserBifur}. The strong resemblance between them indicate the 
power of the digital twin in extrapolating the correct global behavior of the
target system. Moreover, this demonstrates that not only can this approach
extrapolate with various driving amplitudes $A$ (as demonstrated in the main
text), but the approach can also work with varying driving frequencies $\Omega$.

\section{A driven chaotic ecological system} \label{sec:eco}

\begin{figure*}[ht!bp]
\centering
\includegraphics[width=0.9\linewidth]{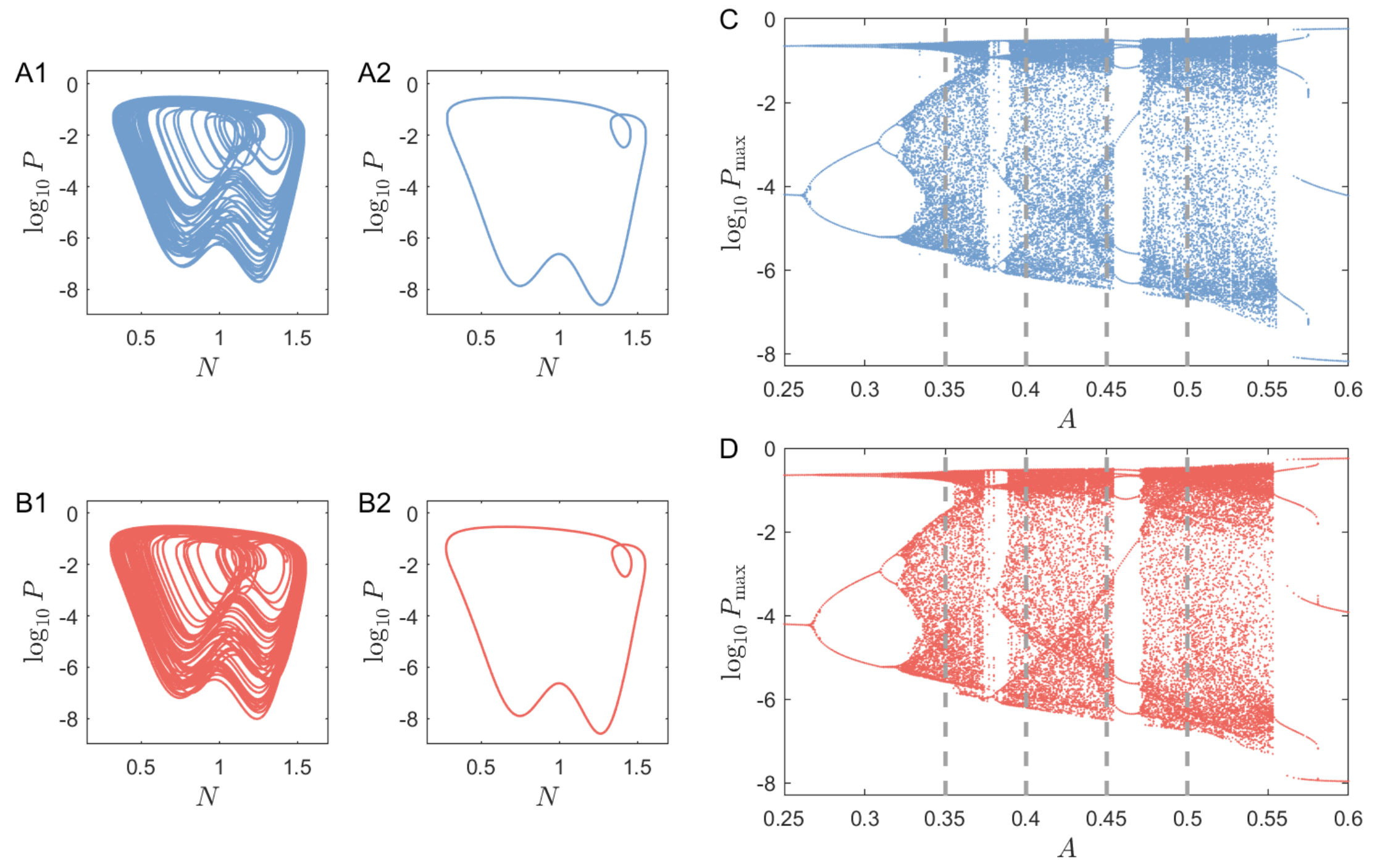}
\caption{\footnotesize
Performance of the digital twin of an ecological system of the blooms 
of phytoplankton with seasonality. The effect of seasonality is modeled by a 
sinusoidal driving signal $f(t)= A\sin(\omega_{{\rm eco}} t)$. (A1, A2) 
Chaotic and periodic attractors of this system in the $(N,\log_{10}P)$ 
plane for $A=0.45$ and $A=0.56$, respectively. (B1, B2) The corresponding 
attractors generated by the digital twin under the same driving signals $f(t)$ 
as in (A1, A2). The digital twin has successfully extrapolated the periodical 
behavior outside the chaotic training region. (C) The ground-truth bifurcation 
diagram of the target system. (D) The digital-twin generated bifurcation 
diagram. In (C) and (D), the four vertical grey dashed lines indicate the 
values of driving amplitudes $A$ used for training the RC network. The
strong resemblance between the two bifurcation diagrams indicates the
power of the digital twin in extrapolating the correct global behavior of 
the target system.}
\label{fig:eco}
\end{figure*}

We study a chaotic driven ecological system that models the annual blooms of 
phytoplankton under seasonal driving~\cite{HBOS:2005}. Seasonality plays a 
crucial role in ecological systems and epidemic spreading of infectious
diseases~\cite{SOH:2007}, which is usually modeled as a simple periodic 
driving force on the system. The dynamical equations of this model in the 
dimensionless form are~\cite{HBOS:2005}:
\begin{align}
\frac{dN}{dt} &= I - f(t) N P -qN, \label{eq:eco1} \\
\frac{dP}{dt} &= f(t) N P - P, \label{eq:eco2}
\end{align}
where $N$ represents the level of the nutrients, $P$ is the biomass of the 
phytoplankton, the Lotka-Volterra term $NP$ models the phytoplankton uptake of 
the nutrients, $I$ represents a small and constant nutrient flow from external
sources, $q$ is the sinking rate of the nutrients to the lower level of the 
water unavailable to the phytoplankton, and $f(t)$ is the seasonality term: 
$f(t)=A\sin(\omega_{{\rm eco}} t)$. The parameter values are~\cite{HBOS:2005}: 
$I=0.02$, $q=0.0012$, and $\omega_{\textrm{eco}}=0.19$.

Climate change can dramatically alter the dynamics of this 
ecosystem~\cite{WS:2012}. We consider the task of forecasting how the system 
behaves if the climate change causes the seasonal fluctuation to be more
extreme. In particular, suppose the training data are measured from the 
system when it behaves normally under a driving signal of relatively small
amplitude, and we wish to predict the dynamical behaviors of the system in the
future when the amplitude of the driving signal becomes larger (due to climate
change). The training parameter setting is as follows. The size of the 
RC network is $D_r=600$ with $D_{\textrm{in}}=D_{\textrm{out}}=2$. 
The time step of the evolution of the network dynamics is $\Delta t =0.1$.
The training and validation lengths for each value of the driving amplitude 
$A$ in the training are $t=1,500$ and $t=500$, respectively. The optimized 
hyperparameters of the RC are $d=350$, $\lambda=0.42$, $k_{\rm in}=0.39$,
$k_c=1.59$, $\alpha =0.131$, and $\beta=1\times10^{-7.5}$.

Figure~\ref{fig:eco} shows the results of our digital twin approach on this
ecological model to learn from the dynamics under a few different values of
the driving amplitude to generate the correct response of the system to a
driving signal of larger amplitude. In 
particular, the training data are collected with the driving amplitude 
$A=0.35, 0.4, 0.45$ and $0.5$, all in the chaotic regions. 
Figures~\ref{fig:eco}(A1) and \ref{fig:eco}(A2) show the true attractors of
the system for $A = 0.45$ and $0.56$, respectively, where the attractor is 
chaotic in the former case (within the training parameter regime) and periodic
in the latter (outside the training regime). The corresponding attractors 
generated by the digital twin are shown in Figs.~\ref{fig:eco}(B1) and 
\ref{fig:eco}(B2). The digital twin can not only
replicate the chaotic behavior in the training data [Fig.~\ref{fig:eco}(B1)]
but also predict the transition to a periodic attractor under a driving signal
with larger amplitudes (more extreme seasonality), as shown in 
Fig.~\ref{fig:eco}(B2). In fact, the digital twin can faithfully produce the
global dynamical behavior of the system, both inside and outside the training
regime, as can be seen from the nice agreement between the ground-truth 
bifurcation diagram in Fig.~\ref{fig:eco}(C) and the diagram generated by the
digital twin in Fig.~\ref{fig:eco}(D). 

\section{Continual forecasting under non-stationary external driving with sparse real-time data} \label{sec:LongTerm}

\begin{figure*}[ht!]
\centering
\includegraphics[width=0.95\linewidth]{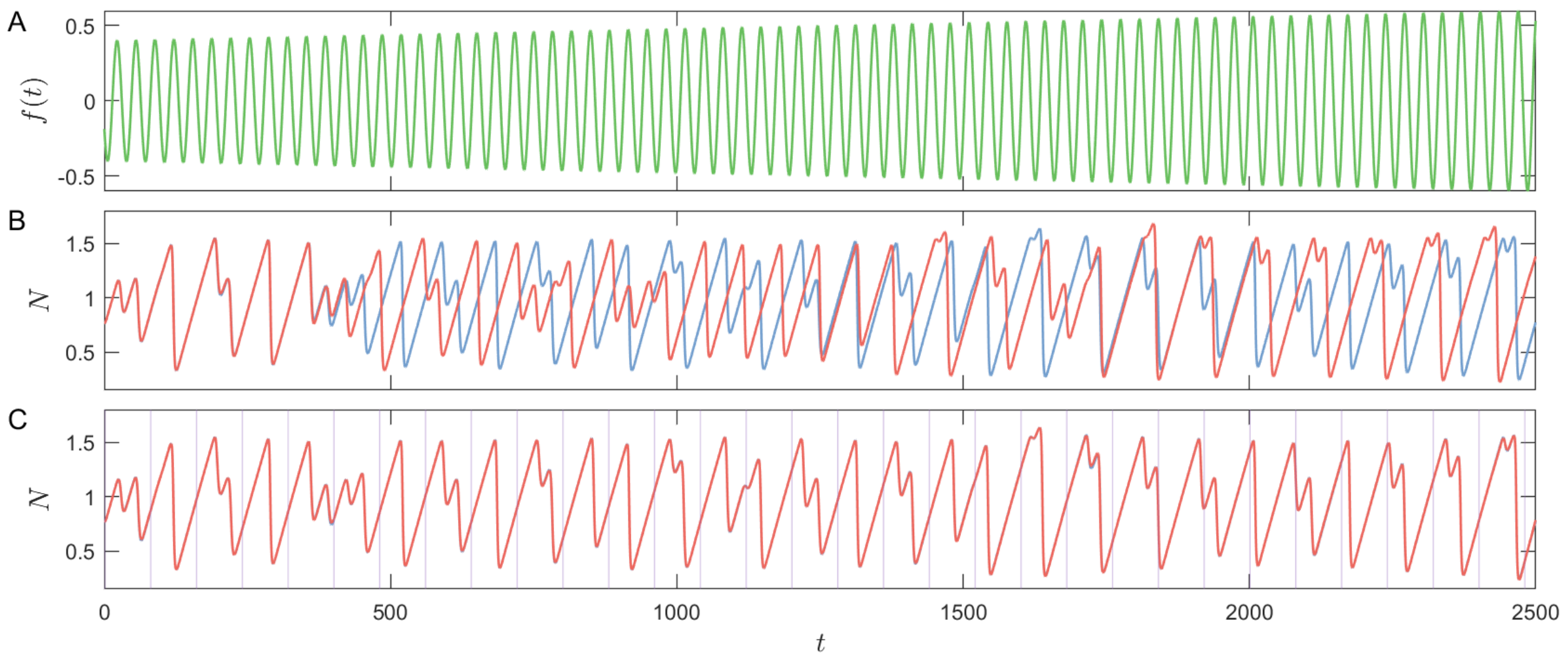}
\caption{\footnotesize Continual forecasting of the chaotic ecological system 
under non-stationary external driving with sparse updates. (A) A non-stationary 
sinusoidal driving signal $f(t)$ whose amplitude increases with time. The task for the digital twin is to forecast the response of the chaotic target system under 
this driving signal for a relatively long term (B) The trajectory generated
by the digital twin (red) in comparison with the true trajectory (blue). For
$0 \le t \alt 400$, the two trajectories match each other with small 
errors, but the digital-twin generated trajectory begins to deviate from the
true trajectory at $t \sim 400$ (due to chaos). (C) With only sparse updates 
from real data at times indicated by the vertical lines (2.5\% of the time 
steps in the given time interval), the digital twin can make relatively 
accurate predictions for a long term, demonstrating the ability to perform 
continual forecasting.}
\label{fig:long}
\end{figure*}

The three examples (Lorenz-96 climate network in the main text, the driven
$\mbox{CO}_2$ laser and the ecological system) have demonstrated that our
RC based digital twin is capable of extrapolating and 
generating the correct statistical features of the dynamical trajectories 
of the target system such as the attractor and bifurcation diagram. That is,
the digital twin can be regarded as a ``twin'' of the target system only on 
a statistical sense. In particular, from random initial conditions the digital
twin can generate an ensemble of trajectories, and the statistics calculated
from the ensemble agree with those of the original system. At the level of 
individual trajectories, if a target system and its digital twin start from 
the same initial condition, the trajectory generated by the twin can stay 
close to the true trajectory only for a short period of time (due to chaos).
However, with infrequent state updates, the trajectory generated by the twin 
can shadow the true trajectory (in principle) for an arbitrarily long period 
of time~\cite{FJZWL:2020}, realizing {\em continual forecasting} of the state
evolution of the target system.

In data assimilation for numerical weather forecasting, the state of the 
model system needs to be updated from time to 
time~\cite{Kalnay:2003,ABN:2016,WPHSGO:2021}. This idea has recently been 
exploited to realize long-term prediction of the state evolution of chaotic 
systems using RC~\cite{FJZWL:2020}. Here we demonstrate that, 
even when the driving signal is non-stationary, the digital twin can still 
generate the correct state evolution of the target system. As a specific 
example, we use the chaotic ecosystem in Eqs.~(\ref{eq:eco1}-\ref{eq:eco2}) 
with the same RC network trained in Sec.~\ref{sec:eco}. 
Figure~\ref{fig:long}(A) shows the non-stationary external driving 
$f(t)=A(t)\sin(\omega_{{\rm eco}}t)$ whose amplitude $A(t)$ increases linearly 
from $A(t=0)=0.4$ to $A(t=2500)=0.6$ in the time interval $[0,2500]$. 
Figure~\ref{fig:long}(B) shows the true (blue) and digital-twin generated (red)
time evolution of the nutrient abundance. Due to chaos, without state updates, 
the two trajectories diverge from each other after a few cycles of oscillation.
However, even with rare state updates, the two trajectories can stay close to 
each other for any arbitrarily long time, as shown in Fig.~\ref{fig:long}(C). 
In particular, there are 800 time steps involved in the time interval 
$[0,2500]$ and the state of the digital twin is updated 20 times, i.e., 
$2.5\%$ of the available time series data. 
We will discuss the results further discussion in the next section.

\begin{figure*}[ht!]
\centering
\includegraphics[width=0.95\linewidth]{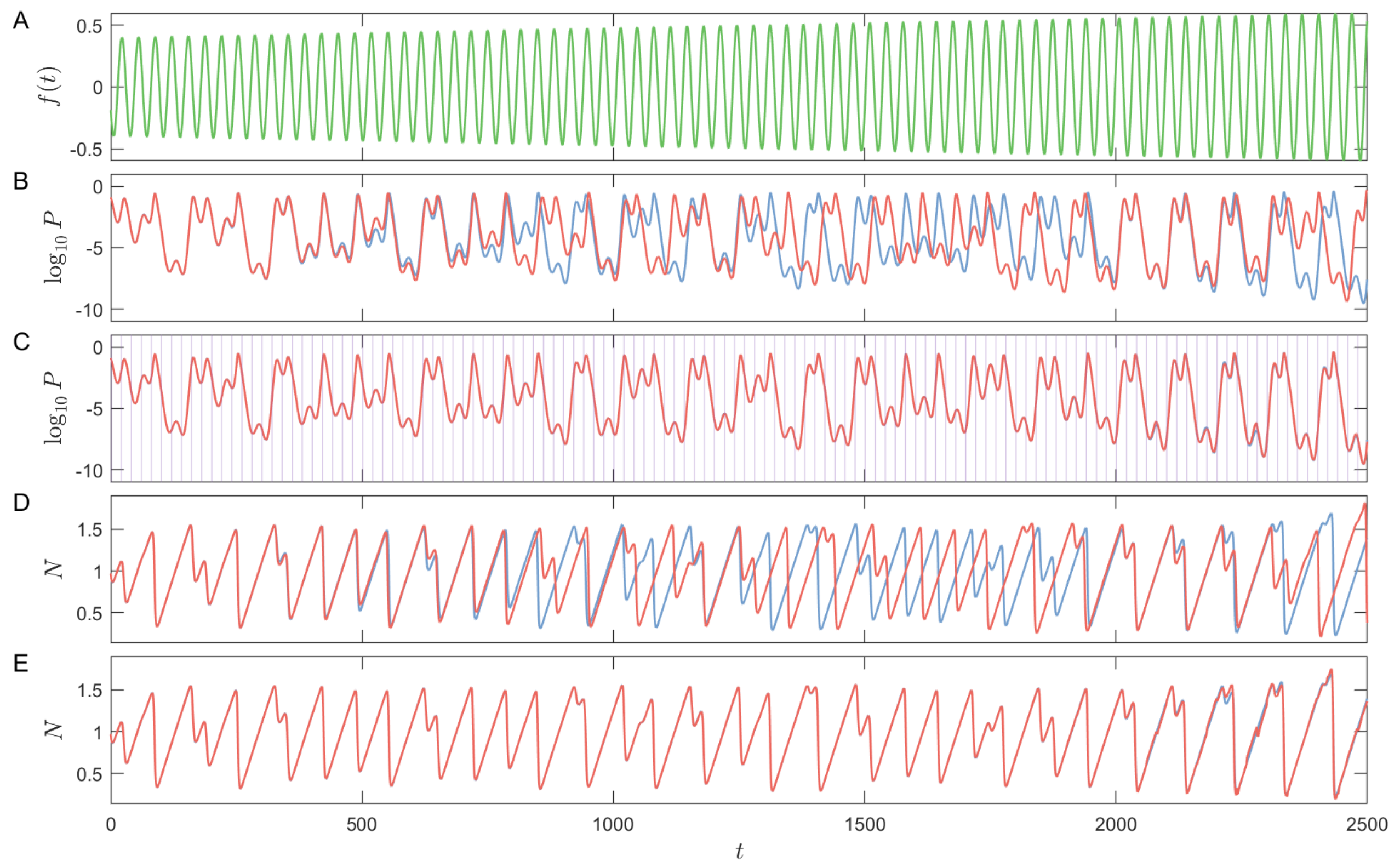}
\caption{\footnotesize
Continual forecasting and monitoring of a hidden dynamical variable 
in the chaotic ecological system under non-stationary external driving with 
sparse updates from the observable. The system is described by 
Eqs.~(\ref{eq:eco1}) and (\ref{eq:eco2}). The dynamical variable $N(t)$ is 
hidden, and the other variable $P(t)$ is externally accessible but only 
sparsely sampled measurement of it can be performed. (A) The non-stationary 
sinusoidal driving signal $f(t)$ with a time-varying amplitude. (B) Digital-twin
generated time evolution of the accessible variable $P(t)$ (red) in comparison 
with the ground truth (blue) in the absence of any state update of $P(t)$. The 
predicted time evolution quickly diverges from the true behavior. (C) With
sparse updates of $P(t)$ at the times indicated by the purple vertical lines 
(10\% of the times steps), the digital twin is able to make an accurate 
forecast of $P(t)$. (D) Digital-twin generated time evolution of the hidden
variable $N(t)$ (red) in comparison with the ground truth (blue) in the 
absence of any state update of $P(t)$. (E) Accurate forecasting of the hidden 
variable $N(t)$ with sparse updates of $P(t)$.}
\label{fig:longHidden}
\end{figure*}

\section{Continual forecasting with hidden dynamical variables} 

% The previous version still got two tasks mixed in a confusing way.
% The first task is independent of the second.
% The second task is on the basis of the first, with much stronger assumptions.
% So it might be better if we separate the descriptions of these two tasks
% given the different assumptions and different scenario settings.
% Continual forecasting of P given hidden N

In real-world scenarios, usually not all the dynamical variables of a target 
system are accessible. It is often the case that only a subset of the dynamical
variables can be measured and the remaining variables are inaccessible or
hidden from the outside world. Can a digital twin still make continual
forecasting in the presence of hidden variables based on the time series data
from the accessible variables? Also, Can the digital twin do this without
knowing that there exists some hidden variables before training? In general,
when there are hidden variables, the reservoir network needs to sense their
existence, encode them in the hidden state of the recurrent layer, and
constantly update them. As such, the recurrent structure of reservoir
computing is necessary, because there must be a place for the machine to store
and restore the implicit information that it has learned from the data.
Compared with the cases where complete information about the dynamical evolution
of all the observable is available, when there are hidden variables, it is
significantly more challenging to predict the evolution of a target system
driven by an non-stationary external signal using sparse observations of the
accessible variables.

As an illustrative example, we again consider the ecosystem described by 
Eqs.~(\ref{eq:eco1}) and (\ref{eq:eco2}). We assume that the dynamical 
variable $N$ (the abundance of the nutrients) is hidden and $P(t)$, the biomass
of the phytoplankton, is externally accessible. Despite the accessibility 
to $P(t)$, we assume that it can be measured only occasionally. That is, only 
sparsely updated data of the variable $P(t)$ is available. It is necessary that
the digital twin is able to learn some equivalent of $N(t)$ as the time 
evolution of $P(t)$ also depends on the value $N(t)$, and to encode the
equivalent in the reservoir network. In an actual application, when the
digital twin is deployed, knowledge about the existence of such a hidden
variable is not required.

Figure~\ref{fig:longHidden} presents a representative resulting trial, where 
Fig.~\ref{fig:longHidden}(A) shows the non-stationary external driving signal 
$f(t)$ (the same as the one in Fig.~\ref{fig:long}(A)).
Figure~\ref{fig:longHidden}(B) shows, when the observable $P(t)$ is not updated
with the real data, the predicted time series (red) $P(t)$ diverges from the
true time series (blue) after about a dozen oscillations. However, if $P(t)$
is updated to the digital twin with the true values at the times indicated by
the purple vertical lines in Fig.~\ref{fig:longHidden}(C), the predicted time
series $P(t)$ matches the ground truth for a much longer time. The results
suggest that the existence of the hidden variable does not significantly
impede the performance of continual forecasting.

The results in Fig.~\ref{fig:longHidden} motivate the following questions.
First, has the reservoir network encoded information about the hidden variable?
Second, suppose it is known that there is a hidden variable and the training 
dataset contains this variable, can its evolution be inferred with only rare 
updates of the observable during continual forecasting? Previous 
results~\cite{LPHGBO:2017,ZP:2018,WYGZS:2019} suggested that reservoir 
computing can be used to infer the hidden variables in a nonlinear dynamical
system. Here we show that, with a segment of the time series of $N(t)$ used
only for training an additional readout layer, our digital twin can forecast
$N(t)$ with only occasional inputs of the observable time series $P(t)$. In 
particular, the additional readout layer for $N(t)$ is used only for extracting
information about $N(t)$ from the reservoir network and its output is never 
injected back to the reservoir. Consequently, whether this additional task
of inferring $N(t)$ is included or not, the trained output layer for $P(t)$ 
and the forecasting results of $P(t)$ are not altered. 

Figure~\ref{fig:longHidden}(D) shows that, when the observable $P(t)$ is not 
updated with the real data, the digital twin can to infer the hidden 
variable $N(t)$ for several oscillations. If $P(t)$ is updated with the true 
value at the times indicated by the purple vertical lines in 
Fig.~\ref{fig:longHidden}(C), the dynamical evolution of the hidden variable 
$N(t)$ can also be accurately predicted for a much longer period of time, as 
shown in Fig.~\ref{fig:longHidden}(E). It is worth emphasizing that during
the whole process of forecasting and monitoring, no information about the 
hidden variable $N(t)$ is required - only sparse data points of the observable 
$P(t)$ are used.

The training and testing settings of the digital twin for the task involving
a hidden variable are as follows. The input dimension of the reservoir is 
$D_{\textrm{in}}=1$ because there is a single observable $\log_{10} P(t)$. 
The output dimension is $D_{\textrm{out}}=2$ with one dimension of the 
observable $\log_{10} P(t+\Delta t)$ in addition to one dimension of the hidden variable 
$N(t+\Delta t)$. Because of the higher memory requirement in dealing with a 
hidden variable, a somewhat larger reservoir network is needed, so we use 
$D_r=1,000$. The times step of the dynamical evolution of the neural network 
is $\Delta t=0.1$. The training and validating lengths for each value of the
driving amplitude in the training are $t=3,500$ and $t=350$, respectively. 
Other optimized hyperparameters of the reservoir are $d=450$, $\lambda=1.15$, 
$k_\textrm{in}=0.32$, $k_c=3.1$, $\alpha=0.077$, $\beta=1\times10^{-8.3}$,
and $\sigma_{\textrm{noise}}=10^{-3.0}$.

It is also worth noting that Figs.~\ref{fig:long} and \ref{fig:longHidden} 
have demonstrated the ability of the digital twin to extrapolate beyond the 
parameter regime of the target system from which the training data are 
obtained. In particular, the digital twin was trained only with time series
under stationary external driving of the amplitude $A=0.35, 0.4, 0.45,$ and 
$0.5$. During the testing phase associated with both Figs.~\ref{fig:long} and 
\ref{fig:longHidden}, the external driving is non-stationary with its amplitude
linearly increasing from $A=0.4$ to $A=0.6$. The second half of the time series
$P(t)$ and $N(t)$ in Figs.~\ref{fig:long} and \ref{fig:longHidden} are thus 
beyond the training parameter regime.

The results in Figs.~\ref{fig:long} and \ref{fig:longHidden} help legitimize
the terminology ``digital twin,'' as the reservoir computers subject to the
external driving are dynamical twin systems that evolve ``in parallel'' to
the corresponding real systems. Even when the target system is only partially 
observable, the digital twin contains both the observable and hidden variables
whose dynamical evolution is encoded in the recurrent neural network in the 
hidden layer. The dynamical evolution of the output is constantly (albeit 
infrequently) corrected by sparse feedback from the real system, so the 
output trajectory of the digital twin shadows the true trajectory of the target
system. Suppose one wishes to monitor a variable in the target system, it is
only necessary to read it from the digital twin instead of making more 
(possibly costly) measurements on the real system.

\section{Digital twins under external driving with varied waveform}

\begin{figure*}[ht!]
\centering
\includegraphics[width=0.8\linewidth]{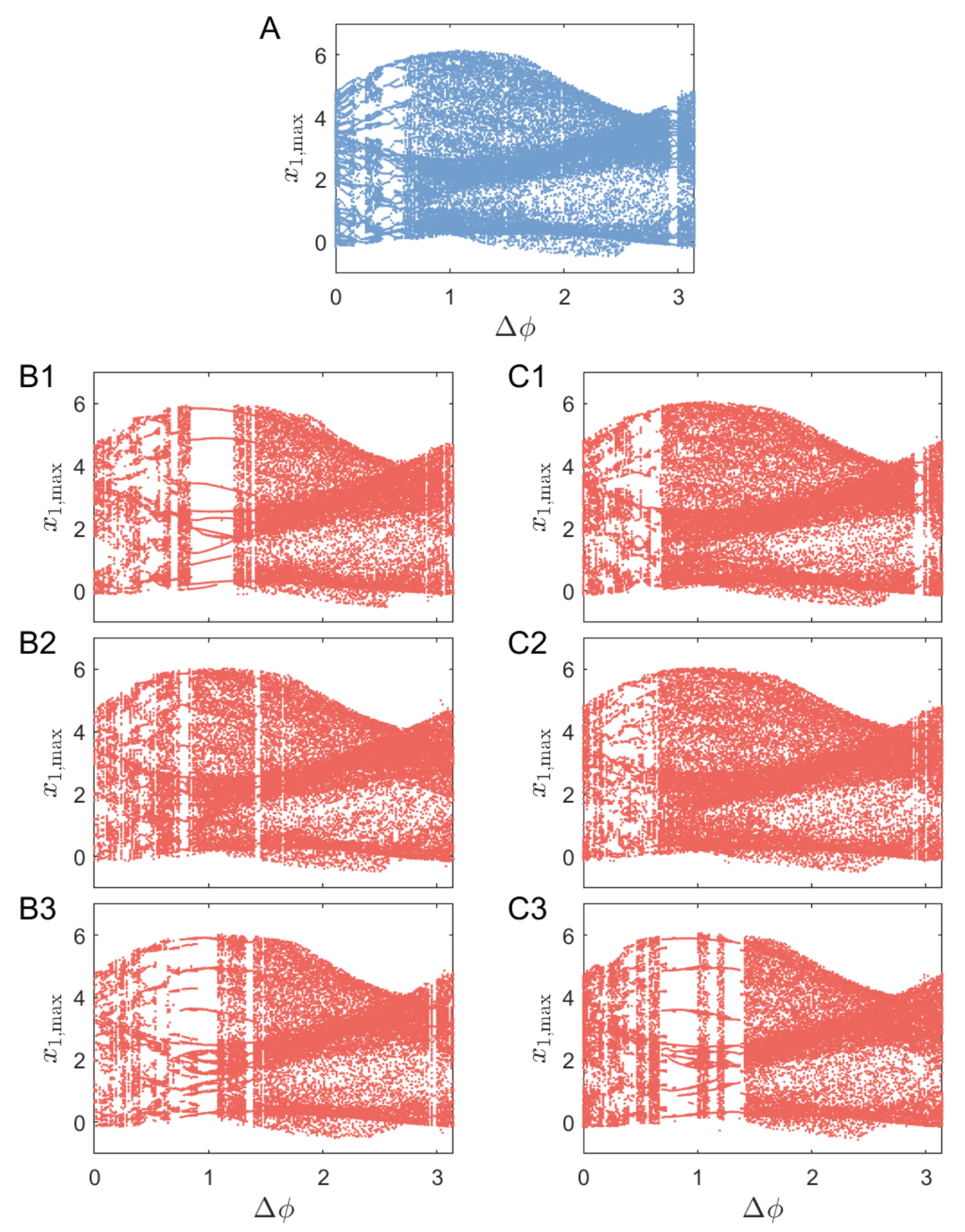}
\caption{\footnotesize
Comparisons of the prediction performance between the noiseless (left)
and noisy (right) cases on the task of predicting under external driving with 
different waveform. The target system is a six-dimensional Lorenz-96 system. 
Panel (A) shows the true bifurcation diagram. Panels (B1-B3) show the prediction
results without any dynamical noise in the training data with three realizations
of the reservoir network. Panels (C1-C3) show the prediction results with
dynamical noise of a strength $\delta_{\text{DB}}=3\times10^{-3}$ in the training
data. The settings are the same as that in Fig.~\ref{fig:phi} in the main text.}
\label{fig:SIPhi}
\end{figure*}

In the main text, it is demonstrated that dynamical noise added to the
driving signal during the training can be beneficial. Figure~\ref{fig:SIPhi}
presents a comparison between the noiseless training and the training
with dynamical noise of a strength $\delta_{\text{DB}}=3\times10^{-3}$ (as
in the main text). The ground-truth bifurcation diagram is shown in 
Fig.~\ref{fig:SIPhi}(A) and three examples with different reservoir neural 
networks for the noiseless (B1, B2, B3) and noisy (C1, C2, C3) training 
schemes are shown. All the settings other than the noise level are the same
as that in Fig.~\ref{fig:phi} in the main text. Though there are still  
fluctuations in the predicted results, adding dynamical noise into the 
training data can produce bifurcation diagrams that are in general closer to 
the ground truth than without noise. 

The results shown in Fig.~\ref{fig:SIPhi} also raises the issue of performance 
fluctuations in the predicted results among different randomly generated RC 
networks~\cite{HR:2020}. It is necessary to train an ensemble of RC networks 
to obtain a statistical quantification of the performance. An example is 
presented in Appendix F, where it is shown that the ensemble average of the 
predicted crisis point is accurate. 

\begin{figure*}[ht!]
\centering
\includegraphics[width=0.8\linewidth]{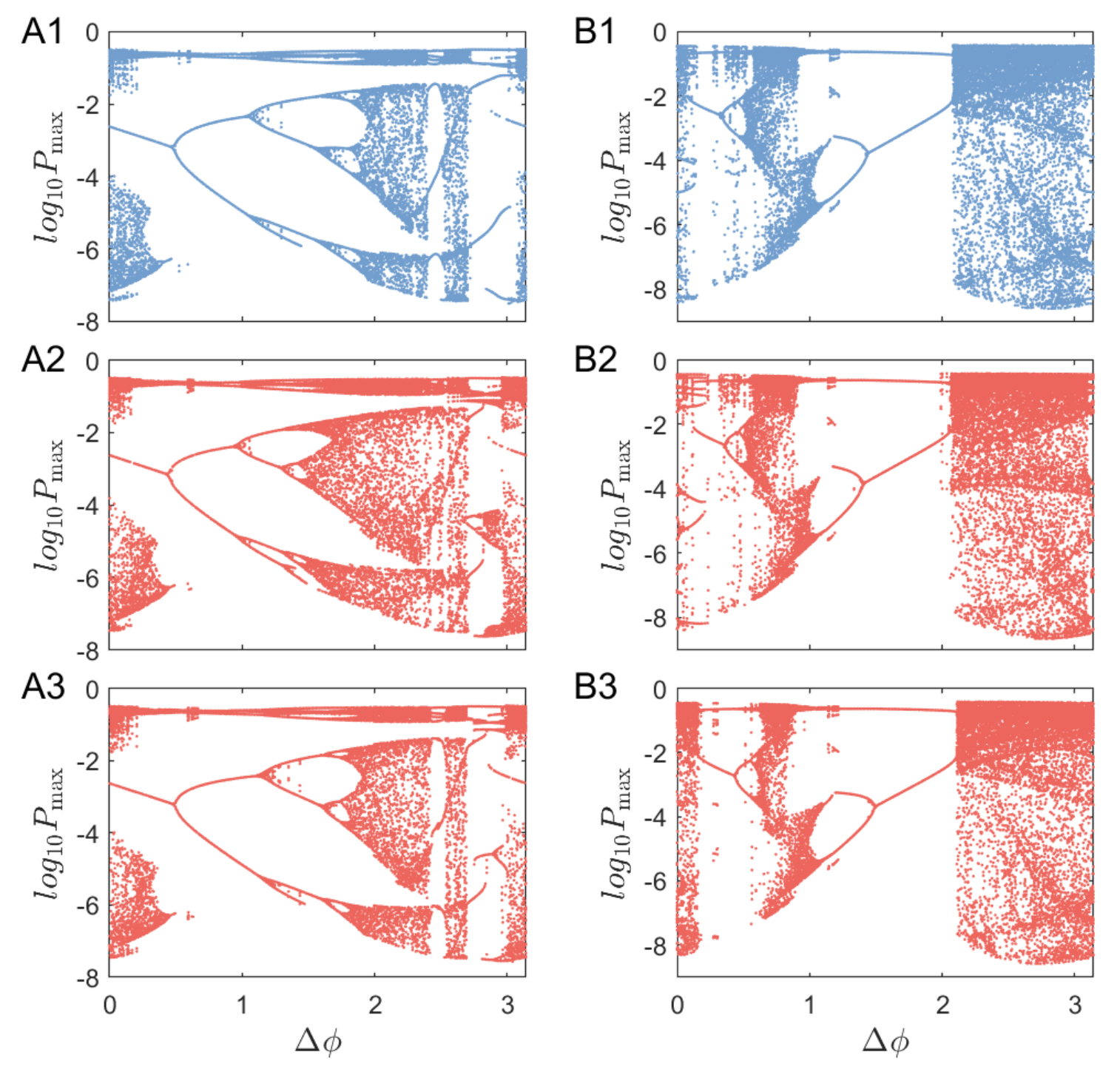}
\caption{\footnotesize 
Performance of the digital twin with the ecological model under driving signals 
with waveform different from the training set. The testing driving signals 
are described by Eq.~\ref{eq:SIPhi} while the training driving signals are 
sinusoidal waves with small dynamical noise. (A1) The real bifurcation diagram 
for $A_{\text{test}}=0.3$. (A2, A3) Predicted bifurcation diagrams for 
$A_{\text{test}}=0.3$ with two random realizations of the reservoir networks. 
(B1-B3) Same as (A1-A3) but with $A_{\text{test}}=0.4$.}
\label{fig:SIEcoPhi}
\end{figure*}

To further demonstrate the beneficial role of noise, we test the additive 
training noise scheme using the ecological system. The training process and 
hyperparameter values of the digital twin are identical to these in 
\ref{sec:eco}. A dynamical noise of amplitude 
$\delta_{\text{DB}}=3\times10^{-4}$ is added to the driving signal $f(t)$ 
during training in the same way as in Fig.~\ref{fig:phi} in the main text. During testing,
the driving signals is altered to
\begin{align}
f_{\text{test}}(t) &=A_{\text{test}}\sin(\omega_{{\rm eco}} t) 
+ \frac{A_{\text{test}}}{2}\sin(\frac{\omega_{{\rm eco}}}{2} t +\Delta \phi) \label{eq:SIPhi}
\end{align}
where $\omega_{{\rm eco}}=0.19$ as in \ref{sec:eco}. Two sets of testing
signals $f_{\text{test}}(t)$ are used, with $A_{\text{test}}=0.3$ and 0.4, 
respectively. Figure~\ref{fig:SIEcoPhi} show the true and predicted bifurcation
diagrams of $\log_{10}P_{max}$ versus $\Delta \phi$ for $A_{\text{test}}=0.3$ 
(left column) and $A_{\text{test}}=0.4$ (right column). It can be seen that
the bifurcation diagrams generated by the digital twin with the aid of 
training noise are remarkably accurate. We also find that, for this ecological
system, the amplitude $\delta_{\text{DB}}$ of the dynamical noise during 
training does not have a significant effect on the predicted bifurcation 
diagram. A plausible reason is that the driving signal $f(t)$ is a 
multiplicative term in the system equations.

\section{Quantitative characterization of performance of digital twin} \label{sec:Q}

In the main text, we demonstrate the performance of the digital twin 
qualitatively based on visually comparing the predicted bifurcation diagram 
with the ground truth. Given the rich bifurcation structure, to quantify the 
similarities between two bifurcation diagrams is difficult. However, 
for a bifurcation diagram, the parameter values at 
which the various bifurcations occur are of great interest, as they define the 
critical points at which characteristic changes in the system can occur. In this
section we focus on the crisis point at which sustained chaotic motion on an
attractor is destroyed and replaced by transient chaos. And, accordingly,
we use two quantities to characterize the performance of the digital twin in
extrapolating the dynamics of the target system: the errors in the predicted
critical bifurcation point and average lifetime of the chaotic transient
after the bifurcation. 

\begin{figure*}[ht!]
\centering
\includegraphics[width=0.5\linewidth]{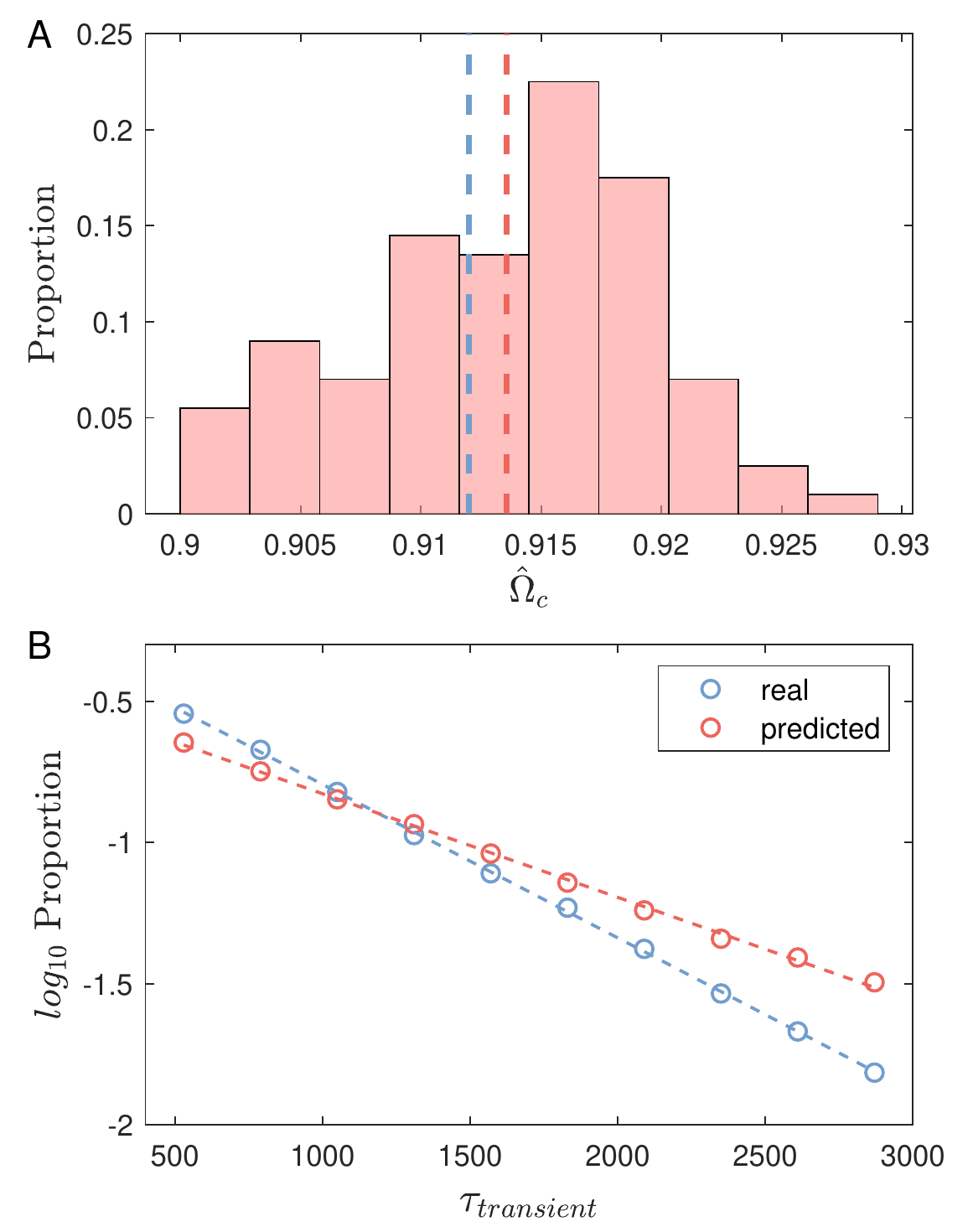}
\caption{\footnotesize
Quantitative performance of the digital twin for a chaotic driven 
laser system. (A) Distribution of the predicted values of the crisis
bifurcation point $\hat\Omega_c$, at which a chaotic attractor is destroyed
and replaced by a non-attracting chaotic invariant set leading to transient 
chaos. The blue and red vertical dashed lines denote the true value 
$\Omega_c \approx 0.912$ and the average predicted value 
$\langle\hat\Omega_c\rangle$, respectively, where 200 random realizations of 
the reservoir neural network are used to generate this distribution. Despite
the fluctuations in the predicted crisis point, the ensemble average value 
of the prediction is quite close to the ground truth. (B) Exponential 
distribution of the lifetime of transient chaos slightly beyond the crisis 
point: true (blue) and predicted (red) behaviors. The predicted distribution
is generated using 100 random reservoir realizations, each with 200 random
initial `warming up'' data.}
\label{fig:SICPCT}
\end{figure*}

\begin{figure*} [ht!]
\centering
\includegraphics[width=0.8\linewidth]{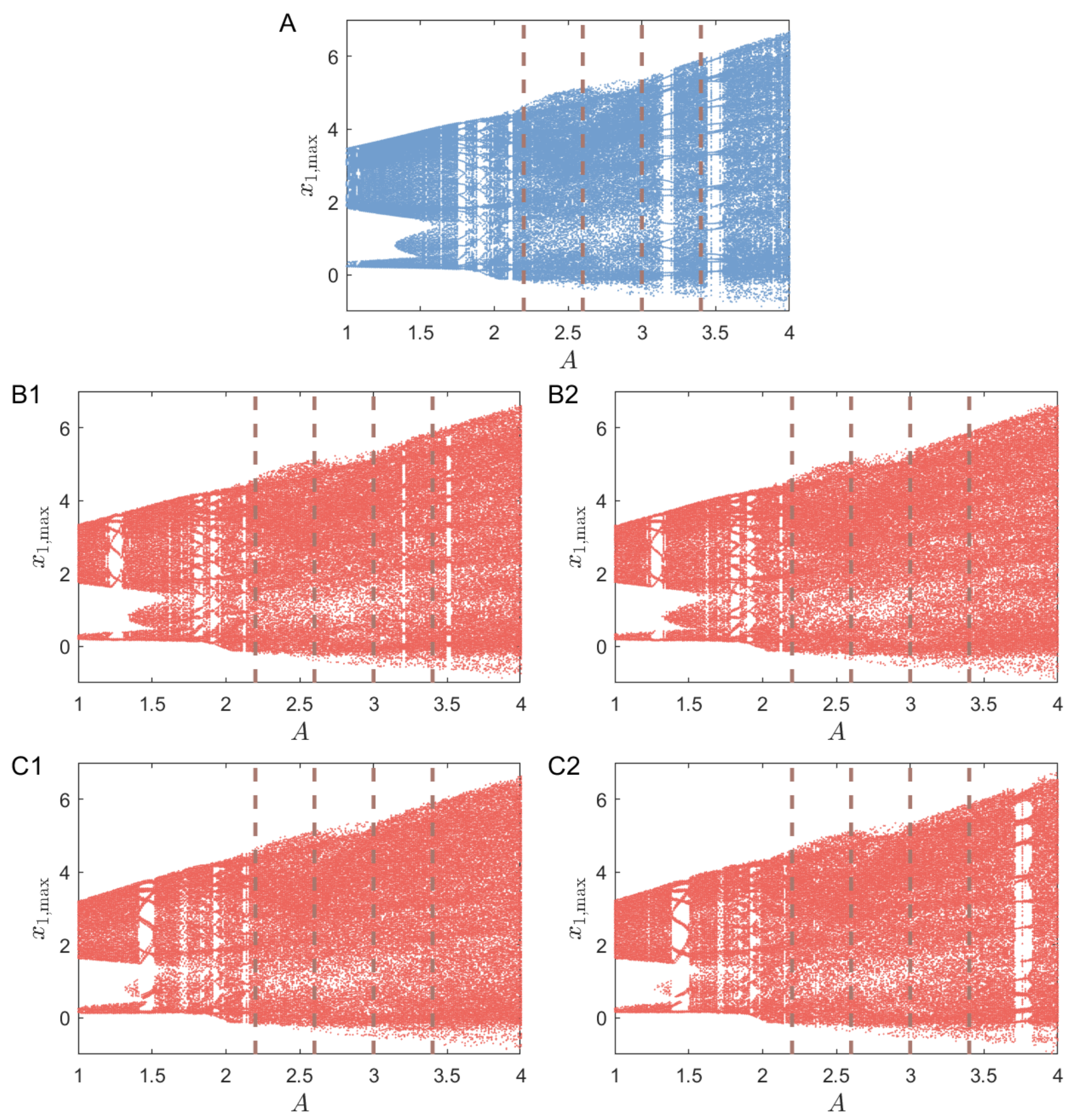}
\caption{\footnotesize
Robustness of digital twin against combined dynamical and observational noises.
The setting is the same as that in Fig.~\ref{fig:m6} in the main text, except with 
additional noises in the training data. (A) A true bifurcation diagram of the 
six-dimensional Lorenz-96 system. (B1, B2) Two examples of the bifurcation 
diagram predicted by the digital twin with training data under dynamical noise 
of amplitude $\sigma_{\text{dyn}}=10^{-2}$ and observational noise of amplitude
$\sigma_{\text{ob}}=10^{-2}$. (C1, C2) Two examples of the predicted bifurcation
diagrams under the two kinds of noise with $\sigma_{\text{dyn}}=10^{-1}$ and 
$\sigma_{\text{ob}}=10^{-1}$. Both the dynamical and observational noises are 
additive Gaussian processes. It can be seen that though larger additional noises
make the predicted details less accurate, the general shapes of the predicted
results are not harmed significantly. The settings 
of the training data and reservoir neural networks are the same as those in 
Fig.~2 in the main text. The dynamical noises are added to the dynamical
equations of the state variables. There is no noise in the sinusoidal external
driving.}
\label{fig:SInoise}
\end{figure*}

As an illustrative example, we take the driven chaotic laser system in 
Appendix~\ref{sec:laser}, where a crisis bifurcation occurs at the critical driving
frequency $\Omega_c \approx 0.912$ at which the chaotic attractor of the system
is destroyed and replaced by a non-attracting chaotic invariant set leading to 
transient chaos. We test to determine if the digital twin can faithfully predict
the crisis point based only on training data from the parameter regime of a
chaotic attractor. Let $\hat\Omega_c$ be the digital-twin predicted critical
point. Figure~\ref{fig:SICPCT}(A) shows the distribution of $\hat\Omega_c$ 
obtained from 200 random realizations of the reservoir neural network. Despite
the fluctuations in the predicted $\hat\Omega_c$, their average value is
$\langle\hat\Omega_c\rangle = 0.914$, which is close to the true value 
$\Omega_c=0.912$. A relative error $\varepsilon_{\Omega}$ of $\hat\Omega_c$ 
can then be defined as
\begin{align}
\varepsilon_{\Omega} = \frac{ |\Omega_c-\hat\Omega_c|}{D(\Omega_c,\{\Omega_\text{train}\})},
\end{align}
where $D(\Omega_c,\{\Omega_\text{train}\})$ denotes the minimal distance from 
$\Omega_c$ to the set of training parameter points $\{\Omega_\text{train}\}$, 
i.e., the difference between $\Omega_c$ and the closest training point. For
the driven laser system, we have 
$D(\Omega_c,\{\Omega_\text{train}\}) \approx 10\%$. 

The second quantity is the lifetime $\tau_{\text{transient}}$ of transient 
chaos after the crisis bifurcation~\cite{KFGL:2021a,KFGL:2021b}, as shown in 
Fig.~\ref{fig:SICPCT}(B). The average transient lifetime is the inverse of 
the slope of the linear regression of predicted data points in 
Fig.~\ref{fig:SICPCT}(B), which is $\langle \tau\rangle \approx 0.8\times 10^3$. 
Compared with the true value $\langle \tau\rangle \approx 1.2\times10^3$, 
we see that the digital twin is able to predict the average chaotic transient
lifetime to within the same order of magnitude. Considering that key to the
transient dynamics is the small escaping region in Fig.~\ref{fig:LaserMain}(D2),
which is sensitive to the inevitable training errors, the performance can 
be deemed as satisfactory.

\section{Robustness of digital twin against combined dynamical/observational noises}

Can our RC based digital twins withstand the influences of 
different types of noises? To address this question, we introduce dynamical 
and observational noises in the training data, which are modeled as additive 
Gaussian noises. Take the six-dimensional Lorenz-96 system in Sec.~IIA in the 
main text as an example. Figure~\ref{fig:SInoise}(A) shows the true bifurcation
diagram under different amplitudes of external driving, where the vertical 
dashed lines specify the training points. Figures~\ref{fig:SInoise}(B1) and 
\ref{fig:SInoise}(B2) show two realizations of the bifurcation diagram 
generated by the digital twin under both dynamical and observational noises 
of amplitudes $\sigma_{\text{dyn}}=10^{-2}$ and $\sigma_{\text{ob}}=10^{-2}$. 
Two bifurcation diagrams for noise amplitudes of an order of magnitude larger: 
$\sigma_{\text{dyn}}=10^{-1}$ and $\sigma_{\text{ob}}=10^{-1}$, are shown
in Figs.~\ref{fig:SInoise}(C1) and \ref{fig:SInoise}(C2). It can be seen that 
the additional noises have little effect on the performance of the digital 
twin in generating the bifurcation diagram.

%\bibliographystyle{naturemag}
%\bibliography{DT}

\begin{thebibliography}{85}%
\makeatletter
\providecommand \@ifxundefined [1]{%
 \@ifx{#1\undefined}
}%
\providecommand \@ifnum [1]{%
 \ifnum #1\expandafter \@firstoftwo
 \else \expandafter \@secondoftwo
 \fi
}%
\providecommand \@ifx [1]{%
 \ifx #1\expandafter \@firstoftwo
 \else \expandafter \@secondoftwo
 \fi
}%
\providecommand \natexlab [1]{#1}%
\providecommand \enquote  [1]{``#1''}%
\providecommand \bibnamefont  [1]{#1}%
\providecommand \bibfnamefont [1]{#1}%
\providecommand \citenamefont [1]{#1}%
\providecommand \href@noop [0]{\@secondoftwo}%
\providecommand \href [0]{\begingroup \@sanitize@url \@href}%
\providecommand \@href[1]{\@@startlink{#1}\@@href}%
\providecommand \@@href[1]{\endgroup#1\@@endlink}%
\providecommand \@sanitize@url [0]{\catcode `\\12\catcode `\$12\catcode
  `\&12\catcode `\#12\catcode `\^12\catcode `\_12\catcode `\%12\relax}%
\providecommand \@@startlink[1]{}%
\providecommand \@@endlink[0]{}%
\providecommand \url  [0]{\begingroup\@sanitize@url \@url }%
\providecommand \@url [1]{\endgroup\@href {#1}{\urlprefix }}%
\providecommand \urlprefix  [0]{URL }%
\providecommand \Eprint [0]{\href }%
\providecommand \doibase [0]{https://doi.org/}%
\providecommand \selectlanguage [0]{\@gobble}%
\providecommand \bibinfo  [0]{\@secondoftwo}%
\providecommand \bibfield  [0]{\@secondoftwo}%
\providecommand \translation [1]{[#1]}%
\providecommand \BibitemOpen [0]{}%
\providecommand \bibitemStop [0]{}%
\providecommand \bibitemNoStop [0]{.\EOS\space}%
\providecommand \EOS [0]{\spacefactor3000\relax}%
\providecommand \BibitemShut  [1]{\csname bibitem#1\endcsname}%
\let\auto@bib@innerbib\@empty
%</preamble>
\bibitem [{\citenamefont {Eric J.~Tuegel}\ \emph {et~al.}(2011)\citenamefont
  {Eric J.~Tuegel}, \citenamefont {Ingraffea}, \citenamefont {Eason},\ and\
  \citenamefont {Spottswood}}]{TIES:2011}%
  \BibitemOpen
  \bibfield  {author} {\bibinfo {author} {\bibfnamefont {E.~J.}\ \bibnamefont
  {Eric J.~Tuegel}}, \bibinfo {author} {\bibfnamefont {A.~R.}\ \bibnamefont
  {Ingraffea}}, \bibinfo {author} {\bibfnamefont {T.~G.}\ \bibnamefont
  {Eason}},\ and\ \bibinfo {author} {\bibfnamefont {S.~M.}\ \bibnamefont
  {Spottswood}},\ }\bibfield  {title} {\bibinfo {title} {Reengineering aircraft
  structural life prediction using a digital twin},\ }\href@noop {} {\bibfield
  {journal} {\bibinfo  {journal} {Int. J. Aerospace Eng.}\ }\textbf {\bibinfo
  {volume} {2011}},\ \bibinfo {pages} {154798} (\bibinfo {year}
  {2011})}\BibitemShut {NoStop}%
\bibitem [{\citenamefont {Tao}\ and\ \citenamefont {Qi}(2019)}]{TQ:2019}%
  \BibitemOpen
  \bibfield  {author} {\bibinfo {author} {\bibfnamefont {F.}~\bibnamefont
  {Tao}}\ and\ \bibinfo {author} {\bibfnamefont {Q.}~\bibnamefont {Qi}},\
  }\bibfield  {title} {\bibinfo {title} {Make more digital twins},\ }\href@noop
  {} {\bibfield  {journal} {\bibinfo  {journal} {Nature}\ }\textbf {\bibinfo
  {volume} {573}},\ \bibinfo {pages} {274} (\bibinfo {year}
  {2019})}\BibitemShut {NoStop}%
\bibitem [{\citenamefont {Rasheed}\ \emph {et~al.}(2020)\citenamefont
  {Rasheed}, \citenamefont {San},\ and\ \citenamefont {Kvamsdal}}]{RSK:2020}%
  \BibitemOpen
  \bibfield  {author} {\bibinfo {author} {\bibfnamefont {A.}~\bibnamefont
  {Rasheed}}, \bibinfo {author} {\bibfnamefont {O.}~\bibnamefont {San}},\ and\
  \bibinfo {author} {\bibfnamefont {T.}~\bibnamefont {Kvamsdal}},\ }\bibfield
  {title} {\bibinfo {title} {Digital twin: Values, challenges and enablers from
  a modeling perspective},\ }\href@noop {} {\bibfield  {journal} {\bibinfo
  {journal} {IEEE Access}\ }\textbf {\bibinfo {volume} {8}},\ \bibinfo {pages}
  {21980} (\bibinfo {year} {2020})}\BibitemShut {NoStop}%
\bibitem [{\citenamefont {Bruynseels}\ \emph {et~al.}(2018)\citenamefont
  {Bruynseels}, \citenamefont {de~Sio},\ and\ \citenamefont {van~den
  Hoven}}]{BSvdH:2018}%
  \BibitemOpen
  \bibfield  {author} {\bibinfo {author} {\bibfnamefont {K.}~\bibnamefont
  {Bruynseels}}, \bibinfo {author} {\bibfnamefont {F.~S.}\ \bibnamefont
  {de~Sio}},\ and\ \bibinfo {author} {\bibfnamefont {J.}~\bibnamefont {van~den
  Hoven}},\ }\bibfield  {title} {\bibinfo {title} {Digital twins in health
  care: {Ethical} implications of an emerging engineering paradigm},\
  }\href@noop {} {\bibfield  {journal} {\bibinfo  {journal} {Front. Gene.}\
  }\textbf {\bibinfo {volume} {9}},\ \bibinfo {pages} {31} (\bibinfo {year}
  {2018})}\BibitemShut {NoStop}%
\bibitem [{\citenamefont {Schwartz}\ \emph {et~al.}(2020)\citenamefont
  {Schwartz}, \citenamefont {Wildenhaus}, \citenamefont {Bucher},\ and\
  \citenamefont {Byrd}}]{SWBB:2020}%
  \BibitemOpen
  \bibfield  {author} {\bibinfo {author} {\bibfnamefont {S.~M.}\ \bibnamefont
  {Schwartz}}, \bibinfo {author} {\bibfnamefont {K.}~\bibnamefont
  {Wildenhaus}}, \bibinfo {author} {\bibfnamefont {A.}~\bibnamefont {Bucher}},\
  and\ \bibinfo {author} {\bibfnamefont {B.}~\bibnamefont {Byrd}},\ }\bibfield
  {title} {\bibinfo {title} {Digital twins and the emerging science of self:
  Implications for digital health experience design and ``small'' data},\
  }\href {https://doi.org/10.3389/fcomp.2020.00031} {\bibfield  {journal}
  {\bibinfo  {journal} {Front. Comp. Sci.}\ }\textbf {\bibinfo {volume} {2}},\
  \bibinfo {pages} {31} (\bibinfo {year} {2020})}\BibitemShut {NoStop}%
\bibitem [{\citenamefont {Laubenbacher}\ \emph {et~al.}(2021)\citenamefont
  {Laubenbacher}, \citenamefont {Sluka},\ and\ \citenamefont
  {Glazier}}]{LSG:2021}%
  \BibitemOpen
  \bibfield  {author} {\bibinfo {author} {\bibfnamefont {R.}~\bibnamefont
  {Laubenbacher}}, \bibinfo {author} {\bibfnamefont {J.~P.}\ \bibnamefont
  {Sluka}},\ and\ \bibinfo {author} {\bibfnamefont {J.~A.}\ \bibnamefont
  {Glazier}},\ }\bibfield  {title} {\bibinfo {title} {Using digital twins in
  viral infection},\ }\href {https://doi.org/10.1126/science.abf3370}
  {\bibfield  {journal} {\bibinfo  {journal} {Science}\ }\textbf {\bibinfo
  {volume} {371}},\ \bibinfo {pages} {1105} (\bibinfo {year}
  {2021})}\BibitemShut {NoStop}%
\bibitem [{\citenamefont {Voosen}(2020)}]{Voosen:2020}%
  \BibitemOpen
  \bibfield  {author} {\bibinfo {author} {\bibfnamefont {P.}~\bibnamefont
  {Voosen}},\ }\bibfield  {title} {\bibinfo {title} {Europe builds ‘digital
  twin’ of earth to hone climate forecasts},\ }\href@noop {} {\bibfield
  {journal} {\bibinfo  {journal} {Science}\ }\textbf {\bibinfo {volume}
  {370}},\ \bibinfo {pages} {16} (\bibinfo {year} {2020})}\BibitemShut
  {NoStop}%
\bibitem [{\citenamefont {Bauer}\ \emph {et~al.}(2021)\citenamefont {Bauer},
  \citenamefont {Stevens},\ and\ \citenamefont {Hazeleger}}]{BSH:2021}%
  \BibitemOpen
  \bibfield  {author} {\bibinfo {author} {\bibfnamefont {P.}~\bibnamefont
  {Bauer}}, \bibinfo {author} {\bibfnamefont {B.}~\bibnamefont {Stevens}},\
  and\ \bibinfo {author} {\bibfnamefont {W.}~\bibnamefont {Hazeleger}},\
  }\bibfield  {title} {\bibinfo {title} {A digital twin of earth for the green
  transition},\ }\href@noop {} {\bibfield  {journal} {\bibinfo  {journal} {Nat.
  Clim. Change}\ }\textbf {\bibinfo {volume} {11}},\ \bibinfo {pages} {80}
  (\bibinfo {year} {2021})}\BibitemShut {NoStop}%
\bibitem [{\citenamefont {Lai}\ and\ \citenamefont {T\'{e}l}(2011)}]{LT:book}%
  \BibitemOpen
  \bibfield  {author} {\bibinfo {author} {\bibfnamefont {Y.-C.}\ \bibnamefont
  {Lai}}\ and\ \bibinfo {author} {\bibfnamefont {T.}~\bibnamefont {T\'{e}l}},\
  }\href@noop {} {\emph {\bibinfo {title} {Transient Chaos - Complex Dynamics
  on Finite Time Scales}}}\ (\bibinfo  {publisher} {Springer},\ \bibinfo
  {address} {New York},\ \bibinfo {year} {2011})\BibitemShut {NoStop}%
\bibitem [{\citenamefont {McCann}\ and\ \citenamefont
  {Yodzis}(1994)}]{MY:1994}%
  \BibitemOpen
  \bibfield  {author} {\bibinfo {author} {\bibfnamefont {K.}~\bibnamefont
  {McCann}}\ and\ \bibinfo {author} {\bibfnamefont {P.}~\bibnamefont
  {Yodzis}},\ }\bibfield  {title} {\bibinfo {title} {Nonlinear dynamics and
  population disappearances},\ }\href@noop {} {\bibfield  {journal} {\bibinfo
  {journal} {Ame. Naturalist}\ }\textbf {\bibinfo {volume} {144}},\ \bibinfo
  {pages} {873} (\bibinfo {year} {1994})}\BibitemShut {NoStop}%
\bibitem [{\citenamefont {Hastings}\ \emph {et~al.}(2018)\citenamefont
  {Hastings}, \citenamefont {Abbott}, \citenamefont {Cuddington}, \citenamefont
  {Francis}, \citenamefont {Gellner}, \citenamefont {Lai}, \citenamefont
  {Morozov}, \citenamefont {Petrivskii}, \citenamefont {Scranton},\ and\
  \citenamefont {Zeeman}}]{HACFGLMPSZ:2018}%
  \BibitemOpen
  \bibfield  {author} {\bibinfo {author} {\bibfnamefont {A.}~\bibnamefont
  {Hastings}}, \bibinfo {author} {\bibfnamefont {K.~C.}\ \bibnamefont
  {Abbott}}, \bibinfo {author} {\bibfnamefont {K.}~\bibnamefont {Cuddington}},
  \bibinfo {author} {\bibfnamefont {T.}~\bibnamefont {Francis}}, \bibinfo
  {author} {\bibfnamefont {G.}~\bibnamefont {Gellner}}, \bibinfo {author}
  {\bibfnamefont {Y.-C.}\ \bibnamefont {Lai}}, \bibinfo {author} {\bibfnamefont
  {A.}~\bibnamefont {Morozov}}, \bibinfo {author} {\bibfnamefont
  {S.}~\bibnamefont {Petrivskii}}, \bibinfo {author} {\bibfnamefont
  {K.}~\bibnamefont {Scranton}},\ and\ \bibinfo {author} {\bibfnamefont
  {M.~L.}\ \bibnamefont {Zeeman}},\ }\bibfield  {title} {\bibinfo {title}
  {Transient phenomena in ecology},\ }\href@noop {} {\bibfield  {journal}
  {\bibinfo  {journal} {Science}\ }\textbf {\bibinfo {volume} {361}},\ \bibinfo
  {pages} {eaat6412} (\bibinfo {year} {2018})}\BibitemShut {NoStop}%
\bibitem [{\citenamefont {Dhamala}\ and\ \citenamefont {Lai}(1999)}]{DL:1999}%
  \BibitemOpen
  \bibfield  {author} {\bibinfo {author} {\bibfnamefont {M.}~\bibnamefont
  {Dhamala}}\ and\ \bibinfo {author} {\bibfnamefont {Y.-C.}\ \bibnamefont
  {Lai}},\ }\bibfield  {title} {\bibinfo {title} {Controlling transient chaos
  in deterministic flows with applications to electrical power systems and
  ecology},\ }\href {https://doi.org/10.1103/PhysRevE.59.1646} {\bibfield
  {journal} {\bibinfo  {journal} {Phys. Rev. E}\ }\textbf {\bibinfo {volume}
  {59}},\ \bibinfo {pages} {1646} (\bibinfo {year} {1999})}\BibitemShut
  {NoStop}%
\bibitem [{\citenamefont {Lai}\ \emph {et~al.}(1999)\citenamefont {Lai},
  \citenamefont {Grebogi},\ and\ \citenamefont {Kurths}}]{LGK:1999}%
  \BibitemOpen
  \bibfield  {author} {\bibinfo {author} {\bibfnamefont {Y.-C.}\ \bibnamefont
  {Lai}}, \bibinfo {author} {\bibfnamefont {C.}~\bibnamefont {Grebogi}},\ and\
  \bibinfo {author} {\bibfnamefont {J.}~\bibnamefont {Kurths}},\ }\bibfield
  {title} {\bibinfo {title} {Modeling of deterministic chaotic systems},\
  }\href {https://doi.org/10.1103/PhysRevE.59.2907} {\bibfield  {journal}
  {\bibinfo  {journal} {Phys. Rev. E}\ }\textbf {\bibinfo {volume} {59}},\
  \bibinfo {pages} {2907} (\bibinfo {year} {1999})}\BibitemShut {NoStop}%
\bibitem [{\citenamefont {Lai}\ and\ \citenamefont {Grebogi}(1999)}]{LG:1999}%
  \BibitemOpen
  \bibfield  {author} {\bibinfo {author} {\bibfnamefont {Y.-C.}\ \bibnamefont
  {Lai}}\ and\ \bibinfo {author} {\bibfnamefont {C.}~\bibnamefont {Grebogi}},\
  }\bibfield  {title} {\bibinfo {title} {Modeling of coupled chaotic
  oscillators},\ }\href {https://doi.org/10.1103/PhysRevLett.82.4803}
  {\bibfield  {journal} {\bibinfo  {journal} {Phys. Rev. Lett.}\ }\textbf
  {\bibinfo {volume} {82}},\ \bibinfo {pages} {4803} (\bibinfo {year}
  {1999})}\BibitemShut {NoStop}%
\bibitem [{\citenamefont {Wang}\ \emph {et~al.}(2011)\citenamefont {Wang},
  \citenamefont {Yang}, \citenamefont {Lai}, \citenamefont {Kovanis},\ and\
  \citenamefont {Grebogi}}]{WYLKG:2011}%
  \BibitemOpen
  \bibfield  {author} {\bibinfo {author} {\bibfnamefont {W.-X.}\ \bibnamefont
  {Wang}}, \bibinfo {author} {\bibfnamefont {R.}~\bibnamefont {Yang}}, \bibinfo
  {author} {\bibfnamefont {Y.-C.}\ \bibnamefont {Lai}}, \bibinfo {author}
  {\bibfnamefont {V.}~\bibnamefont {Kovanis}},\ and\ \bibinfo {author}
  {\bibfnamefont {C.}~\bibnamefont {Grebogi}},\ }\bibfield  {title} {\bibinfo
  {title} {Predicting catastrophes in nonlinear dynamical systems by
  compressive sensing},\ }\href
  {https://doi.org/10.1103/PhysRevLett.106.154101} {\bibfield  {journal}
  {\bibinfo  {journal} {Phys. Rev. Lett.}\ }\textbf {\bibinfo {volume} {106}},\
  \bibinfo {pages} {154101} (\bibinfo {year} {2011})}\BibitemShut {NoStop}%
\bibitem [{\citenamefont {Wang}\ \emph {et~al.}(2016)\citenamefont {Wang},
  \citenamefont {Lai},\ and\ \citenamefont {Grebogi}}]{WLG:2016}%
  \BibitemOpen
  \bibfield  {author} {\bibinfo {author} {\bibfnamefont {W.-X.}\ \bibnamefont
  {Wang}}, \bibinfo {author} {\bibfnamefont {Y.-C.}\ \bibnamefont {Lai}},\ and\
  \bibinfo {author} {\bibfnamefont {C.}~\bibnamefont {Grebogi}},\ }\bibfield
  {title} {\bibinfo {title} {Data based identification and prediction of
  nonlinear and complex dynamical systems},\ }\href
  {https://doi.org/10.1016/j.physrep.2016.06.004} {\bibfield  {journal}
  {\bibinfo  {journal} {Phys. Rep.}\ }\textbf {\bibinfo {volume} {644}},\
  \bibinfo {pages} {1} (\bibinfo {year} {2016})}\BibitemShut {NoStop}%
\bibitem [{\citenamefont {Lai}(2021)}]{Lai:2021}%
  \BibitemOpen
  \bibfield  {author} {\bibinfo {author} {\bibfnamefont {Y.-C.}\ \bibnamefont
  {Lai}},\ }\bibfield  {title} {\bibinfo {title} {Finding nonlinear system
  equations and complex network structures from data: {A} sparse optimization
  approach},\ }\href@noop {} {\bibfield  {journal} {\bibinfo  {journal}
  {Chaos}\ }\textbf {\bibinfo {volume} {31}},\ \bibinfo {pages} {082101}
  (\bibinfo {year} {2021})}\BibitemShut {NoStop}%
\bibitem [{\citenamefont {Jaeger}(2001)}]{Jaeger:2001}%
  \BibitemOpen
  \bibfield  {author} {\bibinfo {author} {\bibfnamefont {H.}~\bibnamefont
  {Jaeger}},\ }\bibfield  {title} {\bibinfo {title} {The “echo state”
  approach to analysing and training recurrent neural networks-with an erratum
  note},\ }\href@noop {} {\bibfield  {journal} {\bibinfo  {journal} {German
  National Research Center for Information Technology GMD Technical Report}\
  }\textbf {\bibinfo {volume} {148}},\ \bibinfo {pages} {13} (\bibinfo {year}
  {2001})}\BibitemShut {NoStop}%
\bibitem [{\citenamefont {Mass}\ \emph {et~al.}(2002)\citenamefont {Mass},
  \citenamefont {Nachtschlaeger},\ and\ \citenamefont {Markram}}]{MNM:2002}%
  \BibitemOpen
  \bibfield  {author} {\bibinfo {author} {\bibfnamefont {W.}~\bibnamefont
  {Mass}}, \bibinfo {author} {\bibfnamefont {T.}~\bibnamefont
  {Nachtschlaeger}},\ and\ \bibinfo {author} {\bibfnamefont {H.}~\bibnamefont
  {Markram}},\ }\bibfield  {title} {\bibinfo {title} {Real-time computing
  without stable states: A new framework for neural computation based on
  perturbations},\ }\href@noop {} {\bibfield  {journal} {\bibinfo  {journal}
  {Neur. Comp.}\ }\textbf {\bibinfo {volume} {14}},\ \bibinfo {pages} {2531}
  (\bibinfo {year} {2002})}\BibitemShut {NoStop}%
\bibitem [{\citenamefont {Jaeger}\ and\ \citenamefont {Haas}(2004)}]{JH:2004}%
  \BibitemOpen
  \bibfield  {author} {\bibinfo {author} {\bibfnamefont {H.}~\bibnamefont
  {Jaeger}}\ and\ \bibinfo {author} {\bibfnamefont {H.}~\bibnamefont {Haas}},\
  }\bibfield  {title} {\bibinfo {title} {Harnessing nonlinearity: Predicting
  chaotic systems and saving energy in wireless communication},\ }\href@noop {}
  {\bibfield  {journal} {\bibinfo  {journal} {Science}\ }\textbf {\bibinfo
  {volume} {304}},\ \bibinfo {pages} {78} (\bibinfo {year} {2004})}\BibitemShut
  {NoStop}%
\bibitem [{\citenamefont {Haynes}\ \emph {et~al.}(2015)\citenamefont {Haynes},
  \citenamefont {Soriano}, \citenamefont {Rosin}, \citenamefont {Fischer},\
  and\ \citenamefont {Gauthier}}]{HSRFG:2015}%
  \BibitemOpen
  \bibfield  {author} {\bibinfo {author} {\bibfnamefont {N.~D.}\ \bibnamefont
  {Haynes}}, \bibinfo {author} {\bibfnamefont {M.~C.}\ \bibnamefont {Soriano}},
  \bibinfo {author} {\bibfnamefont {D.~P.}\ \bibnamefont {Rosin}}, \bibinfo
  {author} {\bibfnamefont {I.}~\bibnamefont {Fischer}},\ and\ \bibinfo {author}
  {\bibfnamefont {D.~J.}\ \bibnamefont {Gauthier}},\ }\bibfield  {title}
  {\bibinfo {title} {Reservoir computing with a single time-delay autonomous
  {Boolean} node},\ }\href {https://doi.org/10.1103/PhysRevE.91.020801}
  {\bibfield  {journal} {\bibinfo  {journal} {Phys. Rev. E}\ }\textbf {\bibinfo
  {volume} {91}},\ \bibinfo {pages} {020801} (\bibinfo {year}
  {2015})}\BibitemShut {NoStop}%
\bibitem [{\citenamefont {Larger}\ \emph {et~al.}(2017)\citenamefont {Larger},
  \citenamefont {Bayl\'on-Fuentes}, \citenamefont {Martinenghi}, \citenamefont
  {Udaltsov}, \citenamefont {Chembo},\ and\ \citenamefont
  {Jacquot}}]{LBMUCJ:2017}%
  \BibitemOpen
  \bibfield  {author} {\bibinfo {author} {\bibfnamefont {L.}~\bibnamefont
  {Larger}}, \bibinfo {author} {\bibfnamefont {A.}~\bibnamefont
  {Bayl\'on-Fuentes}}, \bibinfo {author} {\bibfnamefont {R.}~\bibnamefont
  {Martinenghi}}, \bibinfo {author} {\bibfnamefont {V.~S.}\ \bibnamefont
  {Udaltsov}}, \bibinfo {author} {\bibfnamefont {Y.~K.}\ \bibnamefont
  {Chembo}},\ and\ \bibinfo {author} {\bibfnamefont {M.}~\bibnamefont
  {Jacquot}},\ }\bibfield  {title} {\bibinfo {title} {High-speed photonic
  reservoir computing using a time-delay-based architecture: Million words per
  second classification},\ }\href {https://doi.org/10.1103/PhysRevX.7.011015}
  {\bibfield  {journal} {\bibinfo  {journal} {Phys. Rev. X}\ }\textbf {\bibinfo
  {volume} {7}},\ \bibinfo {pages} {011015} (\bibinfo {year}
  {2017})}\BibitemShut {NoStop}%
\bibitem [{\citenamefont {Pathak}\ \emph {et~al.}(2017)\citenamefont {Pathak},
  \citenamefont {Lu}, \citenamefont {Hunt}, \citenamefont {Girvan},\ and\
  \citenamefont {Ott}}]{PLHGO:2017}%
  \BibitemOpen
  \bibfield  {author} {\bibinfo {author} {\bibfnamefont {J.}~\bibnamefont
  {Pathak}}, \bibinfo {author} {\bibfnamefont {Z.}~\bibnamefont {Lu}}, \bibinfo
  {author} {\bibfnamefont {B.}~\bibnamefont {Hunt}}, \bibinfo {author}
  {\bibfnamefont {M.}~\bibnamefont {Girvan}},\ and\ \bibinfo {author}
  {\bibfnamefont {E.}~\bibnamefont {Ott}},\ }\bibfield  {title} {\bibinfo
  {title} {Using machine learning to replicate chaotic attractors and calculate
  {Lyapunov} exponents from data},\ }\href@noop {} {\bibfield  {journal}
  {\bibinfo  {journal} {Chaos}\ }\textbf {\bibinfo {volume} {27}},\ \bibinfo
  {pages} {121102} (\bibinfo {year} {2017})}\BibitemShut {NoStop}%
\bibitem [{\citenamefont {Lu}\ \emph {et~al.}(2017)\citenamefont {Lu},
  \citenamefont {Pathak}, \citenamefont {Hunt}, \citenamefont {Girvan},
  \citenamefont {Brockett},\ and\ \citenamefont {Ott}}]{LPHGBO:2017}%
  \BibitemOpen
  \bibfield  {author} {\bibinfo {author} {\bibfnamefont {Z.}~\bibnamefont
  {Lu}}, \bibinfo {author} {\bibfnamefont {J.}~\bibnamefont {Pathak}}, \bibinfo
  {author} {\bibfnamefont {B.}~\bibnamefont {Hunt}}, \bibinfo {author}
  {\bibfnamefont {M.}~\bibnamefont {Girvan}}, \bibinfo {author} {\bibfnamefont
  {R.}~\bibnamefont {Brockett}},\ and\ \bibinfo {author} {\bibfnamefont
  {E.}~\bibnamefont {Ott}},\ }\bibfield  {title} {\bibinfo {title} {Reservoir
  observers: Model-free inference of unmeasured variables in chaotic systems},\
  }\href@noop {} {\bibfield  {journal} {\bibinfo  {journal} {Chaos}\ }\textbf
  {\bibinfo {volume} {27}},\ \bibinfo {pages} {041102} (\bibinfo {year}
  {2017})}\BibitemShut {NoStop}%
\bibitem [{\citenamefont {Pathak}\ \emph
  {et~al.}(2018{\natexlab{a}})\citenamefont {Pathak}, \citenamefont {Hunt},
  \citenamefont {Girvan}, \citenamefont {Lu},\ and\ \citenamefont
  {Ott}}]{PHGLO:2018}%
  \BibitemOpen
  \bibfield  {author} {\bibinfo {author} {\bibfnamefont {J.}~\bibnamefont
  {Pathak}}, \bibinfo {author} {\bibfnamefont {B.}~\bibnamefont {Hunt}},
  \bibinfo {author} {\bibfnamefont {M.}~\bibnamefont {Girvan}}, \bibinfo
  {author} {\bibfnamefont {Z.}~\bibnamefont {Lu}},\ and\ \bibinfo {author}
  {\bibfnamefont {E.}~\bibnamefont {Ott}},\ }\bibfield  {title} {\bibinfo
  {title} {Model-free prediction of large spatiotemporally chaotic systems from
  data: A reservoir computing approach},\ }\href
  {https://doi.org/10.1103/PhysRevLett.120.024102} {\bibfield  {journal}
  {\bibinfo  {journal} {Phys. Rev. Lett.}\ }\textbf {\bibinfo {volume} {120}},\
  \bibinfo {pages} {024102} (\bibinfo {year} {2018}{\natexlab{a}})}\BibitemShut
  {NoStop}%
\bibitem [{\citenamefont {Carroll}(2018)}]{Carroll:2018}%
  \BibitemOpen
  \bibfield  {author} {\bibinfo {author} {\bibfnamefont {T.~L.}\ \bibnamefont
  {Carroll}},\ }\bibfield  {title} {\bibinfo {title} {Using reservoir computers
  to distinguish chaotic signals},\ }\href
  {https://doi.org/10.1103/PhysRevE.98.052209} {\bibfield  {journal} {\bibinfo
  {journal} {Phys. Rev. E}\ }\textbf {\bibinfo {volume} {98}},\ \bibinfo
  {pages} {052209} (\bibinfo {year} {2018})}\BibitemShut {NoStop}%
\bibitem [{\citenamefont {Nakai}\ and\ \citenamefont {Saiki}(2018)}]{NS:2018}%
  \BibitemOpen
  \bibfield  {author} {\bibinfo {author} {\bibfnamefont {K.}~\bibnamefont
  {Nakai}}\ and\ \bibinfo {author} {\bibfnamefont {Y.}~\bibnamefont {Saiki}},\
  }\bibfield  {title} {\bibinfo {title} {Machine-learning inference of fluid
  variables from data using reservoir computing},\ }\href
  {https://doi.org/10.1103/PhysRevE.98.023111} {\bibfield  {journal} {\bibinfo
  {journal} {Phys. Rev. E}\ }\textbf {\bibinfo {volume} {98}},\ \bibinfo
  {pages} {023111} (\bibinfo {year} {2018})}\BibitemShut {NoStop}%
\bibitem [{\citenamefont {Roland}\ and\ \citenamefont
  {Parlitz}(2018)}]{ZP:2018}%
  \BibitemOpen
  \bibfield  {author} {\bibinfo {author} {\bibfnamefont {Z.~S.}\ \bibnamefont
  {Roland}}\ and\ \bibinfo {author} {\bibfnamefont {U.}~\bibnamefont
  {Parlitz}},\ }\bibfield  {title} {\bibinfo {title} {Observing spatio-temporal
  dynamics of excitable media using reservoir computing},\ }\href@noop {}
  {\bibfield  {journal} {\bibinfo  {journal} {Chaos}\ }\textbf {\bibinfo
  {volume} {28}},\ \bibinfo {pages} {043118} (\bibinfo {year}
  {2018})}\BibitemShut {NoStop}%
\bibitem [{\citenamefont {Griffith}\ \emph {et~al.}(2019)\citenamefont
  {Griffith}, \citenamefont {Pomerance},\ and\ \citenamefont
  {Gauthier}}]{GPG:2019}%
  \BibitemOpen
  \bibfield  {author} {\bibinfo {author} {\bibfnamefont {A.}~\bibnamefont
  {Griffith}}, \bibinfo {author} {\bibfnamefont {A.}~\bibnamefont
  {Pomerance}},\ and\ \bibinfo {author} {\bibfnamefont {D.~J.}\ \bibnamefont
  {Gauthier}},\ }\bibfield  {title} {\bibinfo {title} {Forecasting chaotic
  systems with very low connectivity reservoir computers},\ }\href@noop {}
  {\bibfield  {journal} {\bibinfo  {journal} {Chaos}\ }\textbf {\bibinfo
  {volume} {29}},\ \bibinfo {pages} {123108} (\bibinfo {year}
  {2019})}\BibitemShut {NoStop}%
\bibitem [{\citenamefont {Jiang}\ and\ \citenamefont {Lai}(2019)}]{JL:2019}%
  \BibitemOpen
  \bibfield  {author} {\bibinfo {author} {\bibfnamefont {J.}~\bibnamefont
  {Jiang}}\ and\ \bibinfo {author} {\bibfnamefont {Y.-C.}\ \bibnamefont
  {Lai}},\ }\bibfield  {title} {\bibinfo {title} {Model-free prediction of
  spatiotemporal dynamical systems with recurrent neural networks: Role of
  network spectral radius},\ }\href
  {https://doi.org/10.1103/PhysRevResearch.1.033056} {\bibfield  {journal}
  {\bibinfo  {journal} {Phys. Rev. Research}\ }\textbf {\bibinfo {volume}
  {1}},\ \bibinfo {pages} {033056} (\bibinfo {year} {2019})}\BibitemShut
  {NoStop}%
\bibitem [{\citenamefont {Tanaka}\ \emph {et~al.}(2019)\citenamefont {Tanaka},
  \citenamefont {Yamane}, \citenamefont {H{\'e}roux}, \citenamefont {Nakane},
  \citenamefont {Kanazawa}, \citenamefont {Takeda}, \citenamefont {Numata},
  \citenamefont {Nakano},\ and\ \citenamefont {Hirose}}]{TYHNKTNNH:2019}%
  \BibitemOpen
  \bibfield  {author} {\bibinfo {author} {\bibfnamefont {G.}~\bibnamefont
  {Tanaka}}, \bibinfo {author} {\bibfnamefont {T.}~\bibnamefont {Yamane}},
  \bibinfo {author} {\bibfnamefont {J.~B.}\ \bibnamefont {H{\'e}roux}},
  \bibinfo {author} {\bibfnamefont {R.}~\bibnamefont {Nakane}}, \bibinfo
  {author} {\bibfnamefont {N.}~\bibnamefont {Kanazawa}}, \bibinfo {author}
  {\bibfnamefont {S.}~\bibnamefont {Takeda}}, \bibinfo {author} {\bibfnamefont
  {H.}~\bibnamefont {Numata}}, \bibinfo {author} {\bibfnamefont
  {D.}~\bibnamefont {Nakano}},\ and\ \bibinfo {author} {\bibfnamefont
  {A.}~\bibnamefont {Hirose}},\ }\bibfield  {title} {\bibinfo {title} {Recent
  advances in physical reservoir computing: A review},\ }\href@noop {}
  {\bibfield  {journal} {\bibinfo  {journal} {Neu. Net.}\ }\textbf {\bibinfo
  {volume} {115}},\ \bibinfo {pages} {100} (\bibinfo {year}
  {2019})}\BibitemShut {NoStop}%
\bibitem [{\citenamefont {Fan}\ \emph {et~al.}(2020)\citenamefont {Fan},
  \citenamefont {Jiang}, \citenamefont {Zhang}, \citenamefont {Wang},\ and\
  \citenamefont {Lai}}]{FJZWL:2020}%
  \BibitemOpen
  \bibfield  {author} {\bibinfo {author} {\bibfnamefont {H.}~\bibnamefont
  {Fan}}, \bibinfo {author} {\bibfnamefont {J.}~\bibnamefont {Jiang}}, \bibinfo
  {author} {\bibfnamefont {C.}~\bibnamefont {Zhang}}, \bibinfo {author}
  {\bibfnamefont {X.}~\bibnamefont {Wang}},\ and\ \bibinfo {author}
  {\bibfnamefont {Y.-C.}\ \bibnamefont {Lai}},\ }\bibfield  {title} {\bibinfo
  {title} {Long-term prediction of chaotic systems with machine learning},\
  }\href {https://doi.org/10.1103/PhysRevResearch.2.012080} {\bibfield
  {journal} {\bibinfo  {journal} {Phys. Rev. Research}\ }\textbf {\bibinfo
  {volume} {2}},\ \bibinfo {pages} {012080} (\bibinfo {year}
  {2020})}\BibitemShut {NoStop}%
\bibitem [{\citenamefont {Zhang}\ \emph {et~al.}(2020)\citenamefont {Zhang},
  \citenamefont {Jiang}, \citenamefont {Qu},\ and\ \citenamefont
  {Lai}}]{ZJQL:2020}%
  \BibitemOpen
  \bibfield  {author} {\bibinfo {author} {\bibfnamefont {C.}~\bibnamefont
  {Zhang}}, \bibinfo {author} {\bibfnamefont {J.}~\bibnamefont {Jiang}},
  \bibinfo {author} {\bibfnamefont {S.-X.}\ \bibnamefont {Qu}},\ and\ \bibinfo
  {author} {\bibfnamefont {Y.-C.}\ \bibnamefont {Lai}},\ }\bibfield  {title}
  {\bibinfo {title} {Predicting phase and sensing phase coherence in chaotic
  systems with machine learning},\ }\href@noop {} {\bibfield  {journal}
  {\bibinfo  {journal} {Chaos}\ }\textbf {\bibinfo {volume} {30}},\ \bibinfo
  {pages} {083114} (\bibinfo {year} {2020})}\BibitemShut {NoStop}%
\bibitem [{\citenamefont {Klos}\ \emph {et~al.}(2020)\citenamefont {Klos},
  \citenamefont {Kossio}, \citenamefont {Goedeke}, \citenamefont {Gilra},\ and\
  \citenamefont {Memmesheimer}}]{KKGGM:2020}%
  \BibitemOpen
  \bibfield  {author} {\bibinfo {author} {\bibfnamefont {C.}~\bibnamefont
  {Klos}}, \bibinfo {author} {\bibfnamefont {Y.~F.~K.}\ \bibnamefont {Kossio}},
  \bibinfo {author} {\bibfnamefont {S.}~\bibnamefont {Goedeke}}, \bibinfo
  {author} {\bibfnamefont {A.}~\bibnamefont {Gilra}},\ and\ \bibinfo {author}
  {\bibfnamefont {R.-M.}\ \bibnamefont {Memmesheimer}},\ }\bibfield  {title}
  {\bibinfo {title} {Dynamical learning of dynamics},\ }\href@noop {}
  {\bibfield  {journal} {\bibinfo  {journal} {Phys. Rev. Lett.}\ }\textbf
  {\bibinfo {volume} {125}},\ \bibinfo {pages} {088103} (\bibinfo {year}
  {2020})}\BibitemShut {NoStop}%
\bibitem [{\citenamefont {Kong}\ \emph
  {et~al.}(2021{\natexlab{a}})\citenamefont {Kong}, \citenamefont {Fan},
  \citenamefont {Grebogi},\ and\ \citenamefont {Lai}}]{KFGL:2021a}%
  \BibitemOpen
  \bibfield  {author} {\bibinfo {author} {\bibfnamefont {L.-W.}\ \bibnamefont
  {Kong}}, \bibinfo {author} {\bibfnamefont {H.-W.}\ \bibnamefont {Fan}},
  \bibinfo {author} {\bibfnamefont {C.}~\bibnamefont {Grebogi}},\ and\ \bibinfo
  {author} {\bibfnamefont {Y.-C.}\ \bibnamefont {Lai}},\ }\bibfield  {title}
  {\bibinfo {title} {Machine learning prediction of critical transition and
  system collapse},\ }\href {https://doi.org/10.1103/PhysRevResearch.3.013090}
  {\bibfield  {journal} {\bibinfo  {journal} {Phys. Rev. Research}\ }\textbf
  {\bibinfo {volume} {3}},\ \bibinfo {pages} {013090} (\bibinfo {year}
  {2021}{\natexlab{a}})}\BibitemShut {NoStop}%
\bibitem [{\citenamefont {Patel}\ \emph {et~al.}(2021)\citenamefont {Patel},
  \citenamefont {Canaday}, \citenamefont {Girvan}, \citenamefont {Pomerance},\
  and\ \citenamefont {Ott}}]{PCGPO:2021}%
  \BibitemOpen
  \bibfield  {author} {\bibinfo {author} {\bibfnamefont {D.}~\bibnamefont
  {Patel}}, \bibinfo {author} {\bibfnamefont {D.}~\bibnamefont {Canaday}},
  \bibinfo {author} {\bibfnamefont {M.}~\bibnamefont {Girvan}}, \bibinfo
  {author} {\bibfnamefont {A.}~\bibnamefont {Pomerance}},\ and\ \bibinfo
  {author} {\bibfnamefont {E.}~\bibnamefont {Ott}},\ }\bibfield  {title}
  {\bibinfo {title} {Using machine learning to predict statistical properties
  of non-stationary dynamical processes: System climate, regime transitions,
  and the effect of stochasticity},\ }\href@noop {} {\bibfield  {journal}
  {\bibinfo  {journal} {Chaos}\ }\textbf {\bibinfo {volume} {31}},\ \bibinfo
  {pages} {033149} (\bibinfo {year} {2021})}\BibitemShut {NoStop}%
\bibitem [{\citenamefont {Kim}\ \emph {et~al.}(2021)\citenamefont {Kim},
  \citenamefont {Lu}, \citenamefont {Nozari}, \citenamefont {Pappas},\ and\
  \citenamefont {Bassett}}]{KLNPB:2021}%
  \BibitemOpen
  \bibfield  {author} {\bibinfo {author} {\bibfnamefont {J.~Z.}\ \bibnamefont
  {Kim}}, \bibinfo {author} {\bibfnamefont {Z.}~\bibnamefont {Lu}}, \bibinfo
  {author} {\bibfnamefont {E.}~\bibnamefont {Nozari}}, \bibinfo {author}
  {\bibfnamefont {G.~J.}\ \bibnamefont {Pappas}},\ and\ \bibinfo {author}
  {\bibfnamefont {D.~S.}\ \bibnamefont {Bassett}},\ }\bibfield  {title}
  {\bibinfo {title} {Teaching recurrent neural networks to infer global
  temporal structure from local examples},\ }\href@noop {} {\bibfield
  {journal} {\bibinfo  {journal} {Nat. Machine Intell.}\ }\textbf {\bibinfo
  {volume} {3}},\ \bibinfo {pages} {316} (\bibinfo {year} {2021})}\BibitemShut
  {NoStop}%
\bibitem [{\citenamefont {Fan}\ \emph {et~al.}(2021)\citenamefont {Fan},
  \citenamefont {Kong}, \citenamefont {Lai},\ and\ \citenamefont
  {Wang}}]{FKLW:2021}%
  \BibitemOpen
  \bibfield  {author} {\bibinfo {author} {\bibfnamefont {H.}~\bibnamefont
  {Fan}}, \bibinfo {author} {\bibfnamefont {L.-W.}\ \bibnamefont {Kong}},
  \bibinfo {author} {\bibfnamefont {Y.-C.}\ \bibnamefont {Lai}},\ and\ \bibinfo
  {author} {\bibfnamefont {X.}~\bibnamefont {Wang}},\ }\bibfield  {title}
  {\bibinfo {title} {Anticipating synchronization with machine learning},\
  }\href@noop {} {\bibfield  {journal} {\bibinfo  {journal} {Phys. Rev.
  Resesearch}\ }\textbf {\bibinfo {volume} {3}},\ \bibinfo {pages} {023237}
  (\bibinfo {year} {2021})}\BibitemShut {NoStop}%
\bibitem [{\citenamefont {Kong}\ \emph
  {et~al.}(2021{\natexlab{b}})\citenamefont {Kong}, \citenamefont {Fan},
  \citenamefont {Grebogi},\ and\ \citenamefont {Lai}}]{KFGL:2021b}%
  \BibitemOpen
  \bibfield  {author} {\bibinfo {author} {\bibfnamefont {L.-W.}\ \bibnamefont
  {Kong}}, \bibinfo {author} {\bibfnamefont {H.}~\bibnamefont {Fan}}, \bibinfo
  {author} {\bibfnamefont {C.}~\bibnamefont {Grebogi}},\ and\ \bibinfo {author}
  {\bibfnamefont {Y.-C.}\ \bibnamefont {Lai}},\ }\bibfield  {title} {\bibinfo
  {title} {Emergence of transient chaos and intermittency in machine
  learning},\ }\href@noop {} {\bibfield  {journal} {\bibinfo  {journal} {J.
  Phys. Complexity}\ }\textbf {\bibinfo {volume} {2}},\ \bibinfo {pages}
  {035014} (\bibinfo {year} {2021}{\natexlab{b}})}\BibitemShut {NoStop}%
\bibitem [{\citenamefont {Bollt}(2021)}]{Bollt:2021}%
  \BibitemOpen
  \bibfield  {author} {\bibinfo {author} {\bibfnamefont {E.}~\bibnamefont
  {Bollt}},\ }\bibfield  {title} {\bibinfo {title} {On explaining the
  surprising success of reservoir computing forecaster of chaos? the universal
  machine learning dynamical system with contrast to var and dmd},\ }\href@noop
  {} {\bibfield  {journal} {\bibinfo  {journal} {Chaos}\ }\textbf {\bibinfo
  {volume} {31}},\ \bibinfo {pages} {013108} (\bibinfo {year}
  {2021})}\BibitemShut {NoStop}%
\bibitem [{\citenamefont {Gauthier}\ \emph {et~al.}(2021)\citenamefont
  {Gauthier}, \citenamefont {Bollt}, \citenamefont {Griffith},\ and\
  \citenamefont {Barbosa}}]{GBGB:2021}%
  \BibitemOpen
  \bibfield  {author} {\bibinfo {author} {\bibfnamefont {D.~J.}\ \bibnamefont
  {Gauthier}}, \bibinfo {author} {\bibfnamefont {E.}~\bibnamefont {Bollt}},
  \bibinfo {author} {\bibfnamefont {A.}~\bibnamefont {Griffith}},\ and\
  \bibinfo {author} {\bibfnamefont {W.~A.}\ \bibnamefont {Barbosa}},\
  }\bibfield  {title} {\bibinfo {title} {Next generation reservoir computing},\
  }\href@noop {} {\bibfield  {journal} {\bibinfo  {journal} {Nat. Commun.}\
  }\textbf {\bibinfo {volume} {12}},\ \bibinfo {pages} {1} (\bibinfo {year}
  {2021})}\BibitemShut {NoStop}%
\bibitem [{\citenamefont {Haluszczynski}\ and\ \citenamefont
  {R{\"a}th}(2021)}]{HR:2021}%
  \BibitemOpen
  \bibfield  {author} {\bibinfo {author} {\bibfnamefont {A.}~\bibnamefont
  {Haluszczynski}}\ and\ \bibinfo {author} {\bibfnamefont {C.}~\bibnamefont
  {R{\"a}th}},\ }\bibfield  {title} {\bibinfo {title} {Controlling nonlinear
  dynamical systems into arbitrary states using machine learning},\ }\href@noop
  {} {\bibfield  {journal} {\bibinfo  {journal} {Scientific reports}\ }\textbf
  {\bibinfo {volume} {11}},\ \bibinfo {pages} {1} (\bibinfo {year}
  {2021})}\BibitemShut {NoStop}%
\bibitem [{\citenamefont {Carroll}(2022)}]{Carroll:2022optimizing}%
  \BibitemOpen
  \bibfield  {author} {\bibinfo {author} {\bibfnamefont {T.~L.}\ \bibnamefont
  {Carroll}},\ }\bibfield  {title} {\bibinfo {title} {Optimizing memory in
  reservoir computers},\ }\href@noop {} {\bibfield  {journal} {\bibinfo
  {journal} {Chaos}\ }\textbf {\bibinfo {volume} {32}},\ \bibinfo {pages}
  {023123} (\bibinfo {year} {2022})}\BibitemShut {NoStop}%
\bibitem [{\citenamefont {Hart}\ \emph {et~al.}(2020)\citenamefont {Hart},
  \citenamefont {Hook},\ and\ \citenamefont {Dawes}}]{HHD:2020}%
  \BibitemOpen
  \bibfield  {author} {\bibinfo {author} {\bibfnamefont {A.}~\bibnamefont
  {Hart}}, \bibinfo {author} {\bibfnamefont {J.}~\bibnamefont {Hook}},\ and\
  \bibinfo {author} {\bibfnamefont {J.}~\bibnamefont {Dawes}},\ }\bibfield
  {title} {\bibinfo {title} {Embedding and approximation theorems for echo
  state networks},\ }\href@noop {} {\bibfield  {journal} {\bibinfo  {journal}
  {Neu. Net.}\ }\textbf {\bibinfo {volume} {128}},\ \bibinfo {pages} {234}
  (\bibinfo {year} {2020})}\BibitemShut {NoStop}%
\bibitem [{cod()}]{codes}%
  \BibitemOpen
  \href@noop {} {}\bibinfo {note} {The codes of this work are shared at
  \url{github.com/lw-kong/Digital_Twin_2021}}\BibitemShut {NoStop}%
\bibitem [{\citenamefont {Herteux}\ and\ \citenamefont
  {R{\"a}th}(2020)}]{HR:2020}%
  \BibitemOpen
  \bibfield  {author} {\bibinfo {author} {\bibfnamefont {J.}~\bibnamefont
  {Herteux}}\ and\ \bibinfo {author} {\bibfnamefont {C.}~\bibnamefont
  {R{\"a}th}},\ }\bibfield  {title} {\bibinfo {title} {Breaking symmetries of
  the reservoir equations in echo state networks},\ }\href@noop {} {\bibfield
  {journal} {\bibinfo  {journal} {Chaos}\ }\textbf {\bibinfo {volume} {30}},\
  \bibinfo {pages} {123142} (\bibinfo {year} {2020})}\BibitemShut {NoStop}%
\bibitem [{\citenamefont {Goldberg}(2006)}]{Goldberg:2006}%
  \BibitemOpen
  \bibfield  {author} {\bibinfo {author} {\bibfnamefont {D.~E.}\ \bibnamefont
  {Goldberg}},\ }\href@noop {} {\emph {\bibinfo {title} {Genetic Algorithms}}}\
  (\bibinfo  {publisher} {Pearson Education India},\ \bibinfo {year}
  {2006})\BibitemShut {NoStop}%
\bibitem [{\citenamefont {Conn}\ \emph {et~al.}(1991)\citenamefont {Conn},
  \citenamefont {Gould},\ and\ \citenamefont {Toint}}]{CGT:1991}%
  \BibitemOpen
  \bibfield  {author} {\bibinfo {author} {\bibfnamefont {A.~R.}\ \bibnamefont
  {Conn}}, \bibinfo {author} {\bibfnamefont {N.~I.}\ \bibnamefont {Gould}},\
  and\ \bibinfo {author} {\bibfnamefont {P.}~\bibnamefont {Toint}},\ }\bibfield
   {title} {\bibinfo {title} {A globally convergent augmented lagrangian
  algorithm for optimization with general constraints and simple bounds},\
  }\href@noop {} {\bibfield  {journal} {\bibinfo  {journal} {SIAM J. Numer.
  Anal.}\ }\textbf {\bibinfo {volume} {28}},\ \bibinfo {pages} {545} (\bibinfo
  {year} {1991})}\BibitemShut {NoStop}%
\bibitem [{\citenamefont {Conn}\ \emph {et~al.}(1997)\citenamefont {Conn},
  \citenamefont {Gould},\ and\ \citenamefont {Toint}}]{CGT:1997}%
  \BibitemOpen
  \bibfield  {author} {\bibinfo {author} {\bibfnamefont {A.}~\bibnamefont
  {Conn}}, \bibinfo {author} {\bibfnamefont {N.}~\bibnamefont {Gould}},\ and\
  \bibinfo {author} {\bibfnamefont {P.}~\bibnamefont {Toint}},\ }\bibfield
  {title} {\bibinfo {title} {A globally convergent lagrangian barrier algorithm
  for optimization with general inequality constraints and simple bounds},\
  }\href@noop {} {\bibfield  {journal} {\bibinfo  {journal} {Math. Comput.}\
  }\textbf {\bibinfo {volume} {66}},\ \bibinfo {pages} {261} (\bibinfo {year}
  {1997})}\BibitemShut {NoStop}%
\bibitem [{\citenamefont {Kennedy}\ and\ \citenamefont
  {Eberhart}(1995)}]{KE:1995}%
  \BibitemOpen
  \bibfield  {author} {\bibinfo {author} {\bibfnamefont {J.}~\bibnamefont
  {Kennedy}}\ and\ \bibinfo {author} {\bibfnamefont {R.}~\bibnamefont
  {Eberhart}},\ }\bibfield  {title} {\bibinfo {title} {Particle swarm
  optimization},\ }in\ \href@noop {} {\emph {\bibinfo {booktitle} {Proceedings
  of ICNN'95-International Conference on Neural Networks}}},\ Vol.~\bibinfo
  {volume} {4}\ (\bibinfo {organization} {IEEE},\ \bibinfo {year} {1995})\ pp.\
  \bibinfo {pages} {1942--1948}\BibitemShut {NoStop}%
\bibitem [{\citenamefont {Mezura-Montes}\ and\ \citenamefont
  {Coello}(2011)}]{MC:2011}%
  \BibitemOpen
  \bibfield  {author} {\bibinfo {author} {\bibfnamefont {E.}~\bibnamefont
  {Mezura-Montes}}\ and\ \bibinfo {author} {\bibfnamefont {C.~A.~C.}\
  \bibnamefont {Coello}},\ }\bibfield  {title} {\bibinfo {title}
  {Constraint-handling in nature-inspired numerical optimization: past, present
  and future},\ }\href@noop {} {\bibfield  {journal} {\bibinfo  {journal}
  {Swarm Evol. Comput.}\ }\textbf {\bibinfo {volume} {1}},\ \bibinfo {pages}
  {173} (\bibinfo {year} {2011})}\BibitemShut {NoStop}%
\bibitem [{\citenamefont {Gelbart}\ \emph {et~al.}(2014)\citenamefont
  {Gelbart}, \citenamefont {Snoek},\ and\ \citenamefont {Adams}}]{GSA:2014}%
  \BibitemOpen
  \bibfield  {author} {\bibinfo {author} {\bibfnamefont {M.~A.}\ \bibnamefont
  {Gelbart}}, \bibinfo {author} {\bibfnamefont {J.}~\bibnamefont {Snoek}},\
  and\ \bibinfo {author} {\bibfnamefont {R.~P.}\ \bibnamefont {Adams}},\
  }\bibfield  {title} {\bibinfo {title} {Bayesian optimization with unknown
  constraints},\ }\href@noop {} {\bibfield  {journal} {\bibinfo  {journal}
  {arXiv preprint arXiv:1403.5607}\ } (\bibinfo {year} {2014})}\BibitemShut
  {NoStop}%
\bibitem [{\citenamefont {Snoek}\ \emph {et~al.}(2012)\citenamefont {Snoek},
  \citenamefont {Larochelle},\ and\ \citenamefont {Adams}}]{SLA:2012}%
  \BibitemOpen
  \bibfield  {author} {\bibinfo {author} {\bibfnamefont {J.}~\bibnamefont
  {Snoek}}, \bibinfo {author} {\bibfnamefont {H.}~\bibnamefont {Larochelle}},\
  and\ \bibinfo {author} {\bibfnamefont {R.~P.}\ \bibnamefont {Adams}},\
  }\bibfield  {title} {\bibinfo {title} {Practical bayesian optimization of
  machine learning algorithms},\ }in\ \href@noop {} {\emph {\bibinfo
  {booktitle} {NeurIPS}}}\ (\bibinfo {year} {2012})\ pp.\ \bibinfo {pages}
  {2951--2959}\BibitemShut {NoStop}%
\bibitem [{\citenamefont {Gutmann}(2001)}]{Gutmann:2001}%
  \BibitemOpen
  \bibfield  {author} {\bibinfo {author} {\bibfnamefont {H.-M.}\ \bibnamefont
  {Gutmann}},\ }\bibfield  {title} {\bibinfo {title} {A radial basis function
  method for global optimization},\ }\href@noop {} {\bibfield  {journal}
  {\bibinfo  {journal} {J. Global Optim.}\ }\textbf {\bibinfo {volume} {19}},\
  \bibinfo {pages} {201} (\bibinfo {year} {2001})}\BibitemShut {NoStop}%
\bibitem [{\citenamefont {Regis}\ and\ \citenamefont
  {Shoemaker}(2007)}]{RS:2007}%
  \BibitemOpen
  \bibfield  {author} {\bibinfo {author} {\bibfnamefont {R.~G.}\ \bibnamefont
  {Regis}}\ and\ \bibinfo {author} {\bibfnamefont {C.~A.}\ \bibnamefont
  {Shoemaker}},\ }\bibfield  {title} {\bibinfo {title} {A stochastic radial
  basis function method for the global optimization of expensive functions},\
  }\href@noop {} {\bibfield  {journal} {\bibinfo  {journal} {INFORMS J.
  Comput.}\ }\textbf {\bibinfo {volume} {19}},\ \bibinfo {pages} {497}
  (\bibinfo {year} {2007})}\BibitemShut {NoStop}%
\bibitem [{\citenamefont {Wang}\ and\ \citenamefont
  {Shoemaker}(2014)}]{WS:2014}%
  \BibitemOpen
  \bibfield  {author} {\bibinfo {author} {\bibfnamefont {Y.}~\bibnamefont
  {Wang}}\ and\ \bibinfo {author} {\bibfnamefont {C.~A.}\ \bibnamefont
  {Shoemaker}},\ }\bibfield  {title} {\bibinfo {title} {A general stochastic
  algorithmic framework for minimizing expensive black box objective functions
  based on surrogate models and sensitivity analysis},\ }\href@noop {}
  {\bibfield  {journal} {\bibinfo  {journal} {arXiv preprint arXiv:1410.6271}\
  } (\bibinfo {year} {2014})}\BibitemShut {NoStop}%
\bibitem [{\citenamefont {Lorenz}(1996)}]{Lorenz:1996}%
  \BibitemOpen
  \bibfield  {author} {\bibinfo {author} {\bibfnamefont {E.~N.}\ \bibnamefont
  {Lorenz}},\ }\bibfield  {title} {\bibinfo {title} {Predictability: A problem
  partly solved},\ }in\ \href@noop {} {\emph {\bibinfo {booktitle} {Proc.
  Seminar on Predictability}}},\ Vol.~\bibinfo {volume} {1}\ (\bibinfo {year}
  {1996})\BibitemShut {NoStop}%
\bibitem [{\citenamefont {Van~den Broeck}\ \emph {et~al.}(1997)\citenamefont
  {Van~den Broeck}, \citenamefont {Parrondo}, \citenamefont {Toral},\ and\
  \citenamefont {Kawai}}]{BPTK:1997}%
  \BibitemOpen
  \bibfield  {author} {\bibinfo {author} {\bibfnamefont {C.}~\bibnamefont
  {Van~den Broeck}}, \bibinfo {author} {\bibfnamefont {J.}~\bibnamefont
  {Parrondo}}, \bibinfo {author} {\bibfnamefont {R.}~\bibnamefont {Toral}},\
  and\ \bibinfo {author} {\bibfnamefont {R.}~\bibnamefont {Kawai}},\ }\bibfield
   {title} {\bibinfo {title} {Nonequilibrium phase transitions induced by
  multiplicative noise},\ }\href@noop {} {\bibfield  {journal} {\bibinfo
  {journal} {Phys. Rev. E}\ }\textbf {\bibinfo {volume} {55}},\ \bibinfo
  {pages} {4084} (\bibinfo {year} {1997})}\BibitemShut {NoStop}%
\bibitem [{\citenamefont {Sussillo}\ and\ \citenamefont
  {Abbott}(2009)}]{SA:2009}%
  \BibitemOpen
  \bibfield  {author} {\bibinfo {author} {\bibfnamefont {D.}~\bibnamefont
  {Sussillo}}\ and\ \bibinfo {author} {\bibfnamefont {L.~F.}\ \bibnamefont
  {Abbott}},\ }\bibfield  {title} {\bibinfo {title} {Generating coherent
  patterns of activity from chaotic neural networks},\ }\href@noop {}
  {\bibfield  {journal} {\bibinfo  {journal} {Neuron}\ }\textbf {\bibinfo
  {volume} {63}},\ \bibinfo {pages} {544} (\bibinfo {year} {2009})}\BibitemShut
  {NoStop}%
\bibitem [{\citenamefont {Kobayashi}\ and\ \citenamefont
  {Sugino}(2019)}]{KS:2019}%
  \BibitemOpen
  \bibfield  {author} {\bibinfo {author} {\bibfnamefont {T.}~\bibnamefont
  {Kobayashi}}\ and\ \bibinfo {author} {\bibfnamefont {T.}~\bibnamefont
  {Sugino}},\ }\bibfield  {title} {\bibinfo {title} {Continual learning
  exploiting structure of fractal reservoir computing},\ }in\ \href@noop {}
  {\emph {\bibinfo {booktitle} {International Conference on Artificial Neural
  Networks}}}\ (\bibinfo {organization} {Springer},\ \bibinfo {year} {2019})\
  pp.\ \bibinfo {pages} {35--47}\BibitemShut {NoStop}%
\bibitem [{\citenamefont {Pathak}\ \emph
  {et~al.}(2018{\natexlab{b}})\citenamefont {Pathak}, \citenamefont {Wikner},
  \citenamefont {Fussell}, \citenamefont {Chandra}, \citenamefont {Hunt},
  \citenamefont {Girvan},\ and\ \citenamefont {Ott}}]{PWFCHGO:2018}%
  \BibitemOpen
  \bibfield  {author} {\bibinfo {author} {\bibfnamefont {J.}~\bibnamefont
  {Pathak}}, \bibinfo {author} {\bibfnamefont {A.}~\bibnamefont {Wikner}},
  \bibinfo {author} {\bibfnamefont {R.}~\bibnamefont {Fussell}}, \bibinfo
  {author} {\bibfnamefont {S.}~\bibnamefont {Chandra}}, \bibinfo {author}
  {\bibfnamefont {B.~R.}\ \bibnamefont {Hunt}}, \bibinfo {author}
  {\bibfnamefont {M.}~\bibnamefont {Girvan}},\ and\ \bibinfo {author}
  {\bibfnamefont {E.}~\bibnamefont {Ott}},\ }\bibfield  {title} {\bibinfo
  {title} {Hybrid forecasting of chaotic processes: Using machine learning in
  conjunction with a knowledge-based model},\ }\href@noop {} {\bibfield
  {journal} {\bibinfo  {journal} {Chaos}\ }\textbf {\bibinfo {volume} {28}},\
  \bibinfo {pages} {041101} (\bibinfo {year} {2018}{\natexlab{b}})}\BibitemShut
  {NoStop}%
\bibitem [{\citenamefont {Chen}\ \emph {et~al.}(2018)\citenamefont {Chen},
  \citenamefont {Rubanova}, \citenamefont {Bettencourt},\ and\ \citenamefont
  {Duvenaud}}]{CRBD:2018}%
  \BibitemOpen
  \bibfield  {author} {\bibinfo {author} {\bibfnamefont {R.~T.}\ \bibnamefont
  {Chen}}, \bibinfo {author} {\bibfnamefont {Y.}~\bibnamefont {Rubanova}},
  \bibinfo {author} {\bibfnamefont {J.}~\bibnamefont {Bettencourt}},\ and\
  \bibinfo {author} {\bibfnamefont {D.~K.}\ \bibnamefont {Duvenaud}},\
  }\bibfield  {title} {\bibinfo {title} {Neural ordinary differential
  equations},\ }\href@noop {} {\bibfield  {journal} {\bibinfo  {journal} {Adv.
  Neu. Info. Proc. Sys.}\ }\textbf {\bibinfo {volume} {31}} (\bibinfo {year}
  {2018})}\BibitemShut {NoStop}%
\bibitem [{\citenamefont {Berkenkamp}\ \emph {et~al.}(2017)\citenamefont
  {Berkenkamp}, \citenamefont {Turchetta}, \citenamefont {Schoellig},\ and\
  \citenamefont {Krause}}]{BTSK:2017}%
  \BibitemOpen
  \bibfield  {author} {\bibinfo {author} {\bibfnamefont {F.}~\bibnamefont
  {Berkenkamp}}, \bibinfo {author} {\bibfnamefont {M.}~\bibnamefont
  {Turchetta}}, \bibinfo {author} {\bibfnamefont {A.~P.}\ \bibnamefont
  {Schoellig}},\ and\ \bibinfo {author} {\bibfnamefont {A.}~\bibnamefont
  {Krause}},\ }\bibfield  {title} {\bibinfo {title} {Safe model-based
  reinforcement learning with stability guarantees},\ }\href@noop {} {\bibfield
   {journal} {\bibinfo  {journal} {arXiv preprint arXiv:1705.08551}\ }
  (\bibinfo {year} {2017})}\BibitemShut {NoStop}%
\bibitem [{\citenamefont {Moerland}\ \emph {et~al.}(2020)\citenamefont
  {Moerland}, \citenamefont {Broekens},\ and\ \citenamefont
  {Jonker}}]{MBJ:2020}%
  \BibitemOpen
  \bibfield  {author} {\bibinfo {author} {\bibfnamefont {T.~M.}\ \bibnamefont
  {Moerland}}, \bibinfo {author} {\bibfnamefont {J.}~\bibnamefont {Broekens}},\
  and\ \bibinfo {author} {\bibfnamefont {C.~M.}\ \bibnamefont {Jonker}},\
  }\bibfield  {title} {\bibinfo {title} {Model-based reinforcement learning: A
  survey},\ }\href@noop {} {\bibfield  {journal} {\bibinfo  {journal} {arXiv
  preprint arXiv:2006.16712}\ } (\bibinfo {year} {2020})}\BibitemShut {NoStop}%
\bibitem [{\citenamefont {Kuramoto}\ and\ \citenamefont
  {Battogtokh}(2002)}]{KB:2002}%
  \BibitemOpen
  \bibfield  {author} {\bibinfo {author} {\bibfnamefont {Y.}~\bibnamefont
  {Kuramoto}}\ and\ \bibinfo {author} {\bibfnamefont {D.}~\bibnamefont
  {Battogtokh}},\ }\bibfield  {title} {\bibinfo {title} {Coexistence of
  coherence and incoherence in nonlocally coupled phase oscillators},\
  }\href@noop {} {\bibfield  {journal} {\bibinfo  {journal} {Nonlin. Phenom.
  Complex Syst.}\ }\textbf {\bibinfo {volume} {5}},\ \bibinfo {pages} {380}
  (\bibinfo {year} {2002})}\BibitemShut {NoStop}%
\bibitem [{\citenamefont {Abrams}\ and\ \citenamefont
  {Strogatz}(2004)}]{AS:2004}%
  \BibitemOpen
  \bibfield  {author} {\bibinfo {author} {\bibfnamefont {D.~M.}\ \bibnamefont
  {Abrams}}\ and\ \bibinfo {author} {\bibfnamefont {S.~H.}\ \bibnamefont
  {Strogatz}},\ }\bibfield  {title} {\bibinfo {title} {Chimera states for
  coupled oscillators},\ }\href {https://doi.org/10.1103/PhysRevLett.93.174102}
  {\bibfield  {journal} {\bibinfo  {journal} {Phys. Rev. Lett.}\ }\textbf
  {\bibinfo {volume} {93}},\ \bibinfo {pages} {174102} (\bibinfo {year}
  {2004})}\BibitemShut {NoStop}%
\bibitem [{\citenamefont {Omelchenko}\ \emph {et~al.}(2011)\citenamefont
  {Omelchenko}, \citenamefont {Maistrenko}, \citenamefont {H\"ovel},\ and\
  \citenamefont {Sch\"oll}}]{OMHS:2011}%
  \BibitemOpen
  \bibfield  {author} {\bibinfo {author} {\bibfnamefont {I.}~\bibnamefont
  {Omelchenko}}, \bibinfo {author} {\bibfnamefont {Y.}~\bibnamefont
  {Maistrenko}}, \bibinfo {author} {\bibfnamefont {P.}~\bibnamefont
  {H\"ovel}},\ and\ \bibinfo {author} {\bibfnamefont {E.}~\bibnamefont
  {Sch\"oll}},\ }\bibfield  {title} {\bibinfo {title} {Loss of coherence in
  dynamical networks: Spatial chaos and chimera states},\ }\href
  {https://doi.org/10.1103/PhysRevLett.106.234102} {\bibfield  {journal}
  {\bibinfo  {journal} {Phys. Rev. Lett.}\ }\textbf {\bibinfo {volume} {106}},\
  \bibinfo {pages} {234102} (\bibinfo {year} {2011})}\BibitemShut {NoStop}%
\bibitem [{\citenamefont {Tinsley}\ \emph {et~al.}(2012)\citenamefont
  {Tinsley}, \citenamefont {Nkomo},\ and\ \citenamefont
  {Showalter}}]{TNS:2012}%
  \BibitemOpen
  \bibfield  {author} {\bibinfo {author} {\bibfnamefont {M.~R.}\ \bibnamefont
  {Tinsley}}, \bibinfo {author} {\bibfnamefont {S.}~\bibnamefont {Nkomo}},\
  and\ \bibinfo {author} {\bibfnamefont {K.}~\bibnamefont {Showalter}},\
  }\bibfield  {title} {\bibinfo {title} {Chimera and phase-cluster states in
  populations of coupled chemical oscillators},\ }\href@noop {} {\bibfield
  {journal} {\bibinfo  {journal} {Nat. Phys.}\ }\textbf {\bibinfo {volume}
  {8}},\ \bibinfo {pages} {662} (\bibinfo {year} {2012})}\BibitemShut {NoStop}%
\bibitem [{\citenamefont {Hagerstrom}\ \emph {et~al.}(2012)\citenamefont
  {Hagerstrom}, \citenamefont {Murphy}, \citenamefont {Roy}, \citenamefont
  {H\"{o}vel}, \citenamefont {Omelchenko},\ and\ \citenamefont
  {Sch{\"o}ll}}]{HMRHOS:2012}%
  \BibitemOpen
  \bibfield  {author} {\bibinfo {author} {\bibfnamefont {A.~M.}\ \bibnamefont
  {Hagerstrom}}, \bibinfo {author} {\bibfnamefont {T.~E.}\ \bibnamefont
  {Murphy}}, \bibinfo {author} {\bibfnamefont {R.}~\bibnamefont {Roy}},
  \bibinfo {author} {\bibfnamefont {P.}~\bibnamefont {H\"{o}vel}}, \bibinfo
  {author} {\bibfnamefont {I.}~\bibnamefont {Omelchenko}},\ and\ \bibinfo
  {author} {\bibfnamefont {E.}~\bibnamefont {Sch{\"o}ll}},\ }\bibfield  {title}
  {\bibinfo {title} {Experimental observation of chimeras in coupled-map
  lattices},\ }\href@noop {} {\bibfield  {journal} {\bibinfo  {journal} {Nat.
  Phys.}\ }\textbf {\bibinfo {volume} {8}},\ \bibinfo {pages} {658} (\bibinfo
  {year} {2012})}\BibitemShut {NoStop}%
\bibitem [{\citenamefont {Omelchenko}\ \emph {et~al.}(2013)\citenamefont
  {Omelchenko}, \citenamefont {Omel'chenko}, \citenamefont {H\"ovel},\ and\
  \citenamefont {Sch\"oll}}]{OOHS:2013}%
  \BibitemOpen
  \bibfield  {author} {\bibinfo {author} {\bibfnamefont {I.}~\bibnamefont
  {Omelchenko}}, \bibinfo {author} {\bibfnamefont {O.~E.}\ \bibnamefont
  {Omel'chenko}}, \bibinfo {author} {\bibfnamefont {P.}~\bibnamefont
  {H\"ovel}},\ and\ \bibinfo {author} {\bibfnamefont {E.}~\bibnamefont
  {Sch\"oll}},\ }\bibfield  {title} {\bibinfo {title} {When nonlocal coupling
  between oscillators becomes stronger: Patched synchrony or multichimera
  states},\ }\href {https://doi.org/10.1103/PhysRevLett.110.224101} {\bibfield
  {journal} {\bibinfo  {journal} {Phys. Rev. Lett.}\ }\textbf {\bibinfo
  {volume} {110}},\ \bibinfo {pages} {224101} (\bibinfo {year}
  {2013})}\BibitemShut {NoStop}%
\bibitem [{\citenamefont {Omelchenko}\ \emph {et~al.}(2015)\citenamefont
  {Omelchenko}, \citenamefont {Zakharova}, \citenamefont {H\"{o}vel},
  \citenamefont {Siebert},\ and\ \citenamefont {Sch\"{o}ll}}]{OZHSS:2015}%
  \BibitemOpen
  \bibfield  {author} {\bibinfo {author} {\bibfnamefont {I.}~\bibnamefont
  {Omelchenko}}, \bibinfo {author} {\bibfnamefont {A.}~\bibnamefont
  {Zakharova}}, \bibinfo {author} {\bibfnamefont {P.}~\bibnamefont
  {H\"{o}vel}}, \bibinfo {author} {\bibfnamefont {J.}~\bibnamefont {Siebert}},\
  and\ \bibinfo {author} {\bibfnamefont {E.}~\bibnamefont {Sch\"{o}ll}},\
  }\bibfield  {title} {\bibinfo {title} {Nonlinearity of local dynamics
  promotes multi-chimeras},\ }\href@noop {} {\bibfield  {journal} {\bibinfo
  {journal} {Chaos}\ }\textbf {\bibinfo {volume} {25}},\ \bibinfo {pages}
  {083104} (\bibinfo {year} {2015})}\BibitemShut {NoStop}%
\bibitem [{\citenamefont {Omelchenko}\ \emph {et~al.}(2018)\citenamefont
  {Omelchenko}, \citenamefont {Omel'chenko}, \citenamefont {Zakharova},\ and\
  \citenamefont {Sch\"oll}}]{OOZS:2018}%
  \BibitemOpen
  \bibfield  {author} {\bibinfo {author} {\bibfnamefont {I.}~\bibnamefont
  {Omelchenko}}, \bibinfo {author} {\bibfnamefont {O.~E.}\ \bibnamefont
  {Omel'chenko}}, \bibinfo {author} {\bibfnamefont {A.}~\bibnamefont
  {Zakharova}},\ and\ \bibinfo {author} {\bibfnamefont {E.}~\bibnamefont
  {Sch\"oll}},\ }\bibfield  {title} {\bibinfo {title} {Optimal design of
  tweezer control for chimera states},\ }\href
  {https://doi.org/10.1103/PhysRevE.97.012216} {\bibfield  {journal} {\bibinfo
  {journal} {Phys. Rev. E}\ }\textbf {\bibinfo {volume} {97}},\ \bibinfo
  {pages} {012216} (\bibinfo {year} {2018})}\BibitemShut {NoStop}%
\bibitem [{\citenamefont {Kong}\ and\ \citenamefont {Lai}(2020)}]{KL:2020}%
  \BibitemOpen
  \bibfield  {author} {\bibinfo {author} {\bibfnamefont {L.-W.}\ \bibnamefont
  {Kong}}\ and\ \bibinfo {author} {\bibfnamefont {Y.-C.}\ \bibnamefont {Lai}},\
  }\bibfield  {title} {\bibinfo {title} {Scaling law of transient lifetime of
  chimera states under dimension-augmenting perturbations},\ }\href
  {https://doi.org/10.1103/PhysRevResearch.2.023196} {\bibfield  {journal}
  {\bibinfo  {journal} {Phys. Rev. Research}\ }\textbf {\bibinfo {volume}
  {2}},\ \bibinfo {pages} {023196} (\bibinfo {year} {2020})}\BibitemShut
  {NoStop}%
\bibitem [{\citenamefont {Dangoisse}\ \emph {et~al.}(1986)\citenamefont
  {Dangoisse}, \citenamefont {Glorieux},\ and\ \citenamefont
  {Hennequin}}]{DGH:1986}%
  \BibitemOpen
  \bibfield  {author} {\bibinfo {author} {\bibfnamefont {D.}~\bibnamefont
  {Dangoisse}}, \bibinfo {author} {\bibfnamefont {P.}~\bibnamefont
  {Glorieux}},\ and\ \bibinfo {author} {\bibfnamefont {D.}~\bibnamefont
  {Hennequin}},\ }\bibfield  {title} {\bibinfo {title} {Laser chaotic
  attractors in crisis},\ }\href@noop {} {\bibfield  {journal} {\bibinfo
  {journal} {Phys. Rev. Lett.}\ }\textbf {\bibinfo {volume} {57}},\ \bibinfo
  {pages} {2657} (\bibinfo {year} {1986})}\BibitemShut {NoStop}%
\bibitem [{\citenamefont {Dangoisse}\ \emph {et~al.}(1987)\citenamefont
  {Dangoisse}, \citenamefont {Glorieux},\ and\ \citenamefont
  {Hennequin}}]{DGH:1987}%
  \BibitemOpen
  \bibfield  {author} {\bibinfo {author} {\bibfnamefont {D.}~\bibnamefont
  {Dangoisse}}, \bibinfo {author} {\bibfnamefont {P.}~\bibnamefont
  {Glorieux}},\ and\ \bibinfo {author} {\bibfnamefont {D.}~\bibnamefont
  {Hennequin}},\ }\bibfield  {title} {\bibinfo {title} {Chaos in a {CO$_2$}
  laser with modulated parameters: experiments and numerical simulations},\
  }\href@noop {} {\bibfield  {journal} {\bibinfo  {journal} {Phys. Rev. A}\
  }\textbf {\bibinfo {volume} {36}},\ \bibinfo {pages} {4775} (\bibinfo {year}
  {1987})}\BibitemShut {NoStop}%
\bibitem [{\citenamefont {Solari}\ \emph {et~al.}(1987)\citenamefont {Solari},
  \citenamefont {Eschenazi}, \citenamefont {Gilmore},\ and\ \citenamefont
  {Tredicce}}]{SEGT:1987}%
  \BibitemOpen
  \bibfield  {author} {\bibinfo {author} {\bibfnamefont {H.~G.}\ \bibnamefont
  {Solari}}, \bibinfo {author} {\bibfnamefont {E.}~\bibnamefont {Eschenazi}},
  \bibinfo {author} {\bibfnamefont {R.}~\bibnamefont {Gilmore}},\ and\ \bibinfo
  {author} {\bibfnamefont {J.~R.}\ \bibnamefont {Tredicce}},\ }\bibfield
  {title} {\bibinfo {title} {Influence of coexisting attractors on the dynamics
  of a laser system},\ }\href@noop {} {\bibfield  {journal} {\bibinfo
  {journal} {Opt. Commun.}\ }\textbf {\bibinfo {volume} {64}},\ \bibinfo
  {pages} {49} (\bibinfo {year} {1987})}\BibitemShut {NoStop}%
\bibitem [{\citenamefont {Schwartz}(1988)}]{Schwartz:1988}%
  \BibitemOpen
  \bibfield  {author} {\bibinfo {author} {\bibfnamefont {I.~B.}\ \bibnamefont
  {Schwartz}},\ }\bibfield  {title} {\bibinfo {title} {Sequential horseshoe
  formation in the birth and death of chaotic attractors},\ }\href@noop {}
  {\bibfield  {journal} {\bibinfo  {journal} {Phys. Rev. Lett.}\ }\textbf
  {\bibinfo {volume} {60}},\ \bibinfo {pages} {1359} (\bibinfo {year}
  {1988})}\BibitemShut {NoStop}%
\bibitem [{\citenamefont {Grebogi}\ \emph {et~al.}(1983)\citenamefont
  {Grebogi}, \citenamefont {Ott},\ and\ \citenamefont {Yorke}}]{GOY:1983}%
  \BibitemOpen
  \bibfield  {author} {\bibinfo {author} {\bibfnamefont {C.}~\bibnamefont
  {Grebogi}}, \bibinfo {author} {\bibfnamefont {E.}~\bibnamefont {Ott}},\ and\
  \bibinfo {author} {\bibfnamefont {J.~A.}\ \bibnamefont {Yorke}},\ }\bibfield
  {title} {\bibinfo {title} {Crises, sudden changes in chaotic attractors and
  chaotic transients},\ }\href@noop {} {\bibfield  {journal} {\bibinfo
  {journal} {Physica D}\ }\textbf {\bibinfo {volume} {7}},\ \bibinfo {pages}
  {181} (\bibinfo {year} {1983})}\BibitemShut {NoStop}%
\bibitem [{\citenamefont {Huppert}\ \emph {et~al.}(2005)\citenamefont
  {Huppert}, \citenamefont {Blasius}, \citenamefont {Olinky},\ and\
  \citenamefont {Stone}}]{HBOS:2005}%
  \BibitemOpen
  \bibfield  {author} {\bibinfo {author} {\bibfnamefont {A.}~\bibnamefont
  {Huppert}}, \bibinfo {author} {\bibfnamefont {B.}~\bibnamefont {Blasius}},
  \bibinfo {author} {\bibfnamefont {R.}~\bibnamefont {Olinky}},\ and\ \bibinfo
  {author} {\bibfnamefont {L.}~\bibnamefont {Stone}},\ }\bibfield  {title}
  {\bibinfo {title} {A model for seasonal phytoplankton blooms},\ }\href@noop
  {} {\bibfield  {journal} {\bibinfo  {journal} {J. Theo. Biol.}\ }\textbf
  {\bibinfo {volume} {236}},\ \bibinfo {pages} {276} (\bibinfo {year}
  {2005})}\BibitemShut {NoStop}%
\bibitem [{\citenamefont {Stone}\ \emph {et~al.}(2007)\citenamefont {Stone},
  \citenamefont {Olinky},\ and\ \citenamefont {Huppert}}]{SOH:2007}%
  \BibitemOpen
  \bibfield  {author} {\bibinfo {author} {\bibfnamefont {L.}~\bibnamefont
  {Stone}}, \bibinfo {author} {\bibfnamefont {R.}~\bibnamefont {Olinky}},\ and\
  \bibinfo {author} {\bibfnamefont {A.}~\bibnamefont {Huppert}},\ }\bibfield
  {title} {\bibinfo {title} {Seasonal dynamics of recurrent epidemics},\
  }\href@noop {} {\bibfield  {journal} {\bibinfo  {journal} {Nature}\ }\textbf
  {\bibinfo {volume} {446}},\ \bibinfo {pages} {533} (\bibinfo {year}
  {2007})}\BibitemShut {NoStop}%
\bibitem [{\citenamefont {Winder}\ and\ \citenamefont
  {Sommer}(2012)}]{WS:2012}%
  \BibitemOpen
  \bibfield  {author} {\bibinfo {author} {\bibfnamefont {M.}~\bibnamefont
  {Winder}}\ and\ \bibinfo {author} {\bibfnamefont {U.}~\bibnamefont
  {Sommer}},\ }\bibfield  {title} {\bibinfo {title} {Phytoplankton response to
  a changing climate},\ }\href@noop {} {\bibfield  {journal} {\bibinfo
  {journal} {Hydrobiologia}\ }\textbf {\bibinfo {volume} {698}},\ \bibinfo
  {pages} {5} (\bibinfo {year} {2012})}\BibitemShut {NoStop}%
\bibitem [{\citenamefont {Kalnay}(2003)}]{Kalnay:2003}%
  \BibitemOpen
  \bibfield  {author} {\bibinfo {author} {\bibfnamefont {E.}~\bibnamefont
  {Kalnay}},\ }\href@noop {} {\emph {\bibinfo {title} {Atmospheric Modeling,
  Data Assimilation and Predictability}}}\ (\bibinfo  {publisher} {Cambridge
  university press},\ \bibinfo {year} {2003})\BibitemShut {NoStop}%
\bibitem [{\citenamefont {Asch}\ \emph {et~al.}(2016)\citenamefont {Asch},
  \citenamefont {Bocquet},\ and\ \citenamefont {Nodet}}]{ABN:2016}%
  \BibitemOpen
  \bibfield  {author} {\bibinfo {author} {\bibfnamefont {M.}~\bibnamefont
  {Asch}}, \bibinfo {author} {\bibfnamefont {M.}~\bibnamefont {Bocquet}},\ and\
  \bibinfo {author} {\bibfnamefont {M.}~\bibnamefont {Nodet}},\ }\href@noop {}
  {\emph {\bibinfo {title} {Data Assimilation: Methods, Algorithms, and
  Applications}}}\ (\bibinfo  {publisher} {SIAM},\ \bibinfo {year}
  {2016})\BibitemShut {NoStop}%
\bibitem [{\citenamefont {Wikner}\ \emph {et~al.}(2021)\citenamefont {Wikner},
  \citenamefont {Pathak}, \citenamefont {Hunt}, \citenamefont {Szunyogh},
  \citenamefont {Girvan},\ and\ \citenamefont {Ott}}]{WPHSGO:2021}%
  \BibitemOpen
  \bibfield  {author} {\bibinfo {author} {\bibfnamefont {A.}~\bibnamefont
  {Wikner}}, \bibinfo {author} {\bibfnamefont {J.}~\bibnamefont {Pathak}},
  \bibinfo {author} {\bibfnamefont {B.~R.}\ \bibnamefont {Hunt}}, \bibinfo
  {author} {\bibfnamefont {I.}~\bibnamefont {Szunyogh}}, \bibinfo {author}
  {\bibfnamefont {M.}~\bibnamefont {Girvan}},\ and\ \bibinfo {author}
  {\bibfnamefont {E.}~\bibnamefont {Ott}},\ }\bibfield  {title} {\bibinfo
  {title} {Using data assimilation to train a hybrid forecast system that
  combines machine-learning and knowledge-based components},\ }\href@noop {}
  {\bibfield  {journal} {\bibinfo  {journal} {Chaos}\ }\textbf {\bibinfo
  {volume} {31}},\ \bibinfo {pages} {053114} (\bibinfo {year}
  {2021})}\BibitemShut {NoStop}%
\bibitem [{\citenamefont {Weng}\ \emph {et~al.}(2019)\citenamefont {Weng},
  \citenamefont {Yang}, \citenamefont {Gu}, \citenamefont {Zhang},\ and\
  \citenamefont {Small}}]{WYGZS:2019}%
  \BibitemOpen
  \bibfield  {author} {\bibinfo {author} {\bibfnamefont {T.}~\bibnamefont
  {Weng}}, \bibinfo {author} {\bibfnamefont {H.}~\bibnamefont {Yang}}, \bibinfo
  {author} {\bibfnamefont {C.}~\bibnamefont {Gu}}, \bibinfo {author}
  {\bibfnamefont {J.}~\bibnamefont {Zhang}},\ and\ \bibinfo {author}
  {\bibfnamefont {M.}~\bibnamefont {Small}},\ }\bibfield  {title} {\bibinfo
  {title} {Synchronization of chaotic systems and their machine-learning
  models},\ }\href {https://doi.org/10.1103/PhysRevE.99.042203} {\bibfield
  {journal} {\bibinfo  {journal} {Phys. Rev. E}\ }\textbf {\bibinfo {volume}
  {99}},\ \bibinfo {pages} {042203} (\bibinfo {year} {2019})}\BibitemShut
  {NoStop}%
\end{thebibliography}

\end{document}